\documentclass{jfm}
\usepackage{graphicx}
\usepackage{epstopdf, epsfig}
\graphicspath{{./Figures/}}
\usepackage{amsmath}
\usepackage{amsfonts}
\usepackage{amssymb}
\usepackage{natbib}

\newcommand{\devtilde}[1]{\widetilde{#1}} 

\newcommand*{\mytriangle}{\mathrel{\scalebox{0.9}{$\blacktriangle$}}}

\newcommand*{\mydiamond}{\mathrel{\scalebox{0.7}{$\blacklozenge$}}}

\DeclareMathOperator\rect{rect}

\newcommand{\velamp}{\boldsymbol{\mathsf{U}}}
\newcommand{\velampx}{{\mathsf{U}_1}}
\newcommand{\velampy}{{\mathsf{U}_2}}
\newcommand{\pressamp}{\mathsf{P}}

\newcommand{\closurevarvelperiodic}{{\boldsymbol{\Phi}^{\star}}}
\newcommand{\closurevarpressperiodic}{{\boldsymbol{\varphi}^{\star}}}
\newcommand{\closurevarvelperiodicf}{{\boldsymbol{\Phi}_f^{\star}}}
\newcommand{\closurevarpressperiodicf}{{\boldsymbol{\varphi}_f^{\star}}}

\newcommand{\devclosurevarvelperiodicf}{{\devtilde{\boldsymbol{\Phi}}_f^{\star}}}
\newcommand{\devclosurevarpressperiodicf}{{\devtilde{\boldsymbol{\varphi}}_f^{\star}}}

\newcommand{\devclosurevarvel}{{\devtilde{\boldsymbol{\Phi}}}}
\newcommand{\devclosurevarpress}{{\devtilde{\boldsymbol{\varphi}}}}
\newcommand{\devclosurevarvelf}{{\devtilde{\boldsymbol{\Phi}}_f}}
\newcommand{\devclosurevarpressf}{{\devtilde{\boldsymbol{\varphi}}_f}}

\def\Rey{\mbox{\it Re}}   
\title[]{A macro-scale description of quasi-periodically developed flow in channels with arrays of in-line square cylinders}
\author{G. Buckinx \aff{1} \aff{2} \aff{3}
        \corresp{\email{geert.buckinx@kuleuven.be}}
}
\affiliation{
\aff{1} Department of Mechanical Engineering, KU Leuven, Celestijnenlaan 300A, 3001 Leuven, Belgium
\aff{2} VITO, Boeretang 200, 2400 Mol, Belgium
\aff{3} EnergyVille, Thor Park, 3600 Genk, Belgium
}
\pubyear{}
\volume{x}
\date{December 2022}

\begin{document}

\maketitle

\begin{abstract}
We present a macro-scale description of quasi-periodically developed flow in channels, which relies on double volume-averaging.
We show that quasi-developed macro-scale flow is characterized by  velocity modes which decay exponentially in the main flow direction.
We prove that the closure force can be represented by an exact permeability tensor consisting of two parts.
The first part, which is due to the developed macro-scale flow, is uniform everywhere, except in the side-wall region, where it is affected by the macro-scale velocity profile and its slip length.
The second part expresses the resistance against the velocity mode, so it decays exponentially as the flow develops.
It satisfies a specific closure problem on a transversal row of the array.
From these properties, we assess the validity of the classical closure problem for the volume-averaged flow equations.
We show that all its underlying assumptions are partly violated by an exponentially vanishing error during flow development.
Furthermore, we show that it modifies the eigenvalues, modes, and onset point of quasi-developed flow, when it is applied to reconstruct the macro-scale flow.
The former theoretical aspects are illustrated for high-aspect-ratio channels with high-porosity arrays of equidistant in-line square cylinders, by means of direct numerical simulation and explicit filtering of the flow.
In particular, we present extensive solutions of the classical closure problem for Reynolds numbers up to 600, porosities between 0.2 and 0.95, and flow directions between 0 and 45 degrees, though the channel height has been kept equal to the cylinder spacing.
These closure solutions are compared with the actual closure force in channels with cylinder arrays of a porosity between 0.75 and 0.94, for Reynolds numbers up to 300.
\end{abstract}


\section{Introduction}
Steady laminar flow in channels containing an array of periodic solid structures has been of interest for different research domains. 
On the one hand, research on the topic has been driven by technological applications like compact heat transfer devices, in which arrays of periodic fins are employed to increase the heat transfer performance.
On the other hand, the topic has been extensively studied as an idealization of the flow through more complex disordered porous media.
Especially the last two decades, the topic has gained renewed interest due the development of microfluidic devices  (\cite{KosarMishraPeles2005}) and ordered microporous materials.
The latter applications often consist of hundreds of circular or square cylinders in a periodic configuration, with a diameter of 10 to 1000 $\mu$m, and a relatively high porosity between $0.3$ and $0.95$ (\cite{SiuHoQuPfefferkorn2007, MohammadiKosar2018}).
Usually, the flow through such arrays of periodic solid structures is confined by the walls of a rectangular channel with a high aspect ratio.


Steady laminar flow in a channel with a large array of periodic solid structures is commonly modelled on a macro-scale level through so-called \textit{porous-medium models}, like the Darcy-Forchheimer equation or the Brinkman equation.
The application of these porous-medium models to describe the flow is mainly motivated by empirical evidence from experiments.
For instance, experimental calibration of the (apparent) permeability in the Darcy-Forchheimer equation has been shown to correlate well the relationship between the overall pressure drop over the channel and the bulk velocity or mass-flow rate through the channel.
In this context, the macro-scale velocity and pressure which appear in the Darcy-Forchheimer equation are thus actually interpreted as the cross-sectional average of the velocity and pressure fields at the inlet and outlet of the channel.

Nevertheless, also formal \textit{upscaling} or \textit{homogenization} methods based on volume averaging of the velocity and pressure fields (\cite{Whitaker1999}) are regularly used as a theoretical and practical framework for the macro-scale description of laminar flow in a channel with periodic solid structures.
In particular, because the former porous-medium models can be theoretically recovered from the Navier-Stokes flow equations through volume averaging, when certain length-scale approximations are invoked.
Moreover, several closure problems have been derived for the volume-averaged Navier-Stokes equations, whose solutions govern the permeability and Forchheimer tensors required in the former porous-medium models.
Especially the classical closure problem proposed by \cite{Whitaker1969, Whitaker1996} is widely used, as it governs the (apparent) permeability tensor for a steady incompressible flow of a Newtonian fluid through a porous medium.
Over the past decades, also closure problems for a variety of other laminar flow regimes have been proposed.
Recent works have treated, for example, the  closure for unsteady incompressible flows (\cite{Lasseux2019}) and the closure for slightly compressible flows in porous media (\cite{Lasseux2016, Lasseux2021}).
These closure problems are an effective means to obtain model reduction, as they can be solved \textit{locally} on a single representative volume of the porous medium, or a geometric unit cell of the periodic array.    

It is well known that steady laminar flow through a channel with periodic solid structures often becomes \textit{periodically developed} after a certain distance from the inlet.
This means that the flow exhibits spatial periodicity over a geometric unit cell of the array. 
The occurrence of periodically developed flow has recently been visualized in channels with arrays of circular and square cylinders, by means of micro-PIV measurements (\cite{RenferBrunschwiler2011, XuPanWu2018}) at low to moderate Reynolds numbers. 
Yet, the earliest experimental observations of periodically developed flow in conventional channels with streamwise-periodic cross sections date back to the work of \cite{PrataSparrow1983}.
When the flow is periodically developed, the flow field satisfies the periodic flow equations formulated by \cite{Patankar1977}, which are mathematically equivalent to the classical closure problem for the volume-averaged Navier-Stokes equations, as proposed by \cite{Whitaker1996}.
Therefore, closure for the macro-scale flow equations is usually obtained by solving the periodic flow equations on a geometric unit cell of the array.

A physically meaningful macro-scale description of periodically developed flow, for which the classical closure problem of \cite{Whitaker1996} becomes exact and so defines a spatially constant  permeability tensor, requires a specific averaging operator for the volume-averaged Navier-Stokes equations, as shown by \cite{BuckinxBaelmans2015}.
This averaging operator is based on a weighting function which represents a double volume average, and was originally introduced by \cite{Quintard1994b} for the homogenization of Stokes flow in ordered porous media.
It has also been used to construct exact and physically meaningful macro-scale descriptions of the periodically developed heat transfer regimes in arrays of periodic solid structures (\cite{BuckinxBaelmans2015b, BuckinxBaelmans2016}).
The use of weighting functions or \textit{filters} to describe the macro-scale flow based on \textit{filtered} Navier-Stokes equations has already been explored by many researchers (\cite{DavitQuintard2017}), since the seminal works of \cite{Marle1965, Marle1967}.
In addition, it has received attention in... 
.


As the region of periodically developed flow in a channel is always preceded by a region of developing flow, the former macro-scale description based on a double-volume-averaging operator is of course no longer exact when the entire flow in the channel is considered.
At present, it is still an open question whether the classical closure problem of \cite{Whitaker1996} is accurate enough to provide an approximative solution of the double-volume-averaged flow equations in the region of developing flow.
So far, it hasn't been investigated whether homogenization of a channel flow by means of a (spatially constant) permeability tensor is possible outside the region of periodically developed flow.

However, it must be noted that the classical closure problem of \cite{Whitaker1996} has been derived as a \textit{local} closure problem, under certain length-scale approximations which are less restrictive than the periodically developed flow equations of \cite{Patankar1977}.
In view of this, it can be applied locally within the developing flow, and under certain conditions,  its local solution in the form of a local  permeability tensor may still be a sufficiently accurate approximation.
If that is the case, it may even allow us to solve the macro-scale flow field over a larger part of the channel.
Nonetheless, empirical evidence or a disproof for the former hypothesis is still lacking, as the influence of flow development on the validity of the local closure problem of \cite{Whitaker1996} has never been addressed.

An obvious reason is that flow development does not occur in the class of disordered porous media, for which many porous-medium models and homogenization methods were originally contrived.
Furthermore, in channels with arrays of periodic solid structures, which are classified as ordered porous media, flow development may occur over a relatively short distance from the inlet and then have little practical relevance.
Another explanation is that flow development is also difficult to study in a general way, since it is strongly affected by the boundary conditions, i.e. the velocity profile at the inlet of the channel and the no-slip condition at the channel walls.
As such, it is affected by the entire geometry of the channel and can only be studied via direct numerical simulation of the detailed flow in the entire channel.
Because of this, the study of flow development in large arrays of solid structures is in many cases not computationally affordable, as it necessitates supercomputing infrastructure. 

Despite these complexities, it has been shown that flow development requires consideration at moderate Reynolds numbers, especially in high-aspect-ratio channels containing high-porosity cylinder arrays, like those employed in microfluidic devices and compact heat transfer devices (\cite{Buckinx2022}).
In addition, it has been shown that in such channels, the flow can be mathematically described as \textit{quasi-periodically developed} over a significant, if not the largest, part of the region of developing flow (\cite{Buckinx2022}).
The occurrence of quasi-periodically developed flow is a more universal feature of flow development in arrays of periodic solid structures, as it is characterized by a single exponential mode, whose shape and eigenvalue do not depend on the specific boundary conditions at the inlet of the channel.

Therefore, the main objective of this work is to give an exact macro-scale description of quasi-periodically developed flow, and assess its influence on the validity of the Whitaker's local closure problem.
In particular, the macro-scale description and the validity of the latter closure problem are explored for channels containing arrays of equidistant in-line square cylinders.
The focus lies on cylinder arrays confined between channel walls with a high aspect ratio and a high porosity, which are representative for a variety of microfluidic devices.

For this type of arrays, solutions of Whitaker's local closure problem have not yet been presented.
Solutions have been presented primarily for two-dimensional arrays of square and circular cylinders, often regarded as an idealized model for more complex disordered porous media.
Early studies on this topic include those of \cite{EdwardsShapiro1990, Ghaddar1995, Koch1997, AmaralSoutoMoyne1997} and \cite{Martin1998}, who all investigated the dependence of the apparent permeability on the Reynolds number, flow direction and geometry of the array.
This dependence was further investigated by \cite{ PapathanasiouDendy2001} to evaluate the validity of the Ergun and Forchheimer correlations for closure, as well as \cite{LasseuxAhmadi2011} and \cite{KhalifaPocherTilton2020}, who focused on the flow regimes in such arrays.
To a lesser extent, also closure solutions for three-dimensional periodic solid structures have been presented, for instance in the works of \cite{FourarRadillaMoyne2004},  \cite{RochaCruz2010}, \cite{Vangeffelen2021} and some other works reviewed by \cite{KhalifaPocherTilton2020}.

For the macro-scale description explicated in this work also the developing flow in the proximity of the channel's side walls is examined, although the local closure problem of \cite{Whitaker1996} is not directly applicable in the side-wall region.
Therefore, an alternative local closure problem for the side-wall region will be derived, which is a correction to the one of \cite{Whitaker1996}.
In the literature, local closure problems for the macro-scale flow in a porous medium near a solid wall have already been explored, for instance in the recent works of \cite{ValdesParada2021b, ValdesParada2021}.
However, they rely on a few assumptions with regard to the flow regime and morphology of the porous medium, which make them inexact and less suitable for the modelling of (quasi-) periodically developed flow in channels.

The remainder of this work is organised as follows.
First, in \S \ref{sec: Channel Domain and Array Geometry}, we set out the channel and array geometry that are the subject of the present study.
We also clarify the boundary conditions chosen in this work for the direct numerical simulation of the developing flow.
In \S \ref{sec: Macro-Scale Flow Equations for Steady Channel Flow}, the macro-scale flow equations for a steady channel flow are briefly reviewed. 
We give special attention to the definition of the closure terms and double-volume-averaging operator. 
The reason is that, in this work, the boundary conditions of the flow need to be taken into account in the averaging procedure, while they have been left out of consideration in the literature.
In \S \ref{sec: Macro-Scale Flow Regions in a Channel}, the different macro-scale flow regions in a channel are identified to facilitate the mathematical notation and interpretation of the results that follow.
The features of quasi-developed macro-scale flow, which are observed after spatial averaging of the quasi-periodically developed flow, are examined in \S \ref{sec: Quasi-Developed Macro-Scale Flow}.
These features include the onset point of quasi-developed macro-scale flow, as well as the macro-scale velocity modes.
Subsequently, in \S \ref{sec: Local Closure for Developed Macro-Scale Flow}, we treat the local closure for developed macro-scale flow.
First, the exact closure solutions for periodically developed flow in arrays of equidistant in-line square cylinders are discussed in \S \ref{subsec: Local Closure for Developed Macro-Scale Flow}, as they serve as the starting point for all other derivations and computational results in this work.
Then, in \S \ref{subsec: Exact Local Closure in the Region of Developed Macro-Scale Flow}, we propose an exact local closure problem for periodically developed flow in the side-wall region.
This closure problem is simplified to obtain an approximate permeability tensor for the side-wall region in \S \ref{subsec: Approximative Local Closure in the Region of Developed Macro-Scale Flow}, which is shown to depend on the profile of the macro-scale velocity and its slip length in the side-wall region.
We comment on the validity of the closure solutions for periodically developed flow in \S \ref{subsec: Validity of the Local Closure Problem for Developed Macro-Scale Flow}.
In \S \ref{sec: Local Closure for Periodically Developed Flow}, the local closure for quasi-developed macro-scale flow is treated.
We start in \S \ref{subsec: Exact Local Closure for Quasi-Developed Macro-Scale Flow} with the formulation of an exact local closure problem for the permeability tensor in quasi-periodically developed flow.
This local closure problem is obtained from the eigenvalue problem that defines quasi-periodically developed flow (\cite{Buckinx2022}), and can be solved on a row of the array.
The classical closure problem, and in particular the approximations that may allow us to apply it in the region of quasi-developed macro-scale flow, are discussed in \S \ref{subsec: Approximative Local Closure for Quasi-Developed Macro-Scale Flow Outside the Side-Wall Region}.
There, we also present some of its solutions for arrays of equidistant in-line square cylinders.
The validity and accuracy of those closure solutions for quasi-periodically developed flow is first analysed from a theoretical point of view in \S \ref{subsec: Validity of the Classical Closure Problem for Quasi-Developed Flow Outside the Side-Wall Region -- Theoretical Considerations}.
To support our theoretical analysis, we conduct a computational study in \S \ref{subsec: Validity of the Classical Closure Problem for Quasi-Developed Flow Outside the Side-Wall Region -- Computational Study}, in which the solutions of the classical closure problem are compared with the actual macro-scale flow in different rectangular channels, all containing an array of equidistant in-line square cylinders with a porosity between 0.75 and 0.94.
The macro-scale flow development is studied by means of direct numerical simulation and explicit filtering of the flow in the channel.
Our theoretical analysis and computational study are extended to the side-wall region of the channel in \S \ref{subsec: Validity of the Classical Closure Problem for Quasi-Developed Flow Inside the Side-Wall Region}.
In \S \ref{sec: Reconstruction of Quasi-Developed Macro-Scale Flow}, we end our work with some computational results which shed light on the suitability of the classical closure problem for reconstructing the macro-scale flow in channels.
Finally, in \S \ref{sec: Conclusions}, we summarize the main the conclusions of this work.

\section{Geometry of the Flow Channel}
\label{sec: Channel Domain and Array Geometry}
We consider the steady laminar flow of an incompressible Newtonian fluid through a straight channel, having a length $L$ and a rectangular cross section of width $W$ and height $H$.
The flow through this channel is described on a fixed, open bounded domain $\Omega \subset \mathbb{R}^3$, which is the disjoint reunion of a fluid region $\Omega_f$ and a  solid region $\Omega_s=\Omega \setminus \Omega_f$.
To locate points within $\Omega$, we introduce a normalized Cartesian vector basis $\left\{\boldsymbol{e}_j\right\}_{j=1,2,3}$ and a corresponding coordinate system $\left\{x_j\right\}_{j=1,2,3}$ such that $\forall \boldsymbol{x} \in \Omega :  
x_1 \in \left(0, L \right) , 
x_2 \in \left(0, W \right) ,
x_3 \in  \left(-H/2,H/2 \right)$.
The inlet and outlet section of the channel then correspond to the domain boundary parts $\Gamma_{\text{in}} = \left\{ \boldsymbol{x} \, \vert \, \boldsymbol{x}  \in \partial \Omega, x_1 =0 \right\}$ and $\Gamma_{\text{out}} = \left\{ \boldsymbol{x} \, \vert \, \boldsymbol{x}  \in \partial \Omega, x_1 =L \right\}$ respectively, while the solid channel walls correspond to 
the boundary part $\Gamma_{\text{wall}} = \partial \Omega \setminus (\Gamma_{\text{in}} \cup \Gamma_{\text{out}})$.
We further distinguish the bottom wall $\Gamma_{\text{bottom}} = \left\{ \boldsymbol{x} \, \vert \, \boldsymbol{x}  \in \partial \Omega, x_3 = -H/2 \right\}$, the top wall $\Gamma_{\text{top}} = \left\{ \boldsymbol{x} \, \vert \, \boldsymbol{x}  \in \partial \Omega, x_3 = H/2 \right\}$, as well as the side walls $\Gamma_{\text{sides}} = \left\{ \boldsymbol{x} \, \vert \, \boldsymbol{x}  \in \partial \Omega, x_2 \in \{0,W\} \right\}$.

The solid region $\Omega_s$ in the channel is assumed to consist of an array of $N_1 \times N_2 $ square solid cylinders of a diameter $d$ and height $H$, separated from each other by a distance $\ell_j$ along each direction $\boldsymbol{e}_j$:
\begin{equation*}
\Omega_s = \left\{ \boldsymbol{x} \, \vert \, \boldsymbol{x} \in \Omega,  
x_1 \in \left[s_0, L-s_N \right], 
\left(
x_j - \left \lfloor{x_j / \ell } \right \rfloor \ell_j
\right) \in \left[\frac{\ell_j -d}{2}, \frac{\ell_j + d}{2}\right] \, \text{for} \, j=1,2
\right\}\,,
\end{equation*}
where $s_0$ and $s_N$ indicate the position of the first and last cylinder row, as $L=s_0 + N_1 \ell_1 +s_N$   and $W=N_2\ell_2$. 
It follows that the porosity in the array equals $\epsilon_f = 1-d^2/(\ell_1 \ell_2)$.
The fluid region $\Omega_s$ has an associated indicator function $\gamma_f$ defined as $\gamma_f(\boldsymbol{x}) = 1 \leftrightarrow \boldsymbol{x} \in \Omega_f, \gamma_f(\boldsymbol{x}) = 0 \leftrightarrow \boldsymbol{x} \notin \Omega_s$. 
The fluid indicator is thus spatially periodic at any position $\boldsymbol{x} \in \Omega$ sufficiently far from the domain boundary $\partial \Omega$:  $\gamma_f(\boldsymbol{x} + \boldsymbol{l}_j) = \gamma_f(\boldsymbol{x})$ with $\boldsymbol{l}_j \triangleq \ell_j \boldsymbol{e}_j$ and $j=1,2$.

In the remainder of this work, the dimensions of the channel and array have been chosen such that they are representative of many microchannels (\cite{RenferBrunschwiler2011, XuPanWu2018}), as well as larger-sized channels encountered in compact heat transfer devices (Ref).
As we focus on the influence of flow development in high-aspect-ratio channels with high-porosity arrays, most computational results are provided in the porosity range $\epsilon_f \in \left[ 0.75, 0.94 \right]$, for a single aspect ratio $W/H=10$ and single height-to-spacing ratio $H/\ell_1=1$.
In addition, we restrict our computational study to equidistant cylinders 
for which $\ell_1=\ell_2$.

The flow velocity $\boldsymbol{u}_f$ and pressure $p_f$  through the channel are determined by direct numerical simulation of the incompressible Navier-Stokes equations on $\Omega_f$ for a parabolic inlet velocity profile and a uniform outlet pressure: $\boldsymbol{u}_f(\boldsymbol{x})= 36 x_2 x_3(W-x_2)(H-x_3)/(WH)^2 \boldsymbol{e}_1$ for $\boldsymbol{x}  \in \Gamma_{\text{in}}$ and $p_f(\boldsymbol{x}) =0$ for $\boldsymbol{x}  \in \Gamma_{\text{out}}$.
The bulk velocity $u_b \triangleq -\int_{\Gamma_{\text{in}}} \boldsymbol{n}\boldsymbol{\cdot} \boldsymbol{u}_f \, d\Gamma/(WH)$ through the channel is thus imposed.
In addition to these boundary conditions, a no-slip condition is presumed at the boundary $\Gamma_{0} \triangleq \Gamma_{\text{wall}} \cup \Gamma_{fs}$, which is the union of the channel wall and the fluid-solid interface $\Gamma_{fs} \triangleq \partial \Omega_f \cap \partial \Omega_s$:
\begin{equation}
\label{eq: no-slip condition wall and solid interface}
\boldsymbol{u}_f(\boldsymbol{x}) = 0 \qquad \mbox{for} ~~ \boldsymbol{x}  \in \Gamma_{0} \,.
\end{equation}

\begin{figure}
\begin{center}
\includegraphics[scale=0.38]{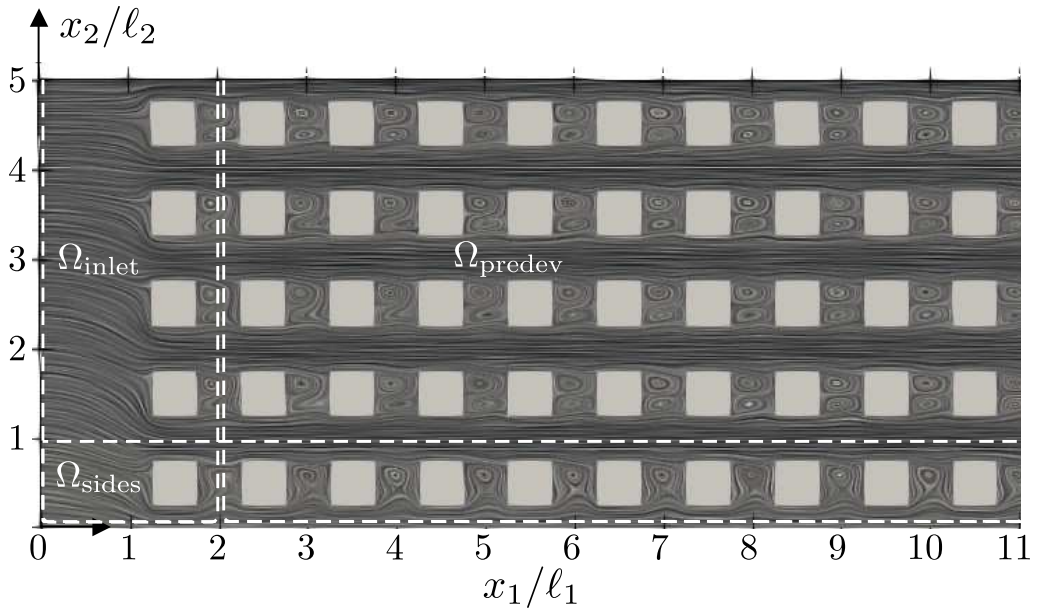}
\caption{Flow velocity field in the mid plane $x_3  = H/2$ of a channel containing an array of in-line square cylinders ($N_1 =20, N_2=10$, $s_0/\ell_1 =1, s_N/\ell_1 = 10$, $H/\ell_1 =1$, $\ell_1/\ell_2 =1$) for $\Rey =50$ and $\epsilon_f = 0.75$. The flow patterns and wakes have been visualized using line integral convolution (LIC) vector field visualization. 
}
\label{fig: Flow Field Re 50 porosity 0.75}
\end{center}
\end{figure}

Figure \ref{fig: Flow Field Re 50 porosity 0.75} illustrates the velocity field that is obtained for the previous boundary conditions, in the mid plane of a channel containing an array of $20 \times 10$ in-line square cylinders, for a Reynolds number $\Rey \triangleq \rho_f u_b 2H/ \mu_f =50$ and a porosity $\epsilon_f=0.75$. 
Because the flow is symmetric with respect to the plane $x_2 = W/2$, only a part of the mid plane is shown.
It can be seen from the flow patterns and the wakes behind each cylinder, that at $x_1/\ell_1 \simeq 11$, the flow has almost become periodically developed.
In this case, the direct numerical simulation was performed on a regularly-sized mesh of about $98$ millon mesh cells, resulting in a computational time of about $6$ hours on $13 $ nodes of each $36$ processors, for a total number of $2500$ discrete time steps until a steady state was observed at time $t=20\ell_1/u_b$.
The flow simulation was started from a uniform zero velocity field as initial condition.
A mesh-refinement study has been carried out, to ensure that the estimated discretisation error on the local velocity profiles was at least below $2.5\%$. 
For all other direct numerical simulations presented in this work, the discretisation error has been estimated to have the same relative magnitude.

For the direct numerical simulation of the flow equations and their boundary conditions, the  software package FEniCSLab was developed within the finite-element framework FEniCS (\cite{FenicsProject2015}).
The package FEniCSLab contains an object-oriented re-implementation of the parallel fractional-step solver of \textit{Oasis} developed by \cite{MortensenValenSendstad2015} for the unsteady incompressible Navier-Stokes equations, and has been modified to allow for variable time stepping and coupled mass and heat transfer between a fluid and a (moving) solid. 
The discretization of the Navier-Stokes equations in FEniCSLab relies on piecewise quadratic Lagrange elements  for the velocity and piecewise linear Lagrange elements  for the pressure, so that almost fourth-order accuracy in velocity and second-order in pressure
accuracy is achieved (\cite{MortensenValenSendstad2015}) on the regularly-sized meshes used in this work.



\section{Macro-Scale Flow Equations for Steady Channel Flow}
\label{sec: Macro-Scale Flow Equations for Steady Channel Flow}
The macro-scale velocity field $\langle \boldsymbol{u}\rangle_m $ and macro-scale pressure field $\langle p \rangle_m $ in the channel are obtained by applying a spatial averaging operator or filter $\langle \; \rangle_m$ to the velocity and pressure distributions $\boldsymbol{u}$ and $p$, which follow from a direct numerical simulation of the Navier-Stokes equations. 
As we consider steady channel flow, they satisfy the following macro-scale Navier-Stokes equations, 
\begin{align}
\rho_f \boldsymbol{\nabla} \boldsymbol{\cdot} \left(
\epsilon_{fm}^{-1} \langle \boldsymbol{u}\rangle_m \langle \boldsymbol{u}\rangle_m \right)  &= 
- \boldsymbol{\nabla}\langle p \rangle_m + 
\mu_f \nabla^2  \langle \boldsymbol{u}\rangle_m
 - \rho_f \boldsymbol{\nabla} \boldsymbol{\cdot} {\boldsymbol{M}} + \boldsymbol{b} \,, 
\label{eq: macro-scale momentum equation}\\
\boldsymbol{\nabla} \boldsymbol{\cdot} \langle
 \boldsymbol{u}\rangle_m  &= 0\,.
\label{eq: macro-scale continuity equation}
\end{align}
Here, $\epsilon_{fm} \triangleq \langle \gamma_f \rangle_m$ is the weighted porosity, and
$\boldsymbol{M} \triangleq \langle \boldsymbol{u}\boldsymbol{u} \rangle_m - \epsilon_{fm}^{-1} \langle \boldsymbol{u}\rangle_m \langle \boldsymbol{u}\rangle_m $ the macro-scale momentum dispersion tensor.
The closure force $\boldsymbol{b}$ results from the no-slip condition (\ref{eq: no-slip condition wall and solid interface}) at the
channel walls and the fluid-solid interface:
\begin{equation}
\label{eq: final form macro-scale no-slip force}
\boldsymbol{b} \triangleq \langle \boldsymbol{n}_{0} \boldsymbol{\cdot}  (-p_f \boldsymbol{I} + \boldsymbol{\tau}_f) \delta_{0} \rangle_m \,.
\end{equation}
We remark that $\boldsymbol{I}$ denotes the identity tensor and $\boldsymbol{n}_{0}$ denotes the normal at $\Gamma_{0}$ (pointing towards $\Omega_s$ at $\Gamma_{fs}$ and pointing outwards $\Omega$ at $\Gamma$), while $\delta_0$ is the Dirac surface indicator of the no-slip surface $\Gamma_{0}$.
The filter operator itself is defined by the convolution product in $\mathbb{R}^3$ with a compact weighting function $m$: $\langle \boldsymbol{\phi} \rangle_m \triangleq m \ast  \boldsymbol{\phi}  $ (\cite{Quintard1994b, BuckinxPhD2017}).

In order that $\langle \boldsymbol{u}\rangle_m $ and  $\langle p \rangle_m $, as well as their governing equations (\ref{eq: macro-scale momentum equation}) and (\ref{eq: macro-scale continuity equation}), be defined on the entire domain $\Omega$, $\boldsymbol{u}$ and $p$ are defined here as extended distributions derived from the velocity $\boldsymbol{u}_f$ and pressure $p_f$ which appear in the original Navier-Stokes equations:
$\boldsymbol{u} = \boldsymbol{u}_f \text{ in } \Omega_f$, $\boldsymbol{u} = 0 \text{ in } \Omega_s$, $\boldsymbol{u} = \boldsymbol{u}_e \text{ in } \mathbb{R}^3 \setminus \Omega$ and $p = p_f \text{ in } \Omega_f$, $p = 0 \text{ in } \Omega_s$, $p = p_e \text{ in } \mathbb{R}^3 \setminus \Omega$ (\cite{Schwarz1978}).
Therefore, also the viscous stress tensor $\boldsymbol{\tau}$ is a distribution given by $\boldsymbol{\tau} = \mu_f \left(\boldsymbol{\nabla}^{\nu} \boldsymbol{u} + \boldsymbol{\nabla}^{\nu} \boldsymbol{u}^{\intercal} \right)$,  where $\boldsymbol{\nabla}^{\nu} $ denotes the gradient operator in the usual sense (\cite{Quintard1994b, Gagnon1970}).
As shown in appendix \ref{app: Form of the Macro-Scale Flow Equations}, a suitable choice of the extensions $\boldsymbol{u}_e$, $p_e$ and $\boldsymbol{\tau}_e$ has been made such that the form of the macro-scale flow equations (\ref{eq: macro-scale momentum equation}) and (\ref{eq: macro-scale continuity equation}) is valid.


In this work, the weighting function that is used to define the macro-scale flow, corresponds to a double volume average:
\begin{equation}
\label{eq: definition weighting function double volume average}
m(\boldsymbol{y}) = \frac{1}{H} \rect\left(\frac{y_3}{H}\right) \displaystyle \prod_{j=1}^{2}\frac{l_j-2\vert y_j\vert}{l_j} \rect\left(\frac{y_j}{2\ell_j}\right) \qquad \mbox{with} ~~ y_j \triangleq\boldsymbol{y} \boldsymbol{\cdot} \boldsymbol{e}_j  \,.
\end{equation}
This weighting function has a compact support or \textit{filter window} given by the local unit cell $\Omega_{\text{unit}}^{2\times 2}(\boldsymbol{x})$, which is defined by
\begin{equation}
\Omega_{\text{unit}}^{n_1\times n_2}(\boldsymbol{x}) \triangleq
\left \lbrace \boldsymbol{r} = \boldsymbol{x} + \boldsymbol{y} \,\vert \, \exists \, c_j \in \left[ -\frac{1}{2},\frac{1}{2} \right]  \Leftrightarrow \boldsymbol{y} = \sum_{j=1}^{2} c_j n_j \boldsymbol{l}_j  +c_3 H\boldsymbol{e}_3 \right \rbrace \,.
\end{equation}
Therefore, this weighting function enables an exact macro-scale description of periodically developed flow (\cite{BuckinxBaelmans2015}), as long as the flow field is periodically similar within each unit cell $\Omega_{\text{unit}}^{n_1\times n_2}(\boldsymbol{x})$ with $n_j \leq 2$.

Because we only evaluate the filtered quantities at the mid plane $x_3 = H/2$ of the channel, the centroid $\boldsymbol{x}$ of the filter window is chosen such that the window does not fall out the channel domain, i.e. $\forall \boldsymbol{r} \in \Omega_{\text{unit}}^{2\times 2}(\boldsymbol{x}): \boldsymbol{r}  \in \Omega $.
It must be remarked that we have chosen the height of the filter window equal to that of the channel, since the macro-scale flow then becomes two-dimensional, due to the no-slip condition at the bottom and top surface of the channel.
Moreover, we remark that the filter based on the weighting function (\ref{eq: definition weighting function double volume average}) is a \textit{separable} filter whose action on the flow is equivalent to height-averaging followed by double volume averaging (\cite{BuckinxPhD2017}).

Further in this work, also the intrinsic averaging operator $\langle \; \rangle^f_m$ and deviation operator $(\, \devtilde{\;}\,)$ (\cite{Gray1975})corresponding to the weighting function (\ref{eq: definition weighting function double volume average}) are frequently used, whose definitions are given by $\langle \boldsymbol{\phi} \rangle^f_m \triangleq \epsilon_{fm}^{-1} \langle \boldsymbol{\phi} \rangle_m $ and  $\devtilde{\boldsymbol{\phi}} \triangleq \boldsymbol{\phi} - \langle \boldsymbol{\phi} \rangle^f_m \gamma_f$.
Each of the previous filter operators has been implemented in FEniCSLab as an explicit finite-element integral operator which can be applied to an arbitrary finite-element function.
This discrete integral operator makes use of the automated quadrature degree estimation algorithms available in UFL (\cite{UFL2014}).
For its parallel point-wise evaluation, a custom interpolation algorithm was written in DOLFIN (\cite{DOLFIN2010}) which is quite similar to the interpolation routines of the software package fenicstools by \cite{fenicstools2017}.

\section{Macro-Scale Flow Regions in a Channel}
\label{sec: Macro-Scale Flow Regions in a Channel}
\begin{figure}
\begin{center}
\includegraphics[scale=0.85]{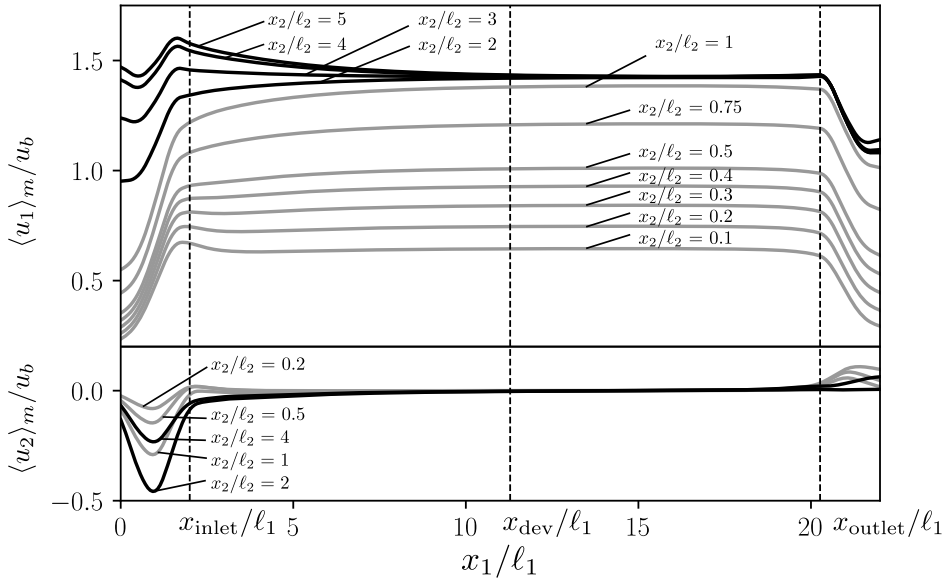}
\caption{Macro-scale velocity components and flow regions in a channel array ($N_1 =20, N_2=10$, $s_0/\ell_1 =1, s_N/\ell_1 = 10$, $H/\ell_1 =1$, $\ell_1/\ell_2 =1$) for $\Rey =50$ and $\epsilon_f = 0.75$.
Grey and black lines correspond to locations inside and outside the side-wall region respectively.}
\label{fig:Macro-Scale Flow Field and Regions Re 50 porosity 0.75}
\end{center}
\end{figure}

From a macro-scale perspective, different flow regions can be identified in a channel containing an array of in-line equidistant square cylinders.
These flow regions are illustrated in figure \ref{fig:Macro-Scale Flow Field and Regions Re 50 porosity 0.75}, which shows the macro-scale velocity field in a channel with an array of $20 \times 10$ cylinders, for a Reynolds number $\Rey =50$. 
The macro-scale velocity components $\langle u_j \rangle^f_m \triangleq \langle \boldsymbol{u} \rangle^f_m \boldsymbol{\cdot} \boldsymbol{e}_j$  in figure \ref{fig:Macro-Scale Flow Field and Regions Re 50 porosity 0.75} have been calculated via explicit filtering of the velocity field that was illustrated in figure \ref{fig: Flow Field Re 50 porosity 0.75}.
The explicit filtering operation for each velocity component took $12$ hours on $3$ nodes of $36$ processors.

The first flow region we identify in figure \ref{fig:Macro-Scale Flow Field and Regions Re 50 porosity 0.75}, is the inlet region, $\Omega_{\text{inlet}} \triangleq \{\boldsymbol{x} \in \Omega \vert x_1 \in (0, x_{\text{inlet}}), x_{\text{inlet}}=s_0+ \ell_1\}$, which extends from the channel inlet to the cross section after the first cylinder row where the weighted porosity $\epsilon_{fm}$ does no longer vary with the coordinate $x_1$ in the main flow direction.

Secondly, there is the outlet region, $\Omega_{\text{outlet}} \triangleq \{ \boldsymbol{x} \in \Omega \vert   x_1 \in (x_{\text{outlet}}, L), x_{\text{outlet}}= L-s_N-\ell_1 \}$, which extends from the channel outlet to
the cross section before the last cylinder row where a gradient of the weighted porosity starts to occur in the main flow direction $x_1$.

Between the inlet region and outlet region, we distinguish the channel's core region, $\Omega_{\text{core}}\triangleq \{ \boldsymbol{x} \in \Omega \vert   x_1 \in \left[x_{\text{inlet}}, x_{\text{outlet}} \right] \}$.
In this region, the weighted porosity $\epsilon_{fm}$ is constant everywhere, except in the region near the side walls, $\Omega_{\text{sides}} \triangleq \{ \boldsymbol{x} \in \Omega \vert   x_2 \in ((0, \ell_2) \cup (W- \ell_2, W)) \}$, where $\epsilon_{fm}$  depends on the coordinate $x_2$ and decreases towards $\Gamma_{\text{sides}}$.

In the channel's core region, typically a periodically developed flow region $\Omega_{\text{periodic}}\triangleq \{ \boldsymbol{x} \in \Omega \vert   x_1 \in \left[x_{\text{periodic}}, x_{\text{end}} \right] \}$ is established, provided that the number of solid cylinders $N_1$ along the main flow direction is sufficiently large with respect to the Reynolds number $\Rey$ and no flow transition caused by vortex shedding occurs.
In the periodically developed flow region, the velocity distribution can be treated as spatially periodic in the main flow direction, i.e. 
\begin{equation}
\label{eq: flow periodicity main flow direction}
\boldsymbol{u}(\boldsymbol{x} + n_1 \boldsymbol{l}_1)= \boldsymbol{u}(\boldsymbol{x})\qquad \mbox{for} ~~ \boldsymbol{x} \in \Omega_{\text{periodic}}\,,
\end{equation}
with $n_1$ some integer.
As a consequence, the macro-scale velocity has the same profile $U_{\text{dev}}(x_2)$ over every cross section in $\Omega_{\text{periodic}}$ which is located at a distance larger than $n_1\ell_1$ from the onset point and end point of flow periodicity.
Hence,  
\begin{equation}
\label{eq: developed macro-scale flow main flow direction}
\langle \boldsymbol{u}  \rangle_m(\boldsymbol{x}) = \boldsymbol{U}_{\text{dev}}(\boldsymbol{x}) \triangleq U_{\text{dev}}(x_2)\boldsymbol{e}_1 
\qquad \mbox{for} ~~
\boldsymbol{x} \in \Omega_{\text{dev}}\,,
\end{equation}
with $\Omega_{\text{dev}} \triangleq \{\boldsymbol{x} \in \Omega \vert x_1 \in (x_{\text{dev}}, x_{\text{end}} - n_1\ell_1 ) \}$ and $ x_{\text{dev}} \triangleq x_{\text{periodic}}+ n_1\ell_1$.
For that reason, the macro-scale velocity is called \textit{developed} in $\Omega_{\text{dev}}$.
We remark that the end point of flow development in good approximation satisfies $x_{\text{end}}\simeq x_{\text{outlet}}$ for the flow conditions and array geometries investigated in this work (\cite{Buckinx2022}).
It is well known (\cite{Patankar1977}) that because of (\ref{eq: flow periodicity main flow direction}), the pressure distribution in $\Omega_{\text{periodic}}$ has the form  
\begin{equation}
\label{eq: developed pressure distribution}
p(\boldsymbol{x}) = (\boldsymbol{\nabla }\mathrm{P}_{\text{dev}} \boldsymbol{\cdot} \boldsymbol{x}) \gamma_f(\boldsymbol{x}) + p^{\star}(\boldsymbol{x})
\qquad \mbox{with} \qquad
p^{\star}(\boldsymbol{x} + n_1 \boldsymbol{l}_1)= p^{\star}(\boldsymbol{x})\,.
\end{equation}
Here, $ \boldsymbol{\nabla }\mathrm{P}_{\text{dev}}$  denotes the constant pressure gradient which drives the periodically developed flow.

The velocity distribution in $\Omega_{\text{periodic}}$ also exhibits transversal periodicity at a distance larger than $\ell_{\text{sides}}$ from $\Gamma_{\text{sides}}$, if the number of solid cylinders $N_2$ in the transversal direction is sufficiently large.
In that case, we thus have 
\begin{equation}
\label{eq: flow periodicity transversal direction}
\boldsymbol{u}(\boldsymbol{x} + n_2 \boldsymbol{l}_2)= \boldsymbol{u}(\boldsymbol{x})
\qquad \mbox{and} \qquad
p^{\star}(\boldsymbol{x} + n_2 \boldsymbol{l}_2)= p^{\star}(\boldsymbol{x})\,,
\end{equation}
for all $\boldsymbol{x} \in \Omega_{\text{periodic}}$ with $ x_2 \in (\ell_{\text{sides}},W-\ell_{\text{sides}})$ and for some integer $n_2$.
As a result, the developed macro-scale velocity field becomes uniform at a distance of $\ell_{\text{sides}} +  n_2 \ell_2$ from $\Gamma_{\text{sides}}$, so that 
\begin{equation}
\boldsymbol{U}_{\text{dev}}(\boldsymbol{x}) = \boldsymbol{U}\triangleq U\boldsymbol{e}_1
\qquad \mbox{for} ~~
\boldsymbol{x} \in \Omega_{\text{uniform}} \,,
\end{equation}
as $\Omega_{\text{uniform}} \triangleq \{  \boldsymbol{x} \in \Omega \vert x_1 \in (x_{\text{dev}}, x_{\text{end}} - n_1 \ell_1 ), x_2 \in (\ell_{\text{sides}} + n_2 \ell_2,W -\ell_{\text{sides}}- n_2 \ell_2 ) \} $.


The last region we identify in $\Omega_{\text{core}}$, is the developing-flow region $\Omega_{\text{predev}}\triangleq \{ \boldsymbol{x} \in \Omega \vert   x_1 \in \left[x_{\text{inlet}}, x_{\text{dev}} \right] \}$, which precedes the periodically developed flow region. 
Now that the different macro-scale flow regions for the filter (\ref{eq: definition weighting function double volume average}) have been defined, we will introduce the region of quasi-developed macro-scale flow. 

\section{Quasi-Developed Macro-Scale Flow}
\label{sec: Quasi-Developed Macro-Scale Flow}
After a certain section in the developing-flow region $\Omega_{\text{predev}}$, the flow can be described as quasi-periodically developed (\cite{Buckinx2022}).
This means that the velocity distribution $\boldsymbol{u}$ converges asymptotically towards a truly periodic velocity distribution $\boldsymbol{u}^{\star}$ along the main flow direction via a single exponential mode:
\begin{equation}
\label{eq: quasi-periodically developed flow}
\boldsymbol{u} = \velamp \exp\left(-\boldsymbol{\lambda} \boldsymbol{\cdot}  \boldsymbol{x}  \right) + \boldsymbol{u}^{\star}\,,
\end{equation} 
where $ \boldsymbol{u}^{\star}(\boldsymbol{x} + n_1\boldsymbol{l}_1) = \boldsymbol{u}^{\star}(\boldsymbol{x})$ and
\begin{equation}
\label{eq: periodic velocity amplitude}
\velamp(\boldsymbol{x} + n_1\boldsymbol{l}_1) = \velamp(\boldsymbol{x})\,.
\end{equation} 
The eigenvalue $\boldsymbol{\lambda} \triangleq \lambda \boldsymbol{e}_1$ and the mode amplitude $ \velamp$ are the solution of the eigenvalue problem given in (\cite{Buckinx2022}).
The region over which the flow is quasi-periodically developed, is denoted by $\Omega_{\text{quasi-periodic}}\triangleq \{ \boldsymbol{x} \in \Omega \vert   x_1 \in \left[x_{\text{quasi-periodic}}, x_{\text{periodic}} \right] \}$ and $x_{\text{quasi-periodic}}$ is called the onset point of quasi-periodically developed flow.
In agreement with (\ref{eq: developed pressure distribution}), the pressure field in $\Omega_{\text{quasi-periodic}}$ is given by
\begin{equation}
\label{eq: quasi periodically developed pressure field}
p_f(\boldsymbol{x}) = \pressamp_f(\boldsymbol{x}) \exp\left(-\boldsymbol{\lambda} \boldsymbol{\cdot}  \boldsymbol{x}  \right) + \boldsymbol{\nabla }\mathrm{P}_{\text{dev}} \boldsymbol{\cdot}\boldsymbol{x} + p^{\star}_f(\boldsymbol{x}) \,,
\end{equation}
where
\begin{equation}
\label{eq: periodic pressure amplitude}
\pressamp(\boldsymbol{x} + n_1\boldsymbol{l}_1) = \pressamp(\boldsymbol{x})\,.
\end{equation} 
The magnitude of the modes $\velamp$ and $\pressamp$ is characterized by a single constant $C_{\velamp} \triangleq \langle \velampx \rangle_{\text{row}}$, which is the row-wise average of $\velampx$.
This constant $C_{\velamp}$ is called the perturbation size and depends on the specific inlet conditions (\cite{Buckinx2022}).

It follows from (\ref{eq: quasi-periodically developed flow}) that the \textit{quasi-developed} macro-scale flow field satisfies
\begin{equation}
\label{eq: quasi-developed macro-scale flow}
\langle \boldsymbol{u} \rangle_m \simeq \boldsymbol{U}_{\text{dev}} + \langle \velamp \rangle_m \exp\left( -\lambda x_1\right) \,,
\end{equation} 
since $\boldsymbol{U}_{\text{dev}} \triangleq \langle \boldsymbol{u}^{\star} \rangle_m $ in agreement with (\ref{eq: flow periodicity main flow direction}) and (\ref{eq: developed macro-scale flow main flow direction}).
Because of (\ref{eq: periodic velocity amplitude}), it holds that 
\begin{equation}
\label{eq: quasi-developed macro-scale flow amplitude}
\frac{\partial }{\partial x_1} \langle \velamp \rangle_m = 0 \,,
\end{equation}
so the amplitude $\langle \velamp \rangle_m$ of the macro-scale velocity mode depends only on the coordinate $x_2$.
The quasi-developed macro-scale pressure field corresponding to (\ref{eq: quasi-developed macro-scale flow}) is given by
\begin{equation}
\label{eq: macro-scale pressure quasi-developed flow}
\langle p \rangle^f_m \simeq 
\boldsymbol{\nabla}\mathrm{P}_{\text{dev}} \boldsymbol{\cdot} \left(\boldsymbol{x}+
\boldsymbol{m} \right)+ 
\langle p^{\star} \rangle^f_m  +
\langle \pressamp \rangle^f_m \exp \left(-\boldsymbol{\lambda} \boldsymbol{\cdot} \boldsymbol{x} \right)\,, 
\end{equation}
if $\boldsymbol{m}$ denotes the first intrinsic spatial moment over the fluid region (\cite{Quintard1994a, DavitQuintard2017}).
As a technical note, we remark that (\ref{eq: quasi-developed macro-scale flow}) and (\ref{eq: macro-scale pressure quasi-developed flow}) are approximations instead of equalities, unless the filter's weighting function is matched to the eigenvalue $\lambda$, as discussed in \citep{BuckinxBaelmans2015b}.
However, for the double volume-averaging operator (\ref{eq: definition weighting function double volume average}), the latter approximations are sufficiently accurate as long as $\lambda\ell_1 \ll 1$.
Otherwise, the filter operator $\langle \; \rangle_m$ should be interpreted as a matched filter.

\begin{figure}
\begin{center}
\includegraphics[scale=0.5]{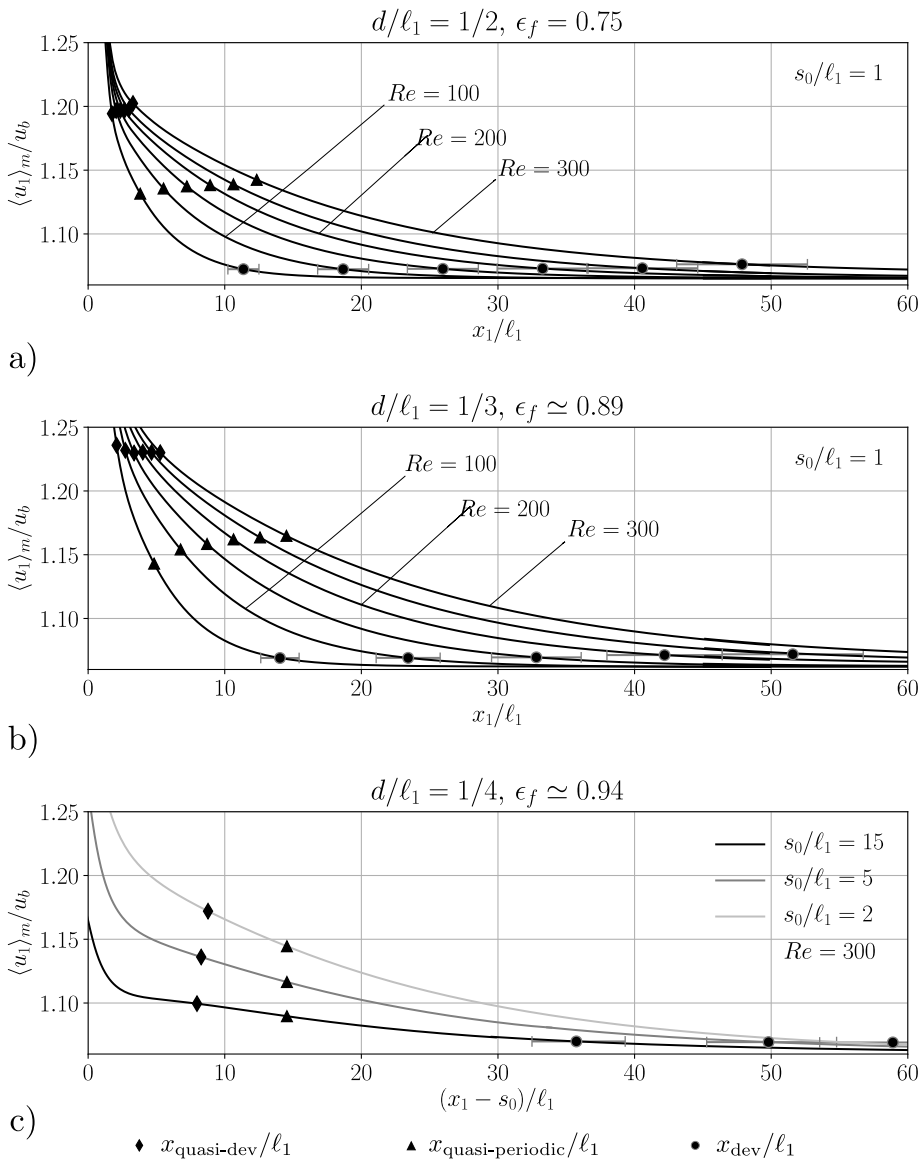}
\caption{Macro-scale velocity profiles along the centreline ($x_2=W/2$) of a channel array ($N_1 =80, N_2=10$, $H/\ell_1 =1$, $\ell_1/\ell_2 =1$, $s_N/\ell_1 = 10$) for different Reynolds numbers $\Rey \in \left\{50, 100, 150, 200, 250, 300 \right\}$ and porosities $\epsilon_f$, as well as different positions $s_0$ of the first cylinder row. The onset point of quasi-periodically developed flow, $x_{\text{quasi-periodic}}/\ell_1$, is shown together with the onset point of quasi-developed macro-scale flow, $x_{\text{quasi-dev}}/\ell_1$, as well as the point of developed flow, $x_{\text{dev}}/\ell_1$. }
\label{fig: Quasi-developed macro-scale flow}
\end{center}
\end{figure}

We may expect that the macro-scale velocity field can be treated as quasi-developed even before the local velocity field has become quasi-periodically developed in the strict sense of (\ref{eq: quasi-periodically developed flow}) and (\ref{eq: periodic velocity amplitude}).
This means that (\ref{eq: quasi-developed macro-scale flow}) and (\ref{eq: quasi-developed macro-scale flow amplitude}) are already an accurate approximation after a certain section $x_1 = x_{\text{quasi-dev}}$ with $x_{\text{quasi-dev}} \leq x_{\text{quasi-periodic}} $.
The reason is that an exponential mode $\boldsymbol{A}(\boldsymbol{x})\exp(-\lambda x_1)$ with $\boldsymbol{A}(\boldsymbol{x}) \simeq \velamp(\boldsymbol{x})$ is already present within the velocity field $\boldsymbol{u}_f$ at the start of the flow development, near $x_1 \simeq x_{\text{inlet}}$, so that after spatial averaging, it may pop up as a dominant mode in the macro-scale velocity field $\langle \boldsymbol{u} \rangle_m$, even before it has become the dominant mode for $\boldsymbol{u}_f$.

The latter expectation is also supported by the numerical evidence in figure \ref{fig: Quasi-developed macro-scale flow}, which shows an exponential evolution of the macro-scale velocity, $\langle u_1 \rangle_m = \langle A_1\rangle_m \exp(-\lambda x_1) + U_{\text{dev},1}$ with $\partial \langle A_1\rangle_m/ \partial x_1 \simeq 0$,  from a section $x_1=x_{\text{quasi-dev}}$ relatively close to the inlet region, as indicated by the markers ($\mydiamond$). 
In a strict sense though, $\partial \langle A_1\rangle_m / \partial x_1 =0$ is only valid when $\langle A_1\rangle_m = \langle \velampx \rangle_m$, which is the case for $x_1 > x_{\text{quasi-periodic}}$, hence after the sections indicated by the markers ($\mytriangle$).
We clarify that the macro-scale velocity profiles in figure \ref{fig: Quasi-developed macro-scale flow} have been obtained by explicit filtering of the velocity fields from our preceding work (\cite{Buckinx2022}).
The explicit filtering operation took $24$ hours on $3 \times 36$ processors for just a single velocity component on each mesh of about $140$ million mesh cells.

In figure \ref{fig: Quasi-developed macro-scale flow} (a,b), one can see that the onset point of quasi-developed macro-scale flow, $x_{\text{quasi-dev}}$, scales in good approximation linearly with the Reynolds number $\Rey$, just like the onset point of quasi-periodically developed flow, $x_{\text{quasi-periodic}}$ (\cite{Buckinx2022}).
For the channel geometries selected for this figure, i.e. for $H=\ell_1=\ell_2=s_0$ and $W/H=N_2=10$, it was found that $x_{\text{quasi-dev}}/\ell_1 \simeq 0.006 \Rey + 1.5$ for $d/\ell_1=1/2$, while $x_{\text{quasi-dev}}/\ell_1 \simeq 0.013 \Rey +1.5$ for $d/\ell_1=1/3$, when $\Rey \in \left\{50, 100, 150, 200, 250, 300 \right\}$.
These linear correlations for $x_{\text{quasi-dev}}/\ell_1$ have been determined numerically by defining $x_{\text{quasi-dev}}$ as the $x_1$-section for which $\langle u_1 \rangle_m - U_{\text{dev},1}$  at $x_2=W/2$ deviates less than $99.9\%$ from the exponential relationship $\langle \velampx \rangle_m \exp(-\lambda x_1)$. 
The relative uncertainty on these correlations is about $10\%$, as the uncertainty on the numerical values for $x_{\text{quasi-dev}}/\ell_1$ in figure \ref{fig: Quasi-developed macro-scale flow}, is within $10\%$ too.

\begin{figure}
\begin{center}
\includegraphics[scale=0.45, trim={0.5cm 0 0.5cm 0}]{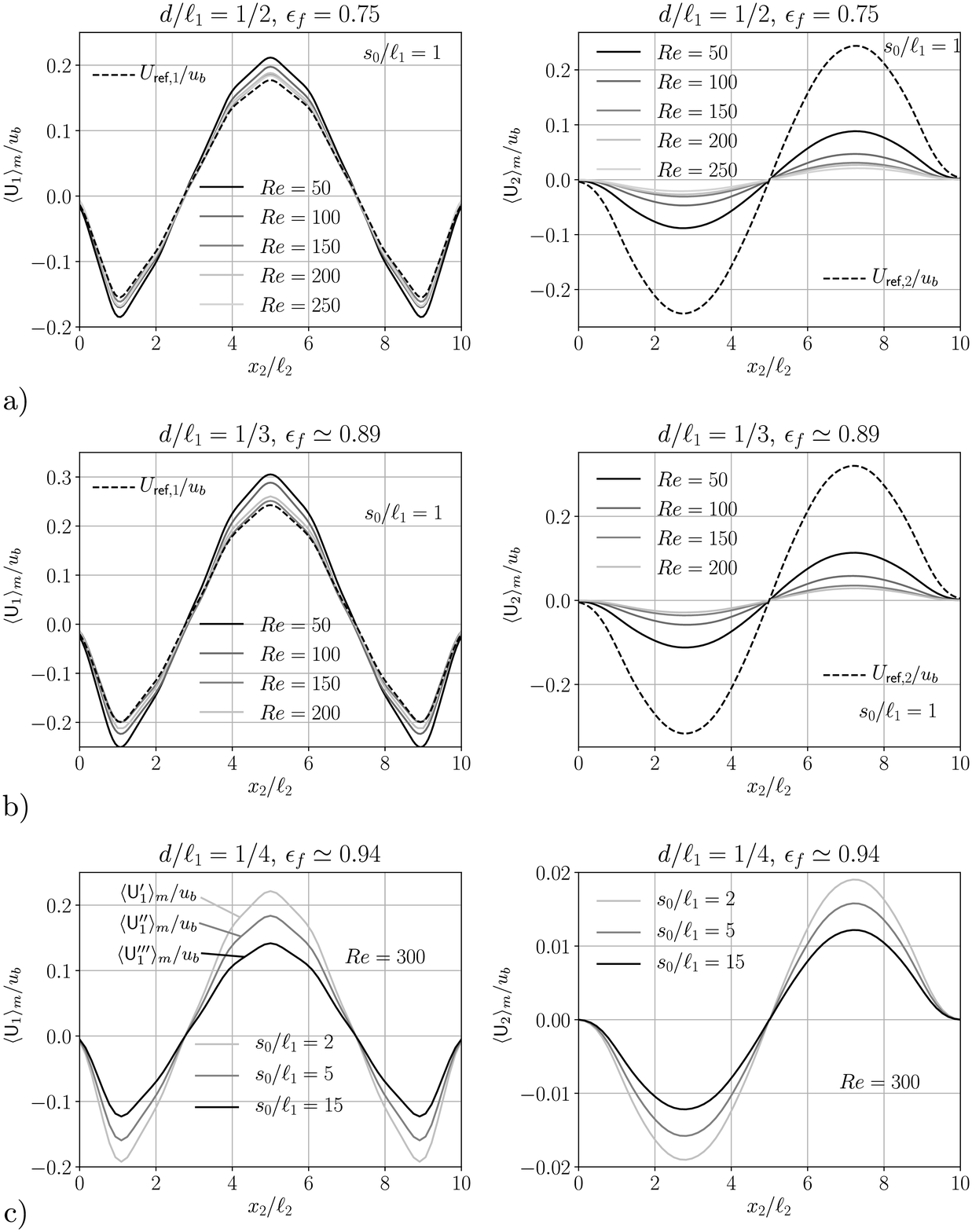}
\caption{Macro-scale velocity modes for quasi-developed flow in a channel array ($N_1 =80, N_2=10$, $H/\ell_1 =1$, $\ell_1/\ell_2 =1$, $s_N/\ell_1 = 10$) for different Reynolds numbers $\Rey \in \left\{50, 100, 150, 200, 250, 300 \right\}$ and porosities $\epsilon_f$, as well as different positions $s_0$ of the first cylinder row.}
\label{fig: Quasi-developed macro-scale flow modes}
\end{center}
\end{figure}

The dimensionless velocity modes $\langle \velamp \rangle_m/u_b$ in the quasi-developed flow region for each of the channel flows depicted in figure \ref{fig: Quasi-developed macro-scale flow} are shown in figure \ref{fig: Quasi-developed macro-scale flow modes}.
Figure \ref{fig: Quasi-developed macro-scale flow modes} (a,b)  demonstrates that the shapes of the dimensionless macro-scale velocity modes at different Reynolds numbers are very similar when the inlet velocity profile and geometry of the channel and array remain unaltered. 
Therefore, their shapes can be represented by a Reynolds-number-independent reference profile $\boldsymbol{U}_{\text{ref}}(x_2)/u_b$ such that 
\begin{equation}
\label{eq: scaling law amplitude macro-scale velocity mode}
\frac{\langle \velamp \rangle_m}{u_b} = \frac{\boldsymbol{U}_{\text{ref}}}{u_b} \left( 1+ \frac{c_2}{\Rey} \right) \boldsymbol{\cdot} \left(\boldsymbol{e}_1 \boldsymbol{e}_1 + \lambda \ell_2 \boldsymbol{e}_2 \boldsymbol{e}_2 \right)\,.
\end{equation}
This correlation form is based on the observation that the mode amplitude scales inversely linear with the Reynolds number: $\langle \velampx \rangle_m/u_b \sim (1+c_2/\Rey)$.
We have for instance $c_2\simeq 9.6$ for $d/\ell_1=1/2$, and $c_2\simeq 13$ for $d/\ell_1=1/3$, if $\ell_1/\ell_2=H/\ell_1=1$, $N_2=10$ and $s_0/\ell_1=1$.
Furthermore, the form of this correlation takes into account that
\begin{equation}
\label{eq: transversal macro-scale velocity mode as integral}
\langle \velampy \rangle_m(x_2) =  \lambda \ell_2 \displaystyle \int_{0}^{x_2/\ell_2}  \langle \velampx \rangle_m(r_2) dr_2
\qquad \mbox{with} \qquad r_2 \triangleq \frac{x_2}{\ell_2}\,,
\end{equation}
due to the fact that $\boldsymbol{\nabla} \boldsymbol{\cdot} \velamp = \boldsymbol{\lambda} \boldsymbol{\cdot} \velamp$, by virtue of (\ref{eq: macro-scale continuity equation}) and (\ref{eq: quasi-developed macro-scale flow}).
We remark that (\ref{eq: transversal macro-scale velocity mode as integral}) implies that only the component $U_{\text{ref,0}}$ and eigenvalue $\lambda$ are essential to reconstruct $\boldsymbol{U}_{\text{ref}}$.

Figure \ref{fig: Quasi-developed macro-scale flow modes} (c) illustrates that the dimensionless macro-scale velocity modes for the same Reynolds number $\Rey$ and same inlet velocity profile are all self-similar, apart from the scaling factor $C_{\velamp}/u_b$.
This scaling factor $C_{\velamp}/u_b$, which determines the absolute value of the macro-scale velocity at the onset point $x_{\text{quasi-periodic}} /\ell_1$, clearly depends on the distance over which the flow is developing and thus the distance $s_0 /\ell_1$ between the channel inlet and the first cylinder row, as one can observe in figure \ref{fig: Quasi-developed macro-scale flow} (c).
The perturbation size $C_{\velamp}/u_b$ obviously increases when the flow has less distance to adapt itself to the array geometry, as the inlet profile has been fixed here.

The macro-scale velocity modes from figure \ref{fig: Quasi-developed macro-scale flow modes} have some features in common with the two-dimensional velocity modes that occur in quasi-developed Poiseuille flow (\cite{Sadri2002, Asai2004}). 
The profile of $\langle \velampx \rangle_m/u_b$ has a similar W-shape over the width of the channel, while the profile of $\langle \velampy \rangle_m/u_b$ has a similar sinusoidal shape.
In addition, the inversely linear relationship between the mode amplitude and the Reynolds number has also been discovered for quasi-developed Poiseuille flow at Reynolds numbers below 500 (\cite{SadriPhD1997}).
A difference, however, is that the modes of the macro-scale velocity do not satisfy a no-slip condition at the side walls of the channel.
Besides, the modes from figure \ref{fig: Quasi-developed macro-scale flow modes} differ in sign with respect to the modes observed in quasi-developed Poiseuille flow.
The sign of the macro-scale velocity modes implies that the macro-scale velocity decreases along the center of the channel when the flow develops, whereas for quasi-developed Poiseuille flow, the velocity at the center of the channel tends to increase when the flow develops.
The different sign of the perturbation size is an outcome of the specific inlet velocity profile and $s_0/\ell_1$-ratio chosen for the direct numerical simulation of the channel flow here.
In figure \ref{fig: Quasi-developed macro-scale flow modes} (a,b), the ratio $s_0/\ell_1$ is small, so that there occurs a velocity peak and  overshoot of the macro-sale velocity in the center of the channel shortly after the flow enters the array.
As figures \ref{fig: Quasi-developed macro-scale flow} and \ref{fig: Quasi-developed macro-scale flow modes} (c) show, this velocity peak decreases when $s_0/\ell_1$ increases.
Eventually, when the distance between the channel inlet and first cylinder row becomes very large, a negative perturbation size, thus an undershoot of the macro-scale velocity at the center of the channel can be expected, in agreement with the experiments for quasi-developed Poiseuille flow \citep{Asai2004}).

\begin{figure}
\begin{center}
\includegraphics[scale=0.48, trim={0cm 0cm 0cm 0}]{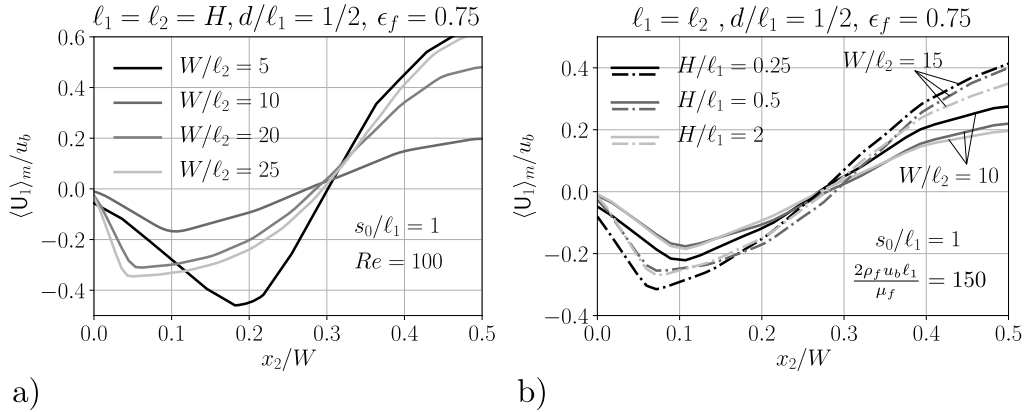}
\caption{Macro-scale velocity modes for quasi-developed flow in a channel array ($N_1 =60$, $\ell_1/\ell_2 =1$, $s_0/\ell_1 = 1$, $s_N/\ell_1 = 10$) for different aspect ratios $W/\ell_2 \in \left\{5, 10, 20, 25 \right\}$ (a) and height ratios $H/\ell_1 \in \left\{0.25, 0.5, 0.25 \right\}$ (b) of the channel.
The porosity $\epsilon_f$ and the Reynolds number based on the cylinder spacing $2\rho_f u_b \ell_1/\mu_f$ have been kept fixed.
}
\label{fig: Quasi-developed macro-scale flow modes: aspect ratio and height}
\end{center}
\end{figure}

The influence of the geometry on the shape of the mode and the perturbation size is demonstrated in figure \ref{fig: Quasi-developed macro-scale flow modes: aspect ratio and height} for a single porosity and a single Reynolds number based on the cylinder spacing.
Only half of the channel is shown, because the mode is symmetric with respect to the center plane $x_1=W/2$.
In figure \ref{fig: Quasi-developed macro-scale flow modes: aspect ratio and height} (a), we can see that a larger aspect ratio $W/H=W/\ell_2$ comes along with a smaller perturbation size for the chosen inlet conditions.
In addition, the location of the minimum of the mode $\langle \velampx \rangle_m$ moves closer to the side walls of the channel, when the aspect ratio increases.
In figure \ref{fig: Quasi-developed macro-scale flow modes: aspect ratio and height} (b), we can see that a larger channel height comes along with a smaller perturbation size until $H/\ell_1 >1$, after which the mode shape and perturbation remain constant.
The influence of the channel height and the aspect ratio at the macro-scale is of course in line with the scaling laws for the velocity mode $\velamp$, which are discussed in (\cite{Buckinx2022}), so they are not treated here again.


In order to obtain an exact local closure problem for quasi-developed macro-scale flow, exact closure solutions for developed flow macro-scale flow need to be achieved first.
Such closure solutions are presented in the next section, for the region of uniform macro-scale flow.
Afterwards, they are extended to include the side-wall region.

\section{Local Closure for Developed Macro-Scale Flow}
\label{sec: Local Closure for Developed Macro-Scale Flow}
\subsection{Exact Local Closure in the Region of Uniform  Macro-Scale Flow}
\label{subsec: Local Closure for Developed Macro-Scale Flow}
In the region $\Omega_{\text{uniform}}$, where the macro-scale velocity $\boldsymbol{U}$ is uniform, also the closure force $\boldsymbol{b}$ adopts a uniform value, which is given by
\begin{equation}
\label{eq: definition uniform closure force}
\boldsymbol{b}^{\star} \triangleq
\langle \boldsymbol{n}_{0} \boldsymbol{\cdot}  (-p^{\star}_f \boldsymbol{I} + \mu_f \boldsymbol{\nabla} \boldsymbol{u}^{\star}_f) \delta_{0} \rangle_m \,.
\end{equation}
Following the derivations from \cite{BuckinxBaelmans2015}, it can be proved that the uniform closure force (\ref{eq: definition uniform closure force}) is exactly represented by a spatially independent apparent permeability tensor $\boldsymbol{K}_{\text{uniform}} \left(\boldsymbol{U} \right)$, which depends on $\boldsymbol{U}$:
\begin{equation}
\label{eq: uniform interfacial force and permeability tensor}
\boldsymbol{b}^{\star}  
= -\mu_f \boldsymbol{K}^{-1}_{\text{uniform}} \boldsymbol{\cdot} \boldsymbol{U}\,.
\end{equation}
The latter apparent permeability tensor is determined by the closure variables $\closurevarvelperiodic$ and $\closurevarpressperiodic$, which define the mappings $\boldsymbol{u}^{\star}(\boldsymbol{r}) = \closurevarvelperiodic(\boldsymbol{r}) \boldsymbol{\cdot} \boldsymbol{U}' $ and $p^{\star}(\boldsymbol{r})= \mu_f \closurevarpressperiodic(\boldsymbol{r}) \boldsymbol{\cdot} \boldsymbol{U}'$:
\begin{equation}
\label{eq: definition permeability tensor uniform macro-scale flow}
\boldsymbol{K}^{-1}_{\text{uniform}} 
\triangleq
\epsilon_{fm}^{-1}
\langle \boldsymbol{n}_{0} \boldsymbol{\cdot}  (\boldsymbol{I}\closurevarpressperiodicf- \boldsymbol{\nabla} \closurevarvelperiodicf ) \delta_{0} \rangle_m 
 \,.
\end{equation}
We note that, apart from the macro-scale velocity $\boldsymbol{U} \triangleq \epsilon_f\boldsymbol{U}'$,  $\boldsymbol{K}^{-1}_{\text{uniform}}$ thus also depends on the fluid properties $\mu_f$ and $\rho_f$, as well as the geometrical parametrization of $\Gamma_{0}$.
Furthermore, since $\langle \closurevarvelperiodic \rangle^f_m$ and $\langle \closurevarpressperiodic \rangle^f_m$ are constant in $\Omega_{\text{uniform}} $, this apparent permeability tensor is equivalent to the one defined by the classical closure problem of \cite{Whitaker1996}:
\begin{equation}
\label{eq: alternative definition permeability tensor uniform macro-scale flow}
\boldsymbol{K}^{-1}_{\text{uniform}} 
=
\epsilon_{fm}^{-1}
\langle \boldsymbol{n}_{0} \boldsymbol{\cdot}  (\boldsymbol{I} \devclosurevarpressperiodicf - \boldsymbol{\nabla} \devclosurevarvelperiodicf  ) \delta_{0} \rangle_m 
 \,.
\end{equation}

\begin{figure}
\begin{center}
\includegraphics[scale=0.8]{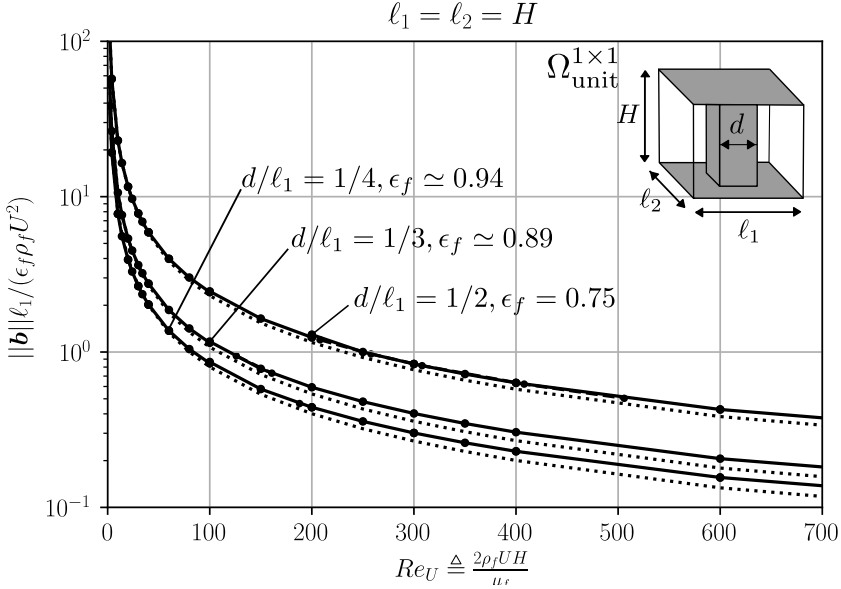}
\caption{Magnitude of the closure force $ \boldsymbol{b}=\boldsymbol{b}^{\star}$, as a function of the macro-scale velocity $U$ and porosity $\epsilon_f$ for a channel with an array of equidistant square in-line cylinders, in the region of uniform macro-scale flow.
The solid lines correspond to the Darcy-Forchheimer relationship $(\ell_1^2/\kappa_D) \Rey_U + \ell_1/\kappa_F$, while the dotted lines correspond to the Darcy relationship $(\ell_1^2/\kappa_D) \Rey_U$.}
\label{fig:Developed closure force uniform macro-scale flow}
\end{center}
\end{figure}

Because the uniform closure force equals the constant macro-scale pressure gradient in $\Omega_{\text{uniform}}$,
\begin{equation}
\label{eq: macro-scale momentum equation uniform}
\boldsymbol{b}^{\star} = \boldsymbol{\nabla}\langle p \rangle_m = \boldsymbol{\nabla }\mathrm{P}_{\text{dev}} \epsilon_{f}\,,
\end{equation}
as shown in (\cite{BuckinxBaelmans2015}), both $\boldsymbol{b}^{\star}$ and $\boldsymbol{K}_{\text{uniform}}$ can be governed as a function of $\boldsymbol{U}$ by solving the periodically developed flow equations given in (\cite{BuckinxBaelmans2015, Buckinx2022}).
Due to the periodicity conditions (\ref{eq: flow periodicity main flow direction}) and (\ref{eq: flow periodicity transversal direction}) in $\Omega_{\text{uniform}}$, the periodically developed flow equations need to be solved on just a single unit cell $\Omega_{\text{unit}}^{n_1 \times n_2}(\boldsymbol{x}) $ with $ \boldsymbol{x} \in \Omega_{\text{uniform}} $.
The result of this classical closure procedure, which has been adopted in many studies (see e.g. Refs), is illustrated in figure \ref{fig:Developed closure force uniform macro-scale flow}, for a channel with an array of equidistant in-line square cylinders.

Figure \ref{fig:Developed closure force uniform macro-scale flow} shows the magnitude $\Vert \boldsymbol{b}^{\star} \Vert $ of the closure force in the region of uniform macro-scale flow, as a function of the macro-scale velocity $U$ and the porosity $\epsilon_f$ of the array.
As mentioned earlier, a single height-to-spacing ratio $H/\ell_1=1$ has been chosen.
The depicted data points, whose estimated accuracy is $1.5\%$ according to our mesh-refinement study, were obtained by numerically solving the periodically developed flow equations on a unit cell $\Omega_{\text{unit}}^{1 \times 1}$, as the actual flow field in the channel is known to become periodic for $n_1=n_2=1$.
Each unit-cell simulation was performed on a mesh of about $1.5$ million cells.
Because we found numerically the same velocity field on $\Omega_{\text{unit}}^{1 \times 1}$ whether $\boldsymbol{\nabla}\mathrm{P}_{\text{dev}}$ or $\boldsymbol{U}$ was imposed, so no flow bifurcations appeared, we argue that the relation between $\boldsymbol{b}$ and $\boldsymbol{U}$ is a one-to-one relationship over the range of Reynolds numbers $\Rey_U \triangleq 2\rho_f U H/\mu_f$ shown in figure \ref{fig:Developed closure force uniform macro-scale flow}.
In agreement with the literature (e.g. \cite{Koch1997, LasseuxAhmadi2011}), this one-to-one relationship
satisfies in good approximation the Darcy-Forchheimer relationship,
\begin{equation}
\label{eq: Darcy-Forchheimer correlation uniform macro-scale flow}
\boldsymbol{b}^{\star} \simeq - \frac{\mu_f}{\kappa_D} \boldsymbol{U} - \frac{\rho_f}{\kappa_F} U \boldsymbol{U} \,,
\end{equation}
so that $\Vert \boldsymbol{b}^{\star} \Vert \ell_1/(\rho_f U^2) \simeq (\ell_1^2/\kappa_D)/ \Rey_U + \ell_1/\kappa_F$ and $\boldsymbol{K}^{-1}_{\text{uniform}} \simeq \left(\kappa_D^{-1} + \rho_fU/(\mu_f\kappa_F)\right) \boldsymbol{I}$.
By means of a least-square fitting procedure, the Darcy coefficient  $\ell_1^2/\kappa_D$ was found to equal $233$ when $d/\ell_1=1/2$, 
$108$ when $d/\ell_1=1/3$, and $80$ when  $d/\ell_1=1/4$ for the geometries and Reynolds numbers in figure \ref{fig:Developed closure force uniform macro-scale flow}.
On the other hand, the corresponding Forchheimer coefficients were found to be much smaller: $\ell_1/\kappa_F= 0.05$ for $d/\ell_1=1/2$, $\ell_1/\kappa_F= 0.035$ for $d/\ell_1=1/3$, and $\ell_1/\kappa_F= 0.03$ for $d/\ell_1=1/2$.
With these values for the Darcy and Forchheimer coefficients, the relationship (\ref{eq: Darcy-Forchheimer correlation uniform macro-scale flow}) deviates no more than $3$ to $5\%$ with respect to the data points depicted in figure \ref{fig:Developed closure force uniform macro-scale flow}, while its mean relative deviation is below $2\%$.
Nevertheless, this relationship actually ignores the occurrence of a weak-inertia regime as discussed in (\cite{LasseuxAhmadi2011}).

%
%

\begin{figure}
\begin{center}
\includegraphics[scale=0.53]{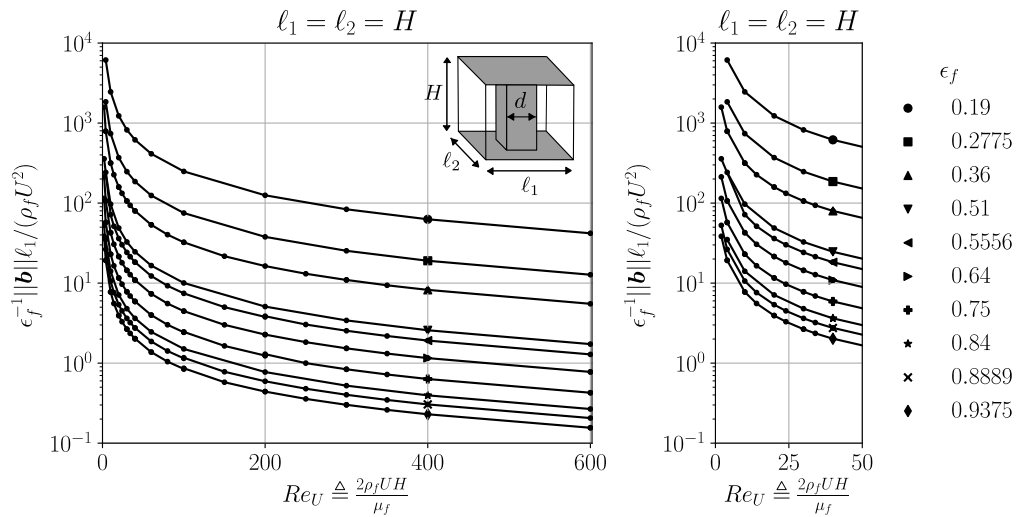}
\caption{Magnitude of the closure force = as a function of the macro-scale velocity $U$ and porosity $\epsilon_f$, for a channel with an array of equidistant square in-line cylinders, in the region of uniform macro-scale flow.}
\label{fig:Developed closure force uniform macro-scale flow: porosity}
\end{center}
\end{figure}

The Darcy and Forchheimer coefficients can be correlated to the porosity of the array via the empirical formulas
\begin{align}
\label{eq: Darcy-Forchheimer coefficients uniform macro-scale flow}
\ell_1^2/\kappa_D = 44.8 \biggl( 1-(1-\epsilon_f)^{0.63} \biggr)^{-3} 
\qquad \mbox{and} \qquad
\ell_1/\kappa_F = 0.13 \left(\frac{1-\epsilon_f}{\epsilon_f^2} \right)\,,
\end{align}
when the cylinder spacing is equal to the channel height: $H=\ell_1=\ell_2$.
These formulas predict the 200 data points for the closure force in figure (\ref{fig:Developed closure force uniform macro-scale flow: porosity}) with a mean relative error of $2\%$ and a maximum relative error below $9\%$.
The estimated discretization error on the data points is below $1.5$ to 2$\%$.
The correlations (\ref{eq: Darcy-Forchheimer coefficients uniform macro-scale flow}) show that if inertia effects at the macro-scale are neglected by setting the Forchheimer coefficient to $\ell_1/\kappa_F=0$, the closure force is underestimated by almost $8\%$ at the highest Reynolds number $\Rey=300$.
Inertia effects at the macro-scale will thus remain rather small for $H/\ell_1<1$, since $\ell_1/\kappa_F$ tends to decrease when the channel height decreases.
Furthermore, for $H/\ell_1>1$, both the Darcy and Forchheimer coefficients are expected to become constant and independent of $H/\ell_1$ (\cite{Vangeffelen2021}).

\subsection{Exact Local Closure in the Region of Developed Macro-Scale Flow}
\label{subsec: Exact Local Closure in the Region of Developed Macro-Scale Flow}
In order to extend the former closure solutions towards the entire region of developed macro-scale flow $\Omega_{\text{dev}}$, we can express the
closure force $\boldsymbol{b}$ explicitly in terms of the periodically developed velocity and pressure fields via (\ref{eq: developed pressure distribution}):
\begin{equation}
\label{eq: no-slip force developed flow region}
\boldsymbol{b}_{\text{dev}} = 
\boldsymbol{b}^{\star} +
\langle p \rangle^f_m   \boldsymbol{\nabla} \epsilon_{fm} 
- \langle p^{\star} \rangle^f_m   \boldsymbol{\nabla} \epsilon_{fm} 
- \epsilon_{fm} \boldsymbol{G}_m \boldsymbol{\cdot}\boldsymbol{\nabla}\mathrm{P}_{\text{dev}} \,. 
\end{equation}
Here, the developed macro-scale pressure field is given by
\begin{equation}
\langle p \rangle^f_m = 
\boldsymbol{\nabla}\mathrm{P}_{\text{dev}} \boldsymbol{\cdot} (\boldsymbol{x} +
\boldsymbol{m}) + 
\langle p^{\star} \rangle^f_m \,, 
\end{equation}
and $\boldsymbol{G}_m \triangleq - \nabla \boldsymbol{m}$ denotes the gradient of the first intrinsic spatial moment over the fluid region $\boldsymbol{m}$ (\cite{BuckinxBaelmans2015, Quintard1994a}).
Additionally, we remark that although $\boldsymbol{b}^{\star}$ in (\ref{eq: no-slip force developed flow region}) is still defined by (\ref{eq: definition uniform closure force}), its value is no longer spatially uniform in $\Omega_{\text{dev}}$, so that (\ref{eq: uniform interfacial force and permeability tensor}) does not hold here.

We will now show that the closure force in the developed flow region (\ref{eq: no-slip force developed flow region}) can be represented by an exact, yet spatially dependent, apparent permeability tensor $\boldsymbol{K}_{\text{dev}} \left(\boldsymbol{U}_{\text{dev}}, x_2\right)$, such that 
\begin{equation}
\label{eq: no-slip force permeability tensor developed flow region}
\boldsymbol{b}_{\text{dev}} = \langle p \rangle^f_m   \boldsymbol{\nabla} \epsilon_{fm} -\mu_f \boldsymbol{K}_{\text{dev}}^{-1} \boldsymbol{\cdot} \boldsymbol{U}_{\text{dev}}\,.
\end{equation}
To this end, we introduce the closure variable $\boldsymbol{\xi}$, which maps at each position in $\Omega_{\text{dev}}$, the uniform macro-scale velocity $\boldsymbol{U}'$ to the actual macro-scale velocity $\boldsymbol{U}'_{\text{dev}} \triangleq \epsilon_{fm}^{-1} \boldsymbol{U}_{\text{dev}}$:
\begin{equation}
\label{eq: shape developed macro-scale velocity profile}
\boldsymbol{U}_{\text{dev}}'(x_2) \triangleq \boldsymbol{\xi} \left(\boldsymbol{U}, x_2\right) \boldsymbol{\cdot} \boldsymbol{U}'
\qquad \mbox{and} \qquad 
\xi\left(x_2\right) \triangleq \frac{U_{\text{dev}}'(x_2)}{U'}\,.
\end{equation}
The closure variable $\boldsymbol{\xi} = \xi \boldsymbol{I}$ thus determines the shape $\xi$ of the macro-scale velocity profile in $\Omega_{\text{dev}}$.
We remark that $\boldsymbol{\xi}$ is solely a function of $\boldsymbol{U}$, if the fluid properties and geometry of $\Gamma_0$ are fixed.
After all, $\boldsymbol{u}_f^{\star}$, $p_f^{\star}$ and hence $\boldsymbol{U}_{\text{dev}}$ in $\Omega_{\text{dev}}$ can be obtained, at least in principle, by solving the periodically developed flow equations  on one or two rows of the array, for a fixed value of $\boldsymbol{U}$ (\cite{Buckinx2022}).

With the aid of $\boldsymbol{\xi}$, we can define $\boldsymbol{K}_{\text{dev}}$ in $\Omega_{\text{dev}}$ in terms of the same closure mapping as the one that was introduced to define $\boldsymbol{K}_{\text{uniform}}$ in $\Omega_{\text{uniform}}$  
(\ref{eq: definition permeability tensor uniform macro-scale flow}):
\begin{equation}
\label{eq: closure mapping entire developed region}
\begin{aligned}
\boldsymbol{u}(\boldsymbol{r}) &= 
\closurevarvelperiodic(\boldsymbol{r}) \boldsymbol{\cdot} \boldsymbol{U}' = 
\closurevarvelperiodic(\boldsymbol{r})
\boldsymbol{\cdot}
\boldsymbol{\xi}^{-1} \left(\boldsymbol{U}, x_2\right)  \boldsymbol{\cdot}\boldsymbol{U}_{\text{dev}}'(x_2)  \,,\\
p^{\star}(\boldsymbol{r}) &= 
\mu_f \closurevarpressperiodic(\boldsymbol{r}) \boldsymbol{\cdot} \boldsymbol{U}' =
\mu_f \closurevarpressperiodic(\boldsymbol{r}) \boldsymbol{\cdot} 
\boldsymbol{\xi}^{-1}\left(\boldsymbol{U}, x_2\right) \boldsymbol{\cdot} \boldsymbol{U}_{\text{dev}}'(x_2) \,.
\end{aligned}
\end{equation}
Substitution of the closure mapping (\ref{eq: closure mapping entire developed region}) in the expression for the closure force yields for the first term on the right-hand side of (\ref{eq: no-slip force developed flow region}):
\begin{equation}
\label{eq: first contribution permeability tensor periodically developed flow}
\boldsymbol{b}^{\star} =
-\mu_f  \epsilon_{fm}^{-1} \langle \boldsymbol{n}_0 \boldsymbol{\cdot} \left( \boldsymbol{I} \closurevarpressperiodicf - \boldsymbol{\nabla} \closurevarvelperiodicf \right) \delta_0\rangle_m \boldsymbol{\cdot}
\boldsymbol{\xi}^{-1} \boldsymbol{\cdot}  \boldsymbol{U}_{\text{dev}}\,.
\end{equation}
For the third term on the right-hand side of (\ref{eq: no-slip force developed flow region}), we obtain
\begin{equation}
\label{eq: second contribution permeability tensor periodically developed flow}
\langle p^{\star} \rangle^f_m   \boldsymbol{\nabla} \epsilon_{fm} =
\mu_f  \epsilon_{fm}^{-1} \boldsymbol{\nabla} \epsilon_{fm} \langle\closurevarpressperiodic \rangle^f_m \boldsymbol{\cdot}
\boldsymbol{\xi}^{-1} \boldsymbol{\cdot}  \boldsymbol{U}_{\text{dev}}\,.
\end{equation}
Furthermore, substitution of (\ref{eq: closure mapping entire developed region}) in the last term of (\ref{eq: no-slip force developed flow region}) results in
\begin{equation}
\label{eq: third contribution permeability tensor periodically developed flow}
\epsilon_{fm} \boldsymbol{G}_m \boldsymbol{\cdot}\boldsymbol{\nabla}\mathrm{P}_{\text{dev}} =
\mu_f \boldsymbol{G}_m \boldsymbol{\cdot} \boldsymbol{K}^{-1}_{\text{uniform}} \boldsymbol{\cdot}
\boldsymbol{\xi}^{-1} \boldsymbol{\cdot}  \boldsymbol{U}_{\text{dev}}\,,
\end{equation}
by virtue of (\ref{eq: uniform interfacial force and permeability tensor}) and (\ref{eq: definition permeability tensor uniform macro-scale flow}).
Finally, we retrieve from (\ref{eq: first contribution permeability tensor periodically developed flow}) - (\ref{eq: third contribution permeability tensor periodically developed flow}) that the apparent permeability tensor in the developed flow region (\ref{eq: no-slip force permeability tensor developed flow region}) is given by
\begin{equation}
\label{eq: exact expresion permeability tensor periodically developed flow}
\boldsymbol{K}_{\text{dev}}^{-1} =
\left( 
\boldsymbol{K}^{-1}_{\text{dev,main}} + 
\boldsymbol{G}_m \boldsymbol{\cdot} \boldsymbol{K}^{-1}_{\text{uniform}}
\right) \boldsymbol{\cdot}
\boldsymbol{\xi}^{-1} \,,
\end{equation}
where $ \boldsymbol{\xi} = \langle \closurevarvelperiodic \rangle^f_m $ and
\begin{equation}
\label{eq: exact expresion Km periodically developed flow}
\boldsymbol{K}^{-1}_{\text{dev,main}}  \triangleq \epsilon_{fm}^{-1} \langle \boldsymbol{n}_0 \boldsymbol{\cdot} \left( \boldsymbol{I} \closurevarpressperiodicf - \boldsymbol{\nabla} \closurevarvelperiodicf \right) \delta_0\rangle_m
+ 
\epsilon_{fm}^{-1} \boldsymbol{\nabla} \epsilon_{fm} \langle\closurevarpressperiodic \rangle^f_m  \,.
\end{equation}
In order to determine $\boldsymbol{K}_{\text{dev,main}}$ and $ \boldsymbol{\xi}$, the closure problem (\ref{eq: closure problem developed region}) from appendix \ref{app: Closure Problem for Developed Macro-Scale Flow} must be solved on (a part of) the side-wall region in $\Omega_{\text{dev}}$.

We note that the functional dependence of $\boldsymbol{K}_{\text{dev}}$ on $\boldsymbol{U}=\epsilon_f \boldsymbol{U}'$ is easily transformed into a functional dependence on $\boldsymbol{U}_{\text{dev}}$ via (\ref{eq: shape developed macro-scale velocity profile}), so the notations $\boldsymbol{K}_{\text{dev}} \left(\boldsymbol{U}, x_2\right)$, $\boldsymbol{K}_{\text{dev}} \left(\boldsymbol{U}', x_2\right)$ and $\boldsymbol{K}_{\text{dev}} \left(\boldsymbol{U}_{\text{dev}}, x_2\right)$ are all equivalent.
Evidently, if $\boldsymbol{x} \in \Omega_{\text{uniform}}$, it holds that $\xi\left(\boldsymbol{U}, x_2 \right)=1$ and $\boldsymbol{G}_m =0$ (\cite{BuckinxBaelmans2015}), so that in $ \Omega_{\text{uniform}}$ we recover again $\boldsymbol{K}_{\text{dev}} \left(\boldsymbol{U}, x_2 \right) = \boldsymbol{K}_{\text{dev,main}}\left(\boldsymbol{U}, x_2 \right)= \boldsymbol{K}_{\text{uniform}} \left(\boldsymbol{U} \right)$, in agreement with (\ref{eq: definition permeability tensor uniform macro-scale flow}).

\subsection{Approximative Local Closure in the Region of Developed Macro-Scale Flow}
\label{subsec: Approximative Local Closure in the Region of Developed Macro-Scale Flow}
Although the exact definition (\ref{eq: exact expresion permeability tensor periodically developed flow}) may be interesting in itself for theoretical reasons, it has limited practical value due to the complexity of the closure problem (\ref{eq: closure problem developed region}).
Nevertheless, it can be used as a starting point for accomplishing approximative closure.
Hereto, we first assume that $\boldsymbol{G}_m \simeq 0$, or
\begin{equation}
\label{eq: main contribution for developed apparent permeability tensor}
\boldsymbol{K}_{\text{dev}} \simeq \boldsymbol{\xi} \boldsymbol{\cdot}
\boldsymbol{K}_{\text{dev,main}} \,,
\end{equation}
since the spatial moment $\boldsymbol{m}$ of the cylinder array in $\Omega_{\text{sides}} $ can be neglected for the double volume-averaging filter of (\ref{eq: definition weighting function double volume average}).
Secondly, we assume that the variation of $\langle \closurevarpressperiodic \rangle^f_m$ with $x_2$ in the  region $\Omega_{\text{dev}} \setminus \Omega_{\text{uniform}}$ is small, so that $\langle \closurevarpressperiodic \rangle^f_m$ can be treated as a constant and be moved within the averaging operator $\langle \; \rangle_m$:
\begin{equation}
\label{eq: constant average mapping field pressure}
\boldsymbol{K}^{-1}_{\text{dev,main}}\simeq
\epsilon_{fm}^{-1} \langle \boldsymbol{n}_0 \boldsymbol{\cdot} ( \boldsymbol{I} \devclosurevarpressperiodicf  - \boldsymbol{\nabla}\closurevarvelperiodicf ) \delta_0\rangle_m \,.
\end{equation}
Also the assumption that $\langle \closurevarpressperiodic \rangle^f_m(x_2)$ is virtually constant, is not so restrictive, as it implies that the constant macro-scale pressure gradient outside the side-wall region is maintained within the side-wall region: $\boldsymbol{\nabla} \langle p \rangle^f_m \simeq \boldsymbol{\nabla} \mathrm{P}_{\text{dev}} $ because of $\boldsymbol{\nabla} \langle p^{\star} \rangle^f_m \simeq 0$ in $\Omega_{\text{dev}}$. 
If we compare (\ref{eq: constant average mapping field pressure}) with 
(\ref{eq: definition permeability tensor uniform macro-scale flow}) and 
(\ref{eq: alternative definition permeability tensor uniform macro-scale flow}), we see that we thus may use the approximation
\begin{equation}
\label{eq: side-wall region permeability approximately uniform}
\boldsymbol{K}_{\text{dev,main}} \simeq
\boldsymbol{K}_{\text{uniform}}
\qquad \mbox{or} \qquad
\boldsymbol{K}_{\text{dev}} \simeq \boldsymbol{\xi} \boldsymbol{\cdot}\boldsymbol{K}_{\text{uniform}} \,,
\end{equation}
provided that the variation of $\boldsymbol{K}_{\text{dev,main}}$ with $x_2$ is much smaller than the variation of $\xi$ with $x_2$ in the side-wall region $\Omega_{\text{sides}}$, or more precisely the region $\Omega_{\text{dev}} \setminus \Omega_{\text{uniform}}$.
This last condition is true as long as $\boldsymbol{b} \simeq \boldsymbol{\nabla} \mathrm{P}_{\text{dev}} $ in $\Omega_{\text{dev}}$, and  
holds for all the flow conditions and array geometries investigated in this work.

\begin{figure}
\begin{center}
\includegraphics[scale=0.47]{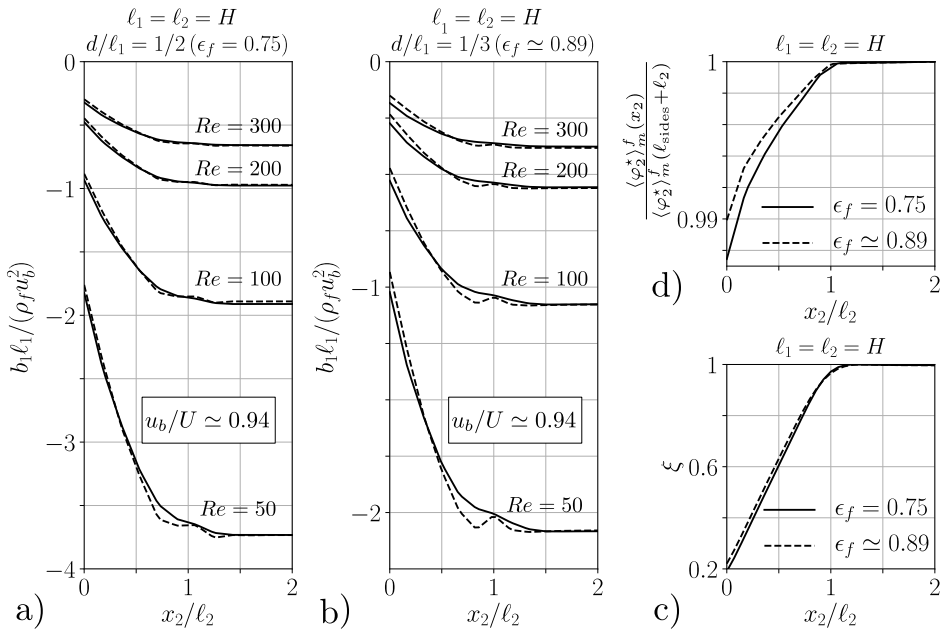}
\caption{
(a,b) Closure force for developed macro-scale flow in the side-wall region of a channel with an array of equidistant in-line cylinders, for different Reynolds numbers $\Rey$ and different porosities $\epsilon_f$.
The solid lines ($-$) represent the component $b_1$ of the actual closure force along the main flow direction.
The dashed lines (-$\,$-) represent the approximation for $b_1$ given by (\ref{eq: side-wall region permeability approximately uniform}).
(c,d) Also the two closure variables $\xi$ and $\langle \closurevarpressperiodic \rangle^f_m$ which determine the permeability tensor in the side-wall region are shown.
}
\label{fig: Developed closure in side-wall region}
\end{center}
\end{figure}

In practise, the approximation (\ref{eq: side-wall region permeability approximately uniform}) is only useful if one knows or can estimate the shape of the developed macro-scale velocity profile $\xi$ a priori.
Our numerical results indicate that in a channel array of square in-line cylinders, $\xi$ varies in a first approximation linearly with the coordinate $x_2$ perpendicular to the the side walls, so that 
\begin{equation}
\label{eq: linear macro-scale velocity shape profile}
\xi\left(\boldsymbol{U}, x_2\right)   \simeq 
\left\{
\begin{aligned}
&\frac{x_2 + \ell_{\text{slip}} }{\ell_{2} + \ell_{\text{slip}} }
&&\qquad \mbox{for} ~~ x_2 \in (0,\ell_{2}) \,, \\ 
& ~~~~~~1
&&\qquad \mbox{for} ~~ x_2 \in (\ell_{2},W-\ell_{2}) \,, \\ 
&\frac{N_2 \ell_2 - x_2 + \ell_{\text{slip}}'}{\ell_{2} + \ell_{\text{slip}}'}
&&\qquad \mbox{for} ~~ x_2 \in (W-\ell_{2},W)  \,.
\end{aligned}
\right.
\end{equation}
Here, the slip length $\ell_{\text{slip}}$ is defined by
\begin{equation}
\label{eq: slip length and slip condition for developed macro-scale flow}
\left.\frac{\partial \langle u_1 \rangle_m}{\partial x_2} \right \vert_{x_2=0} = 
\frac{1}{\ell_{\text{slip}}} \left. \langle u_1  \rangle_m \right \vert_{x_2=0} \,,
\end{equation}
while $\ell_{\text{slip}}'$ is similarly defined for $x_2=W$. 
We note that $\partial \langle u_1 \rangle_m/\partial x_2= d U_{\text{dev}}/dx_2 $. 

Both slip lengths for the velocity profile in (\ref{eq: linear macro-scale velocity shape profile}) are equal when the channel flow exhibits symmetry with respect to the plane $x_2=W/2$.
According to our numerical results, the shape of the velocity profile $\xi$ is virtually independent of the magnitude of the macro-scale velocity, since inertial effects on the macro-scale flow are small: $\xi\left(\boldsymbol{U}, x_2\right) \simeq \xi\left(x_2\right)$.
So, the velocity profile $\xi$ and the slip lengths depend only on the geometry of the cylinder array, just like the velocity profile for fully-developed flow in a channel depends only on the geometry of the channel's cross section.
For symmetric flow in cylinder arrays with $\ell_1=\ell_2= H$ and $\ell_{\text{slip}}=\ell_{\text{slip}}'$, we found that over the Reynolds number range $\Rey \in (25, 300)$ we have $\ell_{\text{slip}}/\ell_1 = 0.24 \pm 0.01$ for $d/\ell_1=1/2$,  $\ell_{\text{slip}}/\ell_1 = 0.30 \pm 0.01$ for $d/\ell_1=1/3$ and $\ell_{\text{slip}}/\ell_1 = 0.31 \pm 0.01$ for $d/\ell_1=1/4$.

As illustrated in figure \ref{fig: Developed closure in side-wall region} (a, b), the combination of the approximations (\ref{eq: side-wall region permeability approximately uniform}) and (\ref{eq: linear macro-scale velocity shape profile}) is quite accurate over the range of investigated flow conditions and geometries shown here, i.e. for $\Rey \in (50, 300)$, $H/\ell_1 \in (0.25, 2)$, $d/\ell_1 \in \left\{1/2, 1/3, 1/4 \right\}$.
The relative error for $b_1\triangleq \boldsymbol{b} \boldsymbol{\cdot} \boldsymbol{e}_1$ due to the approximations is less than $5\%$ almost everywhere.
A maximum relative error of $12\%$ occurs near $x_2 =0$, because there the linear approximation (\ref{eq: linear macro-scale velocity shape profile}) overestimates the smoother actual shape of $\xi$, which is shown in figure \ref{fig: Developed closure in side-wall region} (c). 
If the exact shape of the velocity profile $\xi$ would have been used, an exact reconstruction of the closure force in the main flow direction would have been achieved, since $b_1 = \xi  \boldsymbol{\nabla} \mathrm{P}_{\text{dev}} \boldsymbol{\cdot} \boldsymbol{e}_1$.
Hence, the only approximation made here is that $b_2 \simeq 0$.
However, it can be seen from figure \ref{fig: Developed closure in side-wall region} (d) that the latter approximation, as well as the underlying assumption (\ref{eq: constant average mapping field pressure}) are justified, because $\langle \closurevarpressperiodic  \rangle^f_m$ is indeed almost constant for $\Rey \in (50, 300)$, $H=\ell_1=\ell_2$  and $d/\ell_1 \in \left\{1/2, 1/3, 1/4 \right\}$.
In particular, it is observed that $\langle \closurevarpressperiodic  \rangle^f_m(x_2)\simeq \langle \closurevarpressperiodic  \rangle^f_m(\ell_{\text{sides}} + \ell_2)$ is satisfied within a relative margin of about $1\%$, independently of the Reynolds number $\Rey$.

\begin{figure}
\begin{center}
\includegraphics[scale=0.5]{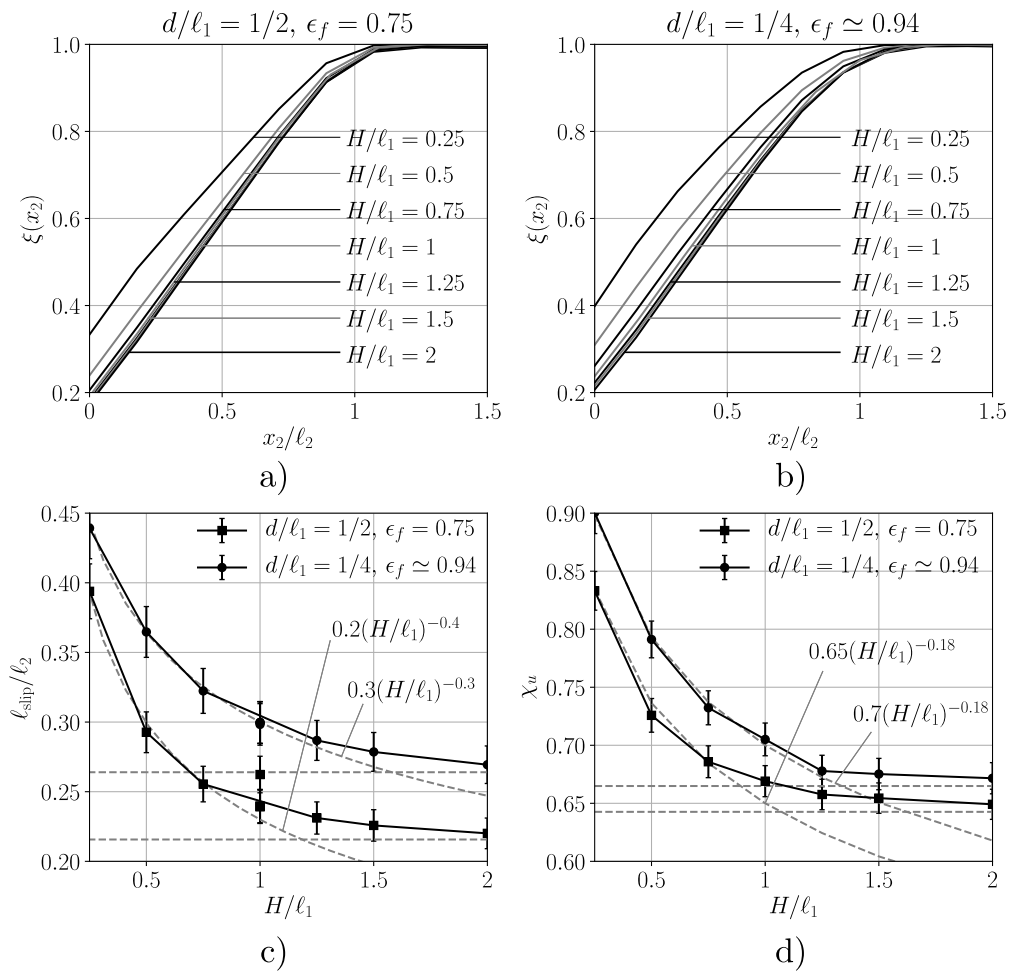}
\caption{(a,b) Shape $\xi$ of the developed macro-scale velocity profile in the side-wall region of a channel with an array of equidistant square in-line cylinders. The velocity profile $\xi$ is shown for different ratios of the channel height $H$ to the cylinder spacing $\ell_1=\ell_2$, as well as for different porosities $\epsilon_f$.
Also the corresponding slip length $\ell_{\text{slip}}$ (c) and displacement factor $\chi_u$ (d) for the flow rate in the side-wall region  are shown.  
}
\label{fig: Macro-scale effects in side-wall region}
\end{center}
\end{figure}

The velocity profile $\xi$, and therefore the permeability tensor in the side-wall region, merely depend on the ratio of the channel height $H$ to the cylinder spacing $\ell_1=\ell_2$ for a fixed porosity $\epsilon_f$.
The relationship between $\xi$ and $H/\ell_2$ is shown in figure \ref{fig: Macro-scale effects in side-wall region} (a,b) for two porosities, $\epsilon_f=0.75$ and $\epsilon_f\simeq 0.94$.
According to this figure, the profile $\xi(x_2)$ becomes almost linear in $x_2$ when the channel height is equal to or greater than the cylinder spacing.
Still, its first derivative $d\xi/dx_2$ is not exactly a constant.
In particular in the neighbourhood of the core region, $x_2/\ell_2 \simeq 1$, the first derivative exhibits a discontinuity, indicating a jump in the macro-scale stress $\mu_f d\langle u_1 \rangle_m/dx_2$ at that location.
At the side wall, the first derivative indicates the slip length: $ \left. d\xi/dx_2 \right \vert_{x_2=0} =1/(\ell_2+ \ell_{\text{slip}})$.

The dependence of the slip length $\ell_{\text{slip}}$ on the ratio of the channel height to cylinder spacing $H/\ell_2$ is shown in more detail in figure \ref{fig: Macro-scale effects in side-wall region} (c).
Just like the velocity profile $\xi$, the slip length clearly becomes independent of the channel height, when the channel height is much larger than the cylinder spacing.
This happens due to the fact that when the cylinders are relatively long, so $H/\ell_2 \gg 1$, the flow patterns around each cylinder are no longer affected by the plate surfaces, nor the distance $H$ between them.

This explains why also the displacement factor $\chi_u$ for the flow rate in the side-wall region becomes independent of $H/\ell_2$ for $H/\ell_2 \gg 1$, as shown in figure \ref{fig: Macro-scale effects in side-wall region} (d).
The latter is defined as the ratio of the mass flow rate through the side-wall region to the mass flow rate through the core region, multiplied with the ratio of the cross-sectional area of the core region to that of the side-wall region.
Therefore, it allows us to calculate the magnitude of the uniform macro-scale velocity in the core of the channel, from the bulk velocity:
\begin{equation}
\label{eq: displacement factor side-wall region}
U = \frac{N_2}{N_2 - 2N_{\text{sides}}(1-\chi_u)} u_b\,,
\end{equation} 
if we accept that $\ell_{\text{sides}}$ is a multiple of the unit cell width $\ell_2$, so that $N_{\text{sides}} \triangleq  1 + \lfloor \ell_{\text{sides}}/\ell_2 \rfloor$.
We remark that the factor $2N_{\text{sides}}$ in (\ref{eq: displacement factor side-wall region}) corresponds to the number of cylinders in a row parallel to the axis $x_2$, located in the region $\Omega_{\text{dev}} \setminus \Omega_{\text{uniform}}$.

The displacement factor $\chi_u$ can be seen to correlate well with the slip length $\ell_{\text{slip}}$, if we compare figure \ref{fig: Macro-scale effects in side-wall region} (c) and (d).
For smaller channel heights, they both obey an empirical power-law scaling with the ratio of the channel height to cylinder spacing, although their exponents differ for the same porosity.  
Under a few assumptions, their mutual relationship can be made explicit.
The first assumption is that $N_{\text{sides}}=1$ so that $\chi_u = \left. \langle u_1 \rangle \right \vert_{x_2=\ell_{\text{sides}}/2} /U$, which is commonly the case as $\ell_{\text{sides}} \simeq \ell_2$.
The second assumption is that the macro-scale velocity has a linear profile $\xi$ and closely matches the volume-averaged velocity in the side-wall region: $\langle u_1 \rangle_m \simeq \langle u_1 \rangle$.
Then, we have $\chi_u \simeq (\ell_{\text{sides}}/2 + \ell_{\text{slip}})/(\ell_{\text{sides}} + \ell_{\text{slip}}) = \left. \xi \right \vert_{x_2=\ell_{\text{sides}}/2}$.
However, this approximation for $\chi_u$ has a typical accuracy of around $20\%$ for the data in figure \ref{fig: Macro-scale effects in side-wall region} (d), because it is only holds when a single volume-averaging operator is used.

Before we close our discussion on the local closure for the developed region $\Omega_{\text{dev}}$, we emphasize that  the permeability tensor $\boldsymbol{K}_{\text{dev}}$ accounts for nearly all macro-scale momentum transport due to gradients of the macro-scale velocity field.
So, the momentum equation in $\Omega_{\text{dev}}$ reduces to $\boldsymbol{\nabla}\langle p \rangle^f_m \simeq 
\mu_f \boldsymbol{K}_{\text{dev}}^{-1} \boldsymbol{\cdot} \boldsymbol{U}'_{\text{dev}}$, even though its exact 
form is $\boldsymbol{\nabla}\langle p \rangle_m = \mu_f d^2 U_{\text{dev}}/dx_2^2 \boldsymbol{e}_1 - \rho_f d \langle u_2 \boldsymbol{u} \rangle_m/d x_2 + \boldsymbol{b}$.
The reason is that based on the estimates $x_2 \sim  \ell_{\text{sides}}$ and $U_{\text{dev}} \sim U \sim u_b$, we typically have $ \Vert \langle u_2 \boldsymbol{u} \rangle_m \Vert / u_b^2 \ll 1$ for laminar channel flows, so that the Brinkmann term $\mu_f d^2 U_{\text{dev}}/dx_2^2$ and the momentum dispersion in the side-wall region are usually negligible at moderate Reynolds numbers $\rho u_b \ell_{\text{sides}} /\mu_f = \Rey \,(\ell_{\text{sides}} / H) \gg 1$.

\subsection{Validity of the Local Closure Problem for Developed Macro-Scale Flow}
\label{subsec: Validity of the Local Closure Problem for Developed Macro-Scale Flow}
As Whitaker's permeability tensor (\ref{eq: uniform interfacial force and permeability tensor}) and its extension in the side-wall region (\ref{eq: no-slip force permeability tensor developed flow region}) are exact once the flow has become periodically developed, the onset point of periodically developed flow, $x_{\text{periodic}}$, is a key parameter to characterize the validity of the preceding local closure models. 
The scaling laws for $x_{\text{periodic}}$ and their relation to the eigenvalue $\lambda$ in the region of quasi-periodically developed flow, have been discussed in (\cite{Buckinx2022}).
Still, the macro-scale flow can often be treated as developed even upstream of the point $x_{\text{periodic}}$.

For instance, if $x_{\text{dev}}$ would have been defined as the $x_1$-section in $\Omega_{\text{predev}}$ for which $\Vert \langle\boldsymbol{u}\rangle_m\Vert=0.99 \Vert \boldsymbol{U}_\text{dev} \Vert$ at $x_2=W/2$, we would have found that $x_{\text{dev}}/\ell_1 \simeq 0.15 \Rey + 4$ for $\Rey \in (50, 300)$ for the flow depicted in 
figure \ref{fig: Quasi-developed macro-scale flow} (a).
In that case, we thus have that $x_{\text{dev}} \simeq 0.9 x_{\text{periodic}}$, according to the definition of $x_{\text{periodic}}$ adopted in (\cite{Buckinx2022}).
Similarly, also for the flow depicted in figure \ref{fig: Quasi-developed macro-scale flow} (b), we then would have found that $x_{\text{dev}}/\ell_1 \simeq 0.19 \Rey + 4.7$ for $\Rey\in (50, 300)$, which is about $10\%$ smaller than $x_{\text{periodic}}/\ell_1$ in that case (\cite{Buckinx2022}).

The approximation $\langle \boldsymbol{u}  \rangle_m \simeq \boldsymbol{U}_{\text{dev}}(\boldsymbol{x}) $ is thus accurate even upstream of the periodically developed flow region, $\Omega_{\text{periodic}}$, as it can be seen from figure \ref{fig:Macro-Scale Flow Field and Regions Re 50 porosity 0.75}.
Therefore, the distinction between $\Omega_{\text{periodic}}$ and $\Omega_{\text{dev}}$ is rather a subtlety from a macro-scale point of view.
As a matter of fact, also the distinction between $\Omega_{\text{periodic}} \setminus \Omega_{\text{sides}}$ and $\Omega_{\text{uniform}}$ appears to be a theoretical subtlety, as
the approximation $\langle \boldsymbol{u}  \rangle_m(\boldsymbol{x}) \simeq \boldsymbol{U} $ 
holds well for $\boldsymbol{x} \in (\Omega_{\text{periodic}} \setminus \Omega_{\text{sides}})$.
The explanation for these observations is two-fold.
Firstly, gradients of the macro-scale velocity in regions like $x_2 \in (\ell_{\text{sides}},\ell_{\text{sides}}+ n_2 \ell_2 )$ or $x_1 \in (x_{\text{periodic}},x_{\text{periodic}} + n_1 \ell_1 )$ occur over a spatial distance smaller than the filter radius and thus tend to be rather small.
Secondly, the distance from the side walls at which the flow displays transversal flow periodicity, $\ell_{\text{sides}}$, has been found to be smaller than the transversal spacing of the cylinders, $\ell_2$, for all the flow conditions investigated in (\cite{Buckinx2022}).

\section{Local Closure for Quasi-Developed Macro-Scale Flow}
\label{sec: Local Closure for Periodically Developed Flow}
\subsection{Exact Local Closure for Quasi-Developed Macro-Scale Flow}
\label{subsec: Exact Local Closure for Quasi-Developed Macro-Scale Flow}

In the region of quasi-periodically developed flow, $\Omega_{\text{quasi-periodic}}$, the quasi-developed closure force is given by
\begin{equation}
\label{eq: quasi-developed closure force}
\boldsymbol{b}_{\text{quasi-dev}} \simeq  \boldsymbol{b}_{\text{dev}} +
\langle \boldsymbol{n}_0 \boldsymbol{\cdot} \left(- \pressamp_f \boldsymbol{I} + \mu_f \boldsymbol{\nabla} \velamp_f \right) \delta_0 \rangle_m \exp \left(-\boldsymbol{\lambda} \boldsymbol{\cdot} \boldsymbol{x} \right) \,,
\end{equation}
as it follows from (\ref{eq: final form macro-scale no-slip force}) after substitution of (\ref{eq: quasi-periodically developed flow}) and (\ref{eq: quasi periodically developed pressure field}).
Again, the approximation symbol in (\ref{eq: quasi-developed closure force}) can be replaced by an equality sign, when a matched filter instead of a double-volume averaging operator is chosen.

Also in $\Omega_{\text{quasi-periodic}}$ there exists an apparent permeability tensor $\boldsymbol{K}_{\text{quasi-dev}} \left(\langle \boldsymbol{u} \rangle_m, \boldsymbol{x} \right)$ to represent the closure force:
\begin{equation}
\label{eq: no-slip force permeability tensor quasi-developed flow region}
\boldsymbol{b}_{\text{quasi-dev}} \simeq \langle p \rangle^f_m   \boldsymbol{\nabla} \epsilon_{fm} -\mu_f \boldsymbol{K}_{\text{quasi-dev}}^{-1} \boldsymbol{\cdot} \langle \boldsymbol{u} \rangle_m \,,
\end{equation}
where the quasi-developed macro-scale pressure field in (\ref{eq: no-slip force permeability tensor quasi-developed flow region}) is given by (\ref{eq: macro-scale pressure quasi-developed flow}).
The apparent permeability tensor for quasi-developed macro-scale flow, $\boldsymbol{K}_{\text{quasi-dev}}$, is spatially dependent, but exact in the case of a matched filter, in the sense that for a matched filter, expressions (\ref{eq: no-slip force permeability tensor quasi-developed flow region}) and (\ref{eq: macro-scale pressure quasi-developed flow}) are no longer approximations.

In order to determine the structure of the tensor $\boldsymbol{K}_{\text{quasi-dev}}$, we introduce a closure mapping which maps the uniform macro-scale velocity in $\Omega_{\text{uniform}}$ to the amplitudes of the velocity and pressure modes in  $\Omega_{\text{quasi-periodic}}$:
\begin{equation}
\label{eq: closure mapping quasi-developed region}
\velamp(\boldsymbol{r}) = 
\boldsymbol{\Psi}(\boldsymbol{r}) \boldsymbol{\cdot} \boldsymbol{U}' \qquad
\mbox{and} \qquad
\pressamp(\boldsymbol{r}) = 
\mu_f \boldsymbol{\psi}(\boldsymbol{r}) \boldsymbol{\cdot} \boldsymbol{U}' \,,
\end{equation}
This mapping exists, as we have shown that both $\velamp$ and $\pressamp$ can be reconstructed from $\boldsymbol{U}'$ from the flow equations on one or two rows of the array, when the flow is quasi-periodically developed (\cite{Buckinx2022}).
The mapping gives rise to a closure problem for the closure variables $\boldsymbol{\Psi}$ and $\boldsymbol{\psi}$ which is included in appendix \ref{app: Closure Problem for Quasi-Developed Macro-Scale Flow}.
This closure problem can still be considered a local closure problem,  although it has to be solved on a transversal row of the array, instead of a single unit cell.
Furthermore, it defines the transformation $\boldsymbol{\zeta}$ from $\boldsymbol{U}'$ to $\langle \velamp \rangle^f_m$:
\begin{equation}
\label{eq: mapping macro-scale velocity mode to constant developed macro-scale velocity}
\langle \velamp \rangle^f_m(x_2) = \boldsymbol{\zeta}(\boldsymbol{U}, x_2)  \boldsymbol{\cdot} \boldsymbol{U}'
\qquad \mbox{with} \qquad 
\boldsymbol{\zeta} \triangleq \langle \boldsymbol{\Psi} \rangle^f_m \,,
\end{equation}
as well as the mapping from $\langle \velamp \rangle^f_m$ to $\langle \boldsymbol{u} \rangle^f_m$:
\begin{equation}
\label{eq: mapping macro-scale velocity mode to constant developed macro-scale velocity quasi-developed}
\langle \boldsymbol{u} \rangle^f_m( \boldsymbol{x}) = \left[
\boldsymbol{\xi}(\boldsymbol{U}, x_2) + \boldsymbol{\zeta}(\boldsymbol{U}, x_2) 
\exp \left(-\boldsymbol{\lambda} \boldsymbol{\cdot} \boldsymbol{x} \right) 
\right] \boldsymbol{\cdot}   \boldsymbol{U}'\,,
\end{equation}
by virtue of (\ref{eq: quasi-developed macro-scale flow}) and (\ref{eq: shape developed macro-scale velocity profile}).

After substitution of the closure mapping (\ref{eq: closure mapping quasi-developed region}), we find that the last term of (\ref{eq: quasi-developed closure force}) can be represented as  
\begin{equation}
\label{eq: representation permeability amplitude quasi-developed macro-scale flow} 
\langle \boldsymbol{n}_0 \boldsymbol{\cdot} \left(- \pressamp_f \boldsymbol{I} + \mu_f \boldsymbol{\nabla} \velamp_f \right) \delta_0 \rangle_m = - \mu_f  \boldsymbol{\mathsf{K}}^{-1} \boldsymbol{\cdot}  \langle \velamp \rangle_m\,,
\end{equation}
where the tensor $\boldsymbol{\mathsf{K}}$ is defined by
\begin{equation}
\boldsymbol{\mathsf{K}}^{-1}  \triangleq
\epsilon_{fm}^{-1} \langle \boldsymbol{n}_0 \boldsymbol{\cdot} \left(\boldsymbol{I}\boldsymbol{\psi}_f  - \boldsymbol{\nabla} \boldsymbol{\Psi}_f \right) \delta_0 \rangle_m  \boldsymbol{\cdot}  \boldsymbol{\zeta}^{-1} \,.
\end{equation}
This result implies that the apparent permeability tensor from (\ref{eq: no-slip force permeability tensor quasi-developed flow region}) is given by
\begin{equation}
\label{eq: exact structure permeability tensor quasi-developed flow}
\boldsymbol{K}_{\text{quasi-dev}}^{-1} \triangleq 
\left[
\boldsymbol{K}_{\text{dev}}^{-1} \boldsymbol{\cdot} \boldsymbol{\xi}   + 
\boldsymbol{\mathsf{K}}^{-1} \boldsymbol{\cdot} \boldsymbol{\zeta} 
\exp \left(-\boldsymbol{\lambda} \boldsymbol{\cdot} \boldsymbol{x} \right) 
\right] 
\boldsymbol{\cdot}
\left[
\boldsymbol{\xi} + \boldsymbol{\zeta} 
\exp \left(-\boldsymbol{\lambda} \boldsymbol{\cdot} \boldsymbol{x} \right) 
\right]^{-1}
 \,,
\end{equation}
as one can verify from (\ref{eq: quasi-developed closure force}) and (\ref{eq: no-slip force permeability tensor developed flow region}).
We thus conclude that the apparent permeability tensor for quasi-developed macro-scale flow, $\boldsymbol{K}_{\text{quasi-dev}}$, consists of two contributions.
On the one hand, it contains a contribution from the apparent permeability tensor for developed macro-scale flow, $\boldsymbol{K}_{\text{dev}}$.
This contribution becomes equal to the apparent permeability tensor for uniform macro-scale flow, outside of the side-wall region: $\boldsymbol{K}_{\text{dev}}^{-1} \boldsymbol{\cdot} \boldsymbol{\xi}= \boldsymbol{K}_{\text{uniform}}^{-1}$.
On the other hand, it contains a contribution from the permeability tensor $\boldsymbol{\mathsf{K}}$, which expresses the resistance against the macro-scale velocity mode $\langle \velamp \rangle_m$ that occurs on top of the developed macro-scale flow, as long as the flow is still developing.
Since both contributions can be determined from the continuity and momentum equations for quasi-periodically developed flow on a transversal row of the array, so can the apparent permeability tensor for quasi-developed macro-scale flow.
However, while both contributions vary only along the coordinate $x_2$ in the transversal direction, the apparent permeability tensor for quasi-developed macro-scale flow also varies along the main flow direction $x_1$.
In particular, $\boldsymbol{K}_{\text{quasi-dev}}$ decays exponentially in the main flow direction at a rate imposed by the eigenvalue $\lambda$.

Finally, we have deduced that $\boldsymbol{K}_{\text{quasi-dev}}$ is affected by the tensor $\boldsymbol{\zeta}$, whose magnitude indicates the relative magnitude of the macro-scale velocity mode, as $\Vert \boldsymbol{\zeta} \boldsymbol{\cdot} \boldsymbol{e}_1 \Vert = \Vert \langle \velamp \rangle^f_m \Vert / U'$ according to (\ref{eq: mapping macro-scale velocity mode to constant developed macro-scale velocity}).
So, $\boldsymbol{K}_{\text{quasi-dev}}$ is affected by the scaling factor 
$C_{\velamp}/U'$, which expresses how strong the quasi-developed macro-scale flow is perturbed from the uniform developed macro-scale flow.
To make this dependency more explicit, we may write $\boldsymbol{\zeta} = \boldsymbol{\zeta}_{\text{ref}} \, C_{\velamp}/U'$, such that $\boldsymbol{\zeta}_{\text{ref}} \boldsymbol{\cdot} \boldsymbol{e}_1 = \langle \velamp \rangle^f_m / C_{\velamp}$.
In that case, we have $\boldsymbol{e}_1 \boldsymbol{\cdot} \langle \boldsymbol{\zeta}_{\text{ref}} \rangle_{\text{row}} \boldsymbol{\cdot} \boldsymbol{e}_1 = 1$ because $C_{\velamp} \triangleq \langle \velamp \rangle_{\text{row}}$, which shows that $\boldsymbol{\zeta}_{\text{ref}}$ does not depend on the relative perturbation size $C_{\velamp}^{+} \triangleq C_{\velamp}/U'$, nor the manner in which the flow develops.
Since the perturbation size often tends to be relatively small, i.e. $\boldsymbol{\zeta} \ll \boldsymbol{I}$ as $ \Vert\langle \velamp \rangle^f_m \Vert \ll  U'$, the following approximation for  $\boldsymbol{K}_{\text{quasi-dev}}$ is usually acceptable:
\begin{equation}
\label{eq: approximate structure permeability tensor quasi-developed flow}
\boldsymbol{K}_{\text{quasi-dev}}^{-1} \simeq
\boldsymbol{K}_{\text{dev}}^{-1}   + 
C_{\velamp}^{+} \boldsymbol{\mathsf{K}}^{-1} \boldsymbol{\cdot} \boldsymbol{\zeta}_{\text{ref}} \boldsymbol{\cdot} \boldsymbol{\xi}^{-1} 
\exp \left(-\boldsymbol{\lambda} \boldsymbol{\cdot} \boldsymbol{x} \right) 
\,.
\end{equation}
The previous expression elucidates that the term $C_{\velamp}^{+} \boldsymbol{\mathsf{K}}^{-1} \boldsymbol{\cdot} \boldsymbol{\zeta}_{\text{ref}}  \boldsymbol{\cdot} \boldsymbol{\xi}^{-1} 
\exp \left(-\boldsymbol{\lambda} \boldsymbol{\cdot} \boldsymbol{x} \right) $ is an asymptotic correction to the apparent permeability tensor from the classical closure problem for developing flow.
This correction term will allow us to analyse the validity of the classical closure problem for quasi-developed macro-scale flow, as well as its closure mapping  (see appendix \ref{app: Closure Mapping for Quasi-Developed Macro-Scale Flow}). 
Yet, before we present this validity analysis, the underlying assumptions and solutions of the classical closure problem are discussed first, in the next subsection.

\subsection{Approximative Local Closure for Quasi-Developed Macro-Scale Flow \\Outside the Side-Wall Region}
\label{subsec: Approximative Local Closure for Quasi-Developed Macro-Scale Flow Outside the Side-Wall Region}  
For a first approximation, the closure force in $\Omega_{\text{quasi-periodic}}$, and possibly even $\Omega_{\text{predev}}$, may be modelled according to the classical closure problem of \cite{Whitaker1996}. 
In that case, the closure force $\boldsymbol{b}$ has an approximative \textit{local} representation of the form
\begin{equation}
\label{eq: definition approx permeability tensor}
\boldsymbol{b}_{\text{approx}} \triangleq - \mu_f \boldsymbol{K}^{-1}_{\text{approx}}\left(\langle \boldsymbol{u} \rangle_m \right) \boldsymbol{\cdot} \langle \boldsymbol{u} \rangle_m \,,
\end{equation}
outside the side-wall region $\Omega_{\text{sides}}$, where $\boldsymbol{\nabla} \epsilon_{fm}=0$.
The apparent permeability tensor $\boldsymbol{K}_{\text{approx}}$ is defined in terms of the closure variables $\devclosurevarvel$ and $\devclosurevarpress$, through the mappings $\devtilde{\boldsymbol{u}}(\boldsymbol{r}) = \devclosurevarvel(\boldsymbol{r}) \boldsymbol{\cdot} \langle \boldsymbol{u} \rangle^f_m(\boldsymbol{x})$ and  $\devtilde{p} (\boldsymbol{r}) = \devclosurevarpress(\boldsymbol{r})\boldsymbol{\cdot} \langle \boldsymbol{u} \rangle^f_m(\boldsymbol{x})$:
\begin{equation}
\label{eq: definition approx permeability tensor deviation mapping}
\boldsymbol{K}^{-1}_{\text{approx}} 
\triangleq
\epsilon_{fm}^{-1}
\langle \boldsymbol{n}_{0} \boldsymbol{\cdot}  (\boldsymbol{I}\devclosurevarpressf - \boldsymbol{\nabla} \devclosurevarvelf  ) \delta_{0} \rangle_m \,.
\end{equation}
The classical closure problem itself, which governs an approximative solution for the deviation fields ($\devtilde{\boldsymbol{u}}$, $\devtilde{p}$) and the closure variables ($\devclosurevarvel$, $\devclosurevarpress$) as a function of $\langle \boldsymbol{u} \rangle_m $, is given by 
\begin{align}
\rho_f \left(\langle \boldsymbol{u} \rangle^f_m + \devtilde{\boldsymbol{u}}_f\right) \boldsymbol{\cdot}  \boldsymbol{\nabla}
\devtilde{\boldsymbol{u}}_f &= 
- \boldsymbol{\nabla} \devtilde{p}_f + 
\mu_f \nabla^2  \devtilde{\boldsymbol{u}}_f
- \epsilon_{fm}^{-1}\boldsymbol{b}  \,, 
\label{eq: deviation momentum equation}\\
\boldsymbol{\nabla}\boldsymbol{\cdot}  \devtilde{\boldsymbol{u}}_f &= 0  \,,
\label{eq: deviation continuity equation} \\
\devtilde{\boldsymbol{u}}_f(\boldsymbol{r}) &= \devtilde{\boldsymbol{u}}_f(\boldsymbol{r} + n_j \boldsymbol{l}_j) \,, \qquad
\devtilde{p}_f(\boldsymbol{r}) = \devtilde{p}_f(\boldsymbol{r} + n_j \boldsymbol{l}_j)
\,, 
\label{eq: periodic deviation fields closure}\\
\devtilde{\boldsymbol{u}} &= - \langle \boldsymbol{u} \rangle^f_m \qquad \mbox{for} ~~ \boldsymbol{r}  \in \Gamma_{0}  \,, 
\label{eq: no-slip condition deviation fields closure}\\
\langle \devtilde{\boldsymbol{u}} \rangle_m &=0 \,,
\label{eq: zero averaged deviation velocity} 
\end{align} 
where it is tacitly assumed that $\boldsymbol{r} \in \Omega_{\text{unit}}^{2\times 2}(\boldsymbol{x})$, $j \in\{1,2\}$ and $ n_j=1$ or $ n_j=2$.

In order to apply the classical closure problem (\ref{eq: deviation momentum equation}) - (\ref{eq: zero averaged deviation velocity}) to obtain the approximate relationship (\ref{eq: definition approx permeability tensor}) between $\langle \boldsymbol{u} \rangle_m$ and $\boldsymbol{b}$ for developing or quasi-developed macro-scale flow, the following assumptions (or approximations) have to be made, in line with Whitaker's original derivation.
First, the macro-scale momentum dispersion source must be negligible with respect to the closure force, so that only the latter appears in the momentum equation (\ref{eq: deviation momentum equation}):
\begin{equation}
\label{eq: closure force dominant wrt dispersion}
\boldsymbol{b} \gg \rho_f \boldsymbol{\nabla} \boldsymbol{\cdot} \boldsymbol{M}\,.
\end{equation} 

Secondly, the closure force $\boldsymbol{b}$ should depend only on the deviation fields $\devtilde{\boldsymbol{u}}$ and  $\devtilde{p}$, or at least, its 
direct dependence on the macro-scale velocity $\langle \boldsymbol{u} \rangle_m$ and pressure $\langle p \rangle_m$ should be of minor importance.
As shown by \cite{Quintard1994b}, for the filter (\ref{eq: definition weighting function double volume average}) this condition is automatically fulfilled, since 
in $\Omega_{\text{predev}} \setminus \Omega_{\text{sides}} $, it holds that
\begin{equation}
\label{eq: closure force deviation fields}
\boldsymbol{b} 
= \langle \boldsymbol{n}_{0} \boldsymbol{\cdot}  (-\devtilde{p}_f \boldsymbol{I} + \mu_f \nabla \devtilde{\boldsymbol{u}}_f) \delta_{0} \rangle_m \,.
\end{equation}
The reason is that in the part of $\Omega_{\text{predev}}$ where $\boldsymbol{\nabla} \epsilon_{fm}$ is zero, also the gradients of all other spatial moments like $\boldsymbol{\nabla}  \boldsymbol{m}$ are zero, due to the properties of the double volume-averaging operator (\ref{eq: definition weighting function double volume average}).

In the third place, the momentum equation (\ref{eq: deviation momentum equation}) incorporates the assumption that
\begin{equation}
\label{eq: macro-scale velocity gradient negligible in closure problem}
 \devtilde{\boldsymbol{u}}_f \boldsymbol{\cdot} \boldsymbol{\nabla}
\devtilde{\boldsymbol{u}}_f  \gg \devtilde{\boldsymbol{u}}_f \boldsymbol{\cdot} \boldsymbol{\nabla}
\langle \boldsymbol{u} \rangle^f_m \,.
\end{equation}

The last assumption behind this closure problem is the periodicity (\ref{eq: periodic deviation fields closure}) of the deviation fields in each unit cell $\Omega_{\text{unit}}^{2 \times 2}$ outside the side-wall region.
Due to the assumed periodicity of the deviation fields, both $\langle \boldsymbol{u} \rangle^f_m$ and $\boldsymbol{b}$ appear as spatially constant vectors in the classical closure problem, and their spatial variation within the unit cell is neglected:
\begin{equation}
\label{eq: no spatial variation of macro-scale velocity and closure force in unit cell}
\langle \boldsymbol{u} \rangle^f_m(\boldsymbol{r}) \simeq \langle \boldsymbol{u} \rangle^f_m(\boldsymbol{x} )
\qquad \mbox{and} \qquad
\boldsymbol{b}(\boldsymbol{r}) \simeq \boldsymbol{b}(\boldsymbol{x}) \,,
\end{equation}
$\forall \boldsymbol{r}  
 \in \Omega_{\text{unit}}^{2 \times 2}(\boldsymbol{x})$, if $\Omega_{\text{unit}}^{2 \times 2}(\boldsymbol{x}) \subset \left( \Omega_{\text{predev}} \setminus \Omega_{\text{sides}} \right) $.
 
Because of these four approximations, the classical closure problem is mathematically equivalent to the periodically developed flow equations \citep{BuckinxBaelmans2015b}:
\begin{equation}
\label{eq: equivalence classical closure problem permeability and uniform macro-scale flow}
\boldsymbol{K}_{\text{approx}}\left(\langle \boldsymbol{u} \rangle_m \right) = \boldsymbol{K}_{\text{uniform}}\left(\langle \boldsymbol{u} \rangle_m \right)\,.
\end{equation}
It thus yields the exact apparent permeability tensor for $\Omega_{\text{uniform}}$ when $\langle \boldsymbol{u} \rangle_m = \boldsymbol{U}$.
Yet, it has to be solved for different directions and magnitudes of $\langle \boldsymbol{u} \rangle_m$,
since the macro-scale velocity in $\Omega_{\text{predev}}$ may vary from point to point, whereas it is uniform in $\Omega_{\text{uniform}}$.

\begin{figure}
\begin{center}
\includegraphics[scale=0.53, trim=0cm 0cm 0cm 0cm, clip=true]{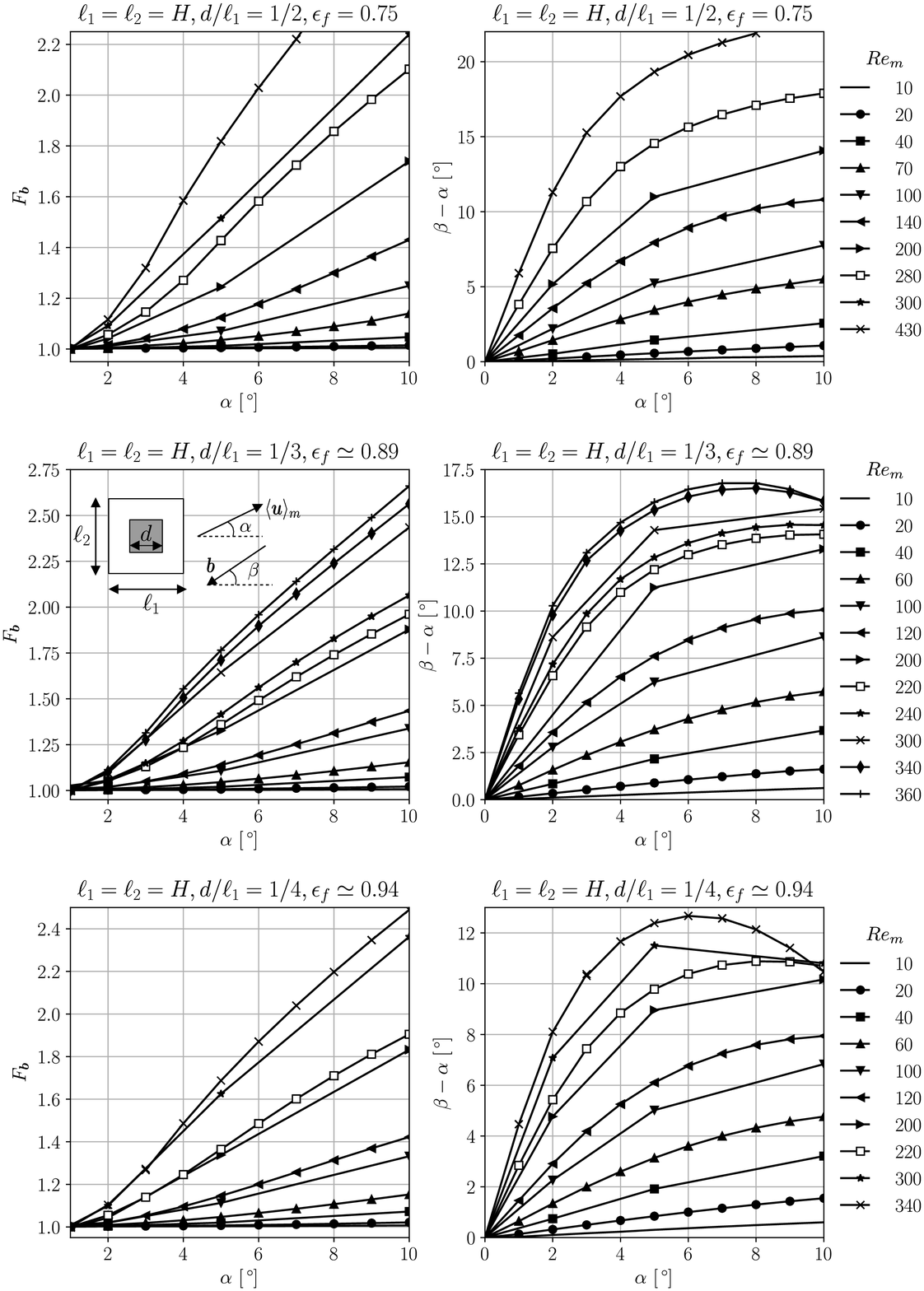}
\caption{
Relative magnitude $F_{\boldsymbol{b}}$ (left) and relative direction $\beta-\alpha$ (right) of the closure force $\boldsymbol{b}$, 
according to the classical closure problem for an array of equidistant in-line square cylinders.
The dependence on the Reynolds number $\Rey_m$ and angle of attack $\alpha$ of the macro-scale velocity $\langle \boldsymbol{u} \rangle_m$ are shown for different porosities $\epsilon_f$, and a fixed height $H=\ell_1$.
}
\label{fig: Closure problem for developing macro-scale flow}
\end{center}
\end{figure}

The numerical solution of the classical closure problem (\ref{eq: deviation momentum equation}) - (\ref{eq: zero averaged deviation velocity}) gives an approximation for the closure force $\boldsymbol{b}(\boldsymbol{x})$ at each point $\boldsymbol{x} \in \left(\Omega_{\text{predev}} \setminus \Omega_{\text{sides}}\right)$, as a function of the local Reynolds number $\Rey_m({\boldsymbol{x}}) \triangleq 2 \rho_f   \Vert \langle\boldsymbol{u} \rangle_m \Vert H/\mu_f$ based on the local macro-scale velocity $\langle\boldsymbol{u} \rangle_m (\boldsymbol{x})$, and the local direction of the macro-scale velocity $\boldsymbol{e}_s (\boldsymbol{x}) \triangleq \langle\boldsymbol{u} \rangle_m/ \Vert \langle\boldsymbol{u} \rangle_m \Vert$.
The local direction of the macro-scale velocity is more conveniently represented by the local angle of attack $\alpha \triangleq \arccos \left(\boldsymbol{e}_1 \boldsymbol{\cdot} \boldsymbol{e}_s \right)$, since the macro-scale flow in the channel is two-dimensional.

In figure \ref{fig: Closure problem for developing macro-scale flow}, the dependence of the magnitude and direction of the closure force $\boldsymbol{b}$ on $\Rey_m$ and $\alpha$, according to the classical closure problem, is illustrated for an array of equidistant in-line square cylinders.
The classical closure problem  was solved on a unit cell $\Omega_{\text{unit}}^{1\times 1}$, because the deviation fields $\devtilde{\boldsymbol{u}}$ and $\devtilde{p}$ are known to become periodic for $n_1=n_2=1$ in the periodically developed flow region. 
The angle of attack $\alpha$ is shown on the horizontal axis, while different Reynolds numbers $\Rey_m$ correspond to different markers, described by the legend for each porosity $\epsilon_f$ on the right.
On the left side of figure \ref{fig: Closure problem for developing macro-scale flow}, the magnitude $\Vert\boldsymbol{b} \Vert$ is expressed by the dimensionless factor $F_{\boldsymbol{b}}(Re_m, \alpha)$, which is defined by $\Vert \boldsymbol{b} \Vert \triangleq F_{\boldsymbol{b}} \Vert\boldsymbol{b}(U, 0) \Vert$.
Here, $\boldsymbol{b}(U, 0)$ denotes the closure force found for $\alpha=0$, in the case of a uniform macro-scale velocity $\langle \boldsymbol{u} \rangle_m = U\boldsymbol{e}_1$ of the same magnitude, as depicted in figure \ref{fig:Developed closure force uniform macro-scale flow}.
On the right side of figure \ref{fig: Closure problem for developing macro-scale flow}, the direction of the closure force is represented by the angle $\beta \triangleq \pi- \arccos \left(\boldsymbol{e}_1 \boldsymbol{\cdot} \boldsymbol{b}/\Vert \boldsymbol{b} \Vert \right)$.

It is observed that for small angles of attack, $\alpha \in (0,10 ^{\circ})$, the magnitude of the closure force increases when the angle of attack increases, especially at higher Reynolds numbers.
In the selected porosity range $\epsilon_f \in (0.75, 0.94)$ and for the chosen channel height $H=\ell_1=\ell_2$, the magnitude of the macro-scale force at an angle $\alpha=10 ^{\circ}$, is more than twice as large as for aligned flow ($\alpha=0$) with the same speed $U$, if the Reynolds number $\Rey_m$ lies above $300$.
On the other hand, if the Reynolds number is below $50$, the dependence of the magnitude of the macro-scale force force  on the angle of attack is rather small, since for any $\alpha \in (0,10 ^{\circ})$, the factor $F_{\boldsymbol{b}}$ is below $1.1$.

It can also be seen that the direction of the closure force deviates stronger from the direction of the macro-scale velocity, when the angle of attack or the Reynolds number becomes higher.
For instance, at a Reynolds number $\Rey_m$ above $100$, the difference in angle between both directions, $\beta -\alpha$, is almost $10 ^{\circ}$ for $\alpha=10 ^{\circ}$.
However, the difference in direction between the closure force and the macro-scale velocity does not increase monotonically with the angle of attack at a certain Reynolds number.
As a matter of fact, a maximum of $\beta -\alpha$ can be identified for each Reynolds number, beyond which the closure force becomes again more parallel to the macro-scale velocity.

\begin{figure}
\begin{center}
\includegraphics[scale=0.45, trim=0cm 0cm 0cm 0cm, clip=true]{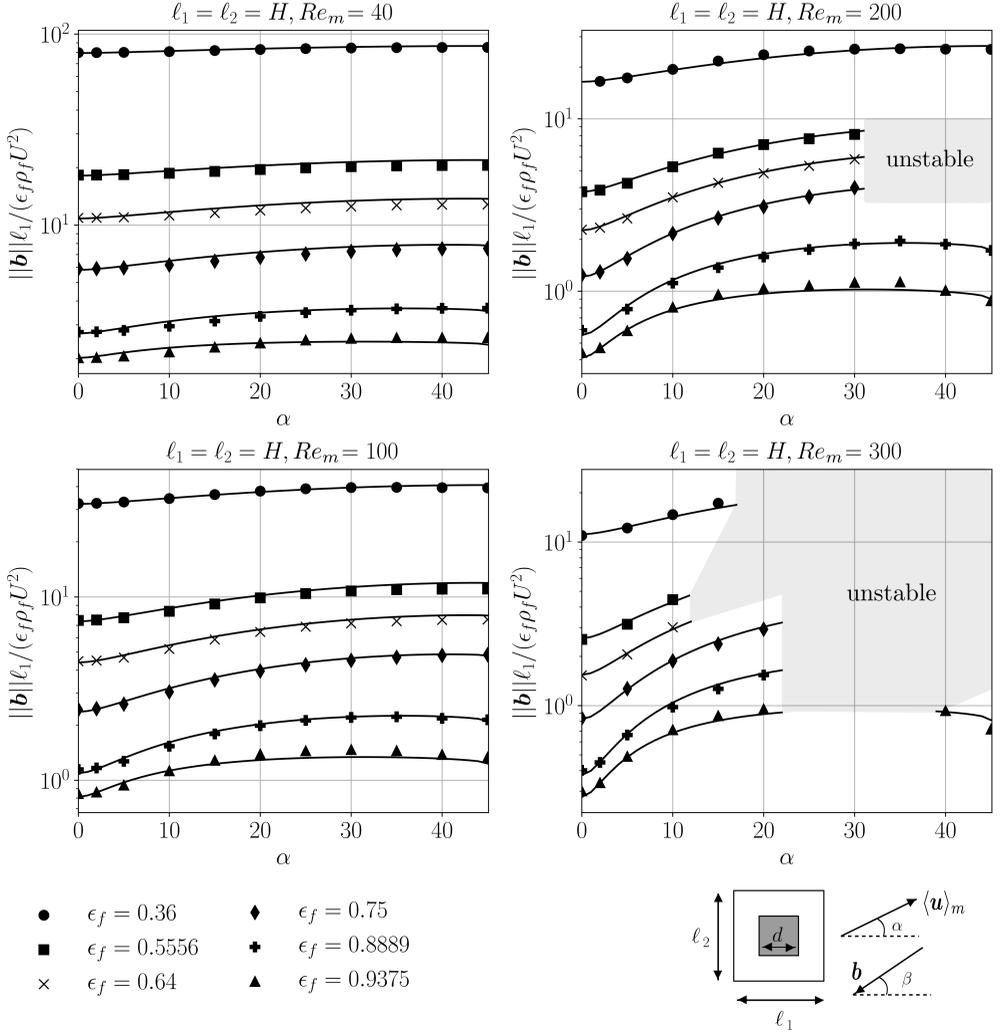}
\caption{
Magnitude of the closure force $\boldsymbol{b}$, according to the classical closure problem for an array of equidistant in-line square cylinders.
The dependence on the porosity $\epsilon_f$ and angle of attack $\alpha$  is shown for selected Reynolds numbers $\Rey_m$, and a fixed height $H=\ell_1$.
The solid lines ($-$) agree with the correlation (\ref{eq: correlation magnitude factor classical closure problem}). 
}
\label{fig: Closure problem for developing macro-scale flow: magnitude - angle}
\end{center}
\end{figure}

In figure \ref{fig: Closure problem for developing macro-scale flow: magnitude - angle}, the dependence of the magnitude of the closure force $\boldsymbol{b}$ on the porosity $\epsilon_f$ is shown for a wide range of angles of attack $\alpha$, but for selected Reynolds numbers $\Rey_m \in\left\{ 40, 100, 200, 300 \right\}$.
At the higher Reynolds numbers $\Rey_m = 200$ and $\Rey_m = 300$, some of the steady solutions of the classical closure problem obtained for larger angles of attack $\alpha$, were found to correspond to unstable solutions of the (time-dependent) periodically developed flow equations.
These unstable solutions have been omitted here, but their parameter range has been indicated by the grey-coloured areas in figure \ref{fig: Closure problem for developing macro-scale flow: magnitude - angle}.

With a mean relative error of $3.5\%$ and a maximum relative error of $10\%$, all of the data points in figure \ref{fig: Closure problem for developing macro-scale flow} (as well as figure \ref{fig: Closure problem for developing macro-scale flow: magnitude - angle}) satisfy the empirical correlation
\begin{equation}
\label{eq: correlation magnitude factor classical closure problem}
F_{\boldsymbol{b}} =1+A_0 \frac{\epsilon_f^2 (1-\epsilon_f)^2}{(1-\epsilon_f)^2 +A_1\vert\sin(2\alpha)\vert} 
\biggl(\frac{\sin^2(2\alpha)}{\sin^2(2\alpha) +A_2} \biggr)\biggl(1+A_3 \epsilon_f^6\sqrt{\vert\cos(2\alpha) \vert} \biggr)\Rey_m^{1.2} \,,
\end{equation}
with $A_0 = 0.022$, $A_1 = 0.01$, $A_2 = 0.15$ and $A_3 = 0.83$.
This correlation has been obtained through a least-square fitting procedure, and reflects that $F_{\boldsymbol{b}} -1 \sim\Rey_m^{1.2}$ is a good approximation over the investigated range of Reynolds numbers.
Further, it is based on the observation that $F_{\boldsymbol{b}} -1 \sim \vert \sin(2\alpha) \vert$ for porosities $\epsilon_f \leq 0.75$.
For higher porosities, the latter form has been corrected into 
$F_{\boldsymbol{b}} -1 \sim \vert \sin(2\alpha) \vert ( 1+A_3 \sqrt{\vert\cos(2\alpha) \vert})$.
The correlation also shows that $F_{\boldsymbol{b}} -1 \sim \alpha^2$ for  $\alpha \rightarrow 0$,  although the numerical uncertainty on this exponent $2$ was found to be quite significant.
Lastly, the correlation has been constructed by matching the approximate asymptotes  $F_{\boldsymbol{b}} -1 \sim \epsilon_f^2$ for $\epsilon_f \rightarrow 0$ and $F_{\boldsymbol{b}} -1 \sim (1-\epsilon_f)^2$ for $\epsilon_f \rightarrow 1$, whose intersection point depends on the angle of attack via the term $A_1\vert\sin(2\alpha)\vert$.
We remark that if the coefficients $A_i$ in the correlation would be optimized for every single porosity value $\epsilon_f$, the maximum relative error of the correlation would be less than $6.5\%$ for that porosity value.

\begin{figure}
\begin{center}
\includegraphics[scale=0.45, trim=0cm 0cm 0cm 0cm, clip=true]{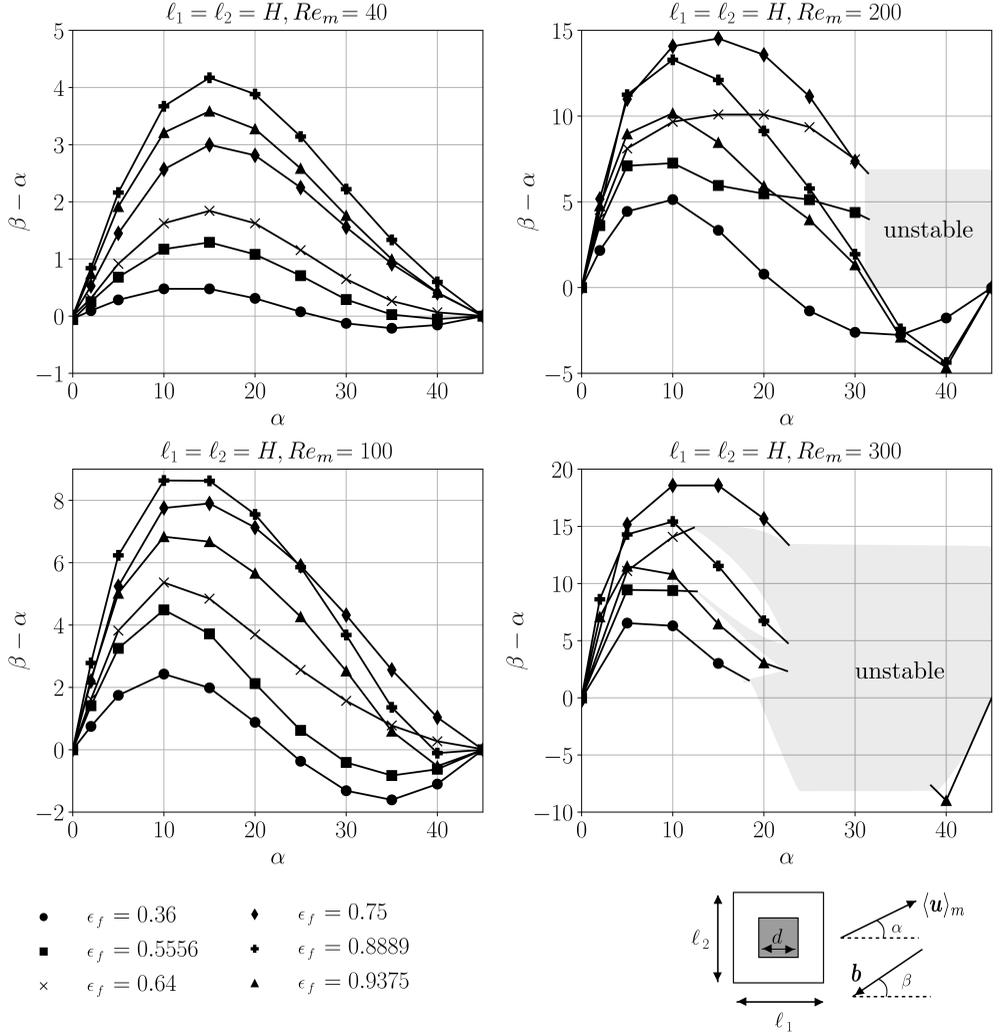}
\caption{
Direction $\beta$ of the closure force $\boldsymbol{b}$, according to the classical closure problem for an array of equidistant in-line square cylinders.
The dependence on the porosity $\epsilon_f$ and angle of attack $\alpha$  is shown for selected Reynolds numbers $\Rey_m$, and a fixed height $H=\ell_1$.
}
\label{fig: Closure problem for developing macro-scale flow: angle - angle}
\end{center}
\end{figure}

The direction $\beta$ of the closure force according to the classical closure problem is given in figure \ref{fig: Closure problem for developing macro-scale flow: angle - angle}.
The range of porosities, angles of attack and Reynolds numbers is the same as in the previous figure.
The relation between $\beta$ and $\alpha$ is quite complex even for a fixed porosity, especially when the Reynolds number $\Rey_m$ is bigger.
But for high porosities, it can be described by the correlation  
\begin{equation}
\frac{\tan \beta}{\tan \alpha} =  1 + B_0 \frac{\Rey_m^2}{\Rey_m + B_1} \left(1-\vert \sin (2 \alpha)\vert \right)\left(1-B_2\Rey_m^{B_3}  \vert \sin (2 \alpha)\vert^{B_4} \right)\,.
\end{equation}
For a porosity $\epsilon_f=0.75$ and angle of attack $\alpha \in (2^{\circ}, 45^{\circ})$, this correlation captures all of the data points from figures \ref{fig: Closure problem for developing macro-scale flow} and \ref{fig: Closure problem for developing macro-scale flow: angle - angle} with a relative accuracy of $4\%$, if $ B_0 = 0.040$,  $ B_1 = 22$, $ B_2 = 0.004$, $ B_3 = 1$ and $B_4=0.5$.
For a porosity $\epsilon_f\simeq 0.89$, the correlation is accurate to within $6\%$ when $\alpha \in ( 2^{\circ}, 35^{\circ})$, if 
$ B_0 = 0.040$,  $ B_1 = 12$, $ B_2 = 0.0075$, $ B_3 = 1$ and $B_4=0.7$.
Moreover, for $\epsilon_f\simeq 0.94$ and $\alpha \in ( 2^{\circ}, 35^{\circ})$, the correlation is accurate to within $7\%$, if $ B_0 = 0.042$,  $ B_1 = 20$, $ B_2 = 0.021$, $ B_3 = 0.8$ and $B_4=0.5$.
On the other hand, for small angles $\alpha \in (0, 2^{\circ})$, the relative accuracy of the latter three correlations reduces to $8\%$.
The correlation also respects that $\alpha=\beta$ for $\alpha=0 ^{\circ}$, $\alpha=45 ^{\circ}$, or $\Rey_m \rightarrow 0$.



\subsection{Validity of the Classical Closure Problem for Quasi-Developed Flow \\Outside the Side-Wall Region -- Theoretical Considerations}
\label{subsec: Validity of the Classical Closure Problem for Quasi-Developed Flow Outside the Side-Wall Region -- Theoretical Considerations}
Although the classical closure problem in $\Omega_{\text{quasi-periodic}}$ has the same mathematical form as the periodically developed flow equations in $\Omega_{\text{uniform}}$, its underlying assumptions (\ref{eq: periodic deviation fields closure}), (\ref{eq: closure force dominant wrt dispersion}), (\ref{eq: macro-scale velocity gradient negligible in closure problem}) and (\ref{eq: no spatial variation of macro-scale velocity and closure force in unit cell}) are less restrictive than true flow periodicity.
The reason is that these assumptions are also justified under certain length-scale conditions which may hold throughout a wider range of flow regimes, as shown by \cite{Whitaker1996}.
Therefore, we might expect that already after some section in the region of quasi-periodically developed flow,
the local approximation (\ref{eq: definition approx permeability tensor}) for the actual closure force (\ref{eq: quasi-developed closure force}) may become relatively accurate.

In view of this expectation, the question arises how well each of the assumptions behind the classical closure problem is satisfied when the flow is still developing in $\Omega_{\text{quasi-periodic}}$.
If we examine the first assumption (\ref{eq: periodic deviation fields closure}), i.e. the periodicity of the deviation fields of the velocity and pressure in the main flow direction, we find that 
\begin{align}
\label{eq: periodicity deviation velocity quasi-periodically developed flow}
\devtilde{\boldsymbol{u}}_f(\boldsymbol{r} + n_1\boldsymbol{l}_1) - 
\devtilde{\boldsymbol{u}}_f(\boldsymbol{r}) &= 
\devtilde{\velamp}_f(\boldsymbol{r})  \exp(-\boldsymbol{\lambda} \boldsymbol{\cdot} \boldsymbol{r}) G_{\lambda}\,,
\\
\label{eq: periodicity deviation pressure quasi-periodically developed flow}
\devtilde{p}_f(\boldsymbol{r} + n_1\boldsymbol{l}_1) - 
\devtilde{p}_f(\boldsymbol{r}) &= 
\devtilde{\pressamp}_f(\boldsymbol{r})  \exp(-\boldsymbol{\lambda} \boldsymbol{\cdot} \boldsymbol{r}) G_{\lambda}
\end{align}
with $G_{\lambda} \triangleq \exp(-\boldsymbol{\lambda} \boldsymbol{\cdot}  n_1\boldsymbol{l}_1) -1 $,
as a consequence of the defining properties of quasi-periodically developed flow (\ref{eq: quasi-periodically developed flow})-(\ref{eq: periodic pressure amplitude}).
We thus see that the periodicity conditions for the classical closure problem are violated by the terms on the right hand side of 
(\ref{eq: periodicity deviation velocity quasi-periodically developed flow}) and (\ref{eq: periodicity deviation pressure quasi-periodically developed flow}), which are proportional to $C_{\velamp} \exp(-\boldsymbol{\lambda} \boldsymbol{\cdot} \boldsymbol{r}) G_{\lambda}$, since $\devtilde{\velamp}_f \sim C_{\velamp}$ and $\devtilde{\pressamp}_f \sim C_{\velamp}$.

A similar conclusion is found with respect to the assumption that the variation of the macro-scale velocity and closure force within the unit cell can be ignored (\ref{eq: no spatial variation of macro-scale velocity and closure force in unit cell}).
Along the main flow direction, we have for instance
\begin{equation}
\label{eq: no spatial variation of macro-scale velocity in quasi-developed macro-scale flow}
\langle \boldsymbol{u} \rangle^f_m(\boldsymbol{r} + n_1\boldsymbol{l}_1) -
\langle \boldsymbol{u} \rangle^f_m(\boldsymbol{r}) = 
\langle \velamp \rangle^f_m(\boldsymbol{r})   \exp(-\boldsymbol{\lambda} \boldsymbol{\cdot} \boldsymbol{r}) G_{\lambda}\,,
\end{equation}
by virtue of (\ref{eq: quasi-developed macro-scale flow}) and (\ref{eq: quasi-developed macro-scale flow amplitude}).
In addition, we have
\begin{equation}
\boldsymbol{b} (\boldsymbol{r} + n_1\boldsymbol{l}_1) -
\boldsymbol{b} (\boldsymbol{r}) = 
- \mu_f  \boldsymbol{\mathsf{K}}^{-1}(\boldsymbol{r}) \boldsymbol{\cdot}  \langle \velamp \rangle_m(\boldsymbol{r})   \exp(-\boldsymbol{\lambda} \boldsymbol{\cdot} \boldsymbol{r}) G_{\lambda}
\end{equation}
due to (\ref{eq: representation permeability amplitude quasi-developed macro-scale flow}).
Hence, the biggest spatial variations of the macro-scale velocity and closure force within the unit cell, which are ignored in the classical closure problem, are also proportional to $C_{\velamp} \exp(-\boldsymbol{\lambda} \boldsymbol{\cdot} \boldsymbol{r}) G_{\lambda}$, as $\langle \velamp \rangle^f_m  \sim C_{\velamp}$.

The third assumption, which implies that the gradient of the macro-scale velocity within the unit cell is negligible (\ref{eq: macro-scale velocity gradient negligible in closure problem}), can be evaluated based on the same criterion as just derived to evaluate the variation of the macro-scale velocity within the unit cell (\ref{eq: no spatial variation of macro-scale velocity in quasi-developed macro-scale flow}).
However, in line with (\ref{eq: quasi-periodically developed flow}) and (\ref{eq: quasi-developed macro-scale flow}), it also requires that
\begin{equation}
\label{eq: negligible gradient of deviation amplitude in unit cell}
-\lambda  \langle \velamp \rangle^f_m \ll \boldsymbol{e}_1 \boldsymbol{\cdot} \frac{\partial \devtilde{\velamp}_f}{\partial \boldsymbol{r} } - \lambda \devtilde{\velamp}_f\,.
\end{equation}
This condition is expected to be automatically satisfied when $\lambda \ell_1\ll 1$, hence as long as the double-volume averaging operator $\langle \; \rangle_m$ has the same properties as a matched filter with respect to the mode $\exp(-\boldsymbol{\lambda} \boldsymbol{\cdot} \boldsymbol{r})$.
The argument as to why $\lambda \ell_1\ll 1$ is a sufficient condition for (\ref{eq: negligible gradient of deviation amplitude in unit cell}) and thus (\ref{eq: macro-scale velocity gradient negligible in closure problem}), is that we may estimate $\langle \velamp \rangle^f_m$ and $\devtilde{\velamp}_f$ in (\ref{eq: negligible gradient of deviation amplitude in unit cell}) to have the same order of magnitude, i.e. $O(\langle \velamp \rangle^f_m) = O(\devtilde{\velamp}_f) = O(\velamp_f)$, since  $\velamp_f(\boldsymbol{x}) = 0$ for $\boldsymbol{x} \in \Gamma_0$, while we have $O(\boldsymbol{r})=O(\ell_1)$ for $\boldsymbol{r} \in \Omega_{\text{unit}}^{2\times 2}$.
As a side note, we add that when $\lambda \ell_1\ll 1$, it holds that $G_{\lambda} \simeq - n_1\lambda \ell_1$.

The last assumption to evaluate is whether the macro-scale momentum dispersion source can be neglected when the flow is quasi-periodically developed, that is (\ref{eq: closure force dominant wrt dispersion}).
The macro-scale momentum dispersion source in $\Omega_{\text{quasi-periodic}}$ is given by
\begin{align}
\label{eq: momentum dispersion source quasi-developed macro-scale flow}
\boldsymbol{M} &= \boldsymbol{M}^{\star} + \left(
\langle \boldsymbol{u}^{\star} \velamp \rangle_m -
\langle \boldsymbol{u}^{\star} \rangle^f_m \langle \velamp \rangle_m + 
\langle  \velamp \boldsymbol{u}^{\star} \rangle_m -
\langle \velamp \rangle^f_m \langle \boldsymbol{u}^{\star} \rangle_m 
\right)\exp(-\boldsymbol{\lambda} \boldsymbol{\cdot} \boldsymbol{r} ) \\
&\simeq \boldsymbol{M}^{\star} + \left(
\langle \devtilde{\boldsymbol{u}}^{\star} \velamp \rangle_m +
\langle  \velamp \devtilde{\boldsymbol{u}}^{\star}  \rangle_m 
\right)\exp(-\boldsymbol{\lambda} \boldsymbol{\cdot} \boldsymbol{r} )
\end{align}
where $\boldsymbol{M}^{\star} \triangleq \langle   \boldsymbol{u}^{\star}  \boldsymbol{u}^{\star} \rangle_m - \langle \boldsymbol{u}^{\star}
 \rangle^f_m \langle \boldsymbol{u}^{\star} \rangle_m$.
Note that (\ref{eq: momentum dispersion source quasi-developed macro-scale flow}) has been obtained by neglecting the small advective contributions of the velocity terms which are proportional to $\exp (-2\lambda x_1)$, since
only the mode $\exp (-\lambda x_1)$ determines the asymptotic convergence of $\boldsymbol{u}$ towards $\boldsymbol{u}^{\star}$ in $\Omega_{\text{quasi-periodic}}$.
As $\boldsymbol{M}^{\star}$ is divergence-free outside of $\Omega_{\text{sides}}$, we can deduce that the approximation $  \boldsymbol{\nabla} \boldsymbol{\cdot} \boldsymbol{M} \simeq 0$ neglects the contribution of the second term on the right hand side of (\ref{eq: momentum dispersion source quasi-developed macro-scale flow}).
This contribution is again proportional to $C_{\velamp} \exp(-\boldsymbol{\lambda} \boldsymbol{\cdot} \boldsymbol{r})$, but it appears to be very small.
According to our numerical simulations, the closure term $\rho_f \boldsymbol{\nabla} \boldsymbol{\cdot} \boldsymbol{M}$ is at least an order of magnitude smaller than the closure force $\boldsymbol{b}$ over the entire core region of the channel, $\Omega_{\text{core}}$.
Moreover, even near the channel inlet and outlet, we have observed that macro-scale momentum dispersion is of minor importance for the boundary conditions studied in this work.

In summary, we conclude that all assumptions behind the classical closure problem are either fulfilled, or violated by an error which is proportional to $C_{\velamp} \exp(-\boldsymbol{\lambda} \boldsymbol{\cdot} \boldsymbol{r})$ in $\Omega_{\text{quasi-periodic}}$.
This explains why the classical closure problem leads to a modelling error for the closure force in $\Omega_{\text{quasi-periodic}}$,
\begin{equation}
\label{eq: modelling error closure force quasi-developed macro-scale flow}
\boldsymbol{b}(\boldsymbol{r}) -  \boldsymbol{b}_{\text{approx}}(\boldsymbol{r}) \simeq -\mu_f 
\left[ \boldsymbol{\mathsf{K}}^{-1}(\boldsymbol{r}) - 
\boldsymbol{K}^{-1}_{\text{uniform}}
\right] \boldsymbol{\cdot} \langle \velamp \rangle_m(\boldsymbol{r})   \exp(-\boldsymbol{\lambda} \boldsymbol{\cdot} \boldsymbol{r})\,,
\end{equation}
which also scales with $C_{\velamp} \exp(-\boldsymbol{\lambda} \boldsymbol{\cdot} \boldsymbol{r})$ and thus diminishes in the main flow direction.
At least, this is true if the dependence of the permeability tensor $\boldsymbol{K}_{\text{approx}}$ on the macro-scale velocity is sufficiently weak, i.e. when $\boldsymbol{K}_{\text{approx}} \left(\langle \boldsymbol{u} \rangle_m \right) \simeq \boldsymbol{K}_{\text{uniform}} \left( \boldsymbol{U} \right)$.
This tends to be the case when the Forchheimer coefficient is much smaller than the Darcy coefficient, or when the perturbation size $C_{\velamp}$ is small.

From the previous analysis, we learn that the classical closure problem will hold with good accuracy over the entire region of quasi-periodically developed flow, under two circumstances.
The first circumstance is when the flow develops in such a manner that the relative perturbation size  $C_{\velamp}^{+} \triangleq C_{\velamp}/U'$ is rather small.
After all, we see that $\boldsymbol{b}_{\text{approx}} \rightarrow \boldsymbol{b}$ as $C_{\velamp} \rightarrow 0$, from (\ref{eq: modelling error closure force quasi-developed macro-scale flow}).
This circumstance is rather obvious, as it implies that the macro-scale flow can be treated as developed over the entire region of quasi-periodically developed flow.
The second circumstance is when the apparent permeability tensors $\boldsymbol{\mathsf{K}}$ and $\boldsymbol{K}_{\text{uniform}}$ match each other closely.
In general, however, this will never be exactly the case, because $\boldsymbol{\mathsf{K}}$ and $\boldsymbol{K}_{\text{uniform}}$ are governed by two mathematically very different closure problems.
Nevertheless, when the difference between $\boldsymbol{\mathsf{K}}$ and $\boldsymbol{K}_{\text{uniform}}$ is small enough with respect to $C_{\velamp}^{+} $, the modelling error (\ref{eq: modelling error closure force quasi-developed macro-scale flow}) will be negligible even for larger perturbations.
Therefore, it is possible that the classical closure problem yields an accurate approximation for the closure force, not just over the entire region of quasi-periodically developed flow, but even more upstream where the macro-scale flow can be treated as approximately quasi-developed.
By this we mean at some point after the section $x_1 = x_{\text{quasi-dev}}$, with $x_{\text{quasi-dev}} \leq x_{\text{quasi-periodic}}$.

Under other circumstances, the classical closure problem will hold
at best over a part of $\Omega_{\text{quasi-periodic}}$.
This understanding brings us to the question for which section in $\Omega_{\text{quasi-periodic}}$, or which section after the point $x_{\text{quasi-dev}}$, the relative error between the actual permeability tensor and its approximation from the classical closure problem equals some prescribed value $\varepsilon_{\boldsymbol{K}}$, defined as
\begin{equation}
\varepsilon_{\boldsymbol{K}} \triangleq \Vert \boldsymbol{I} - \boldsymbol{K}_{\text{approx}} \boldsymbol{\cdot} \boldsymbol{K}^{-1}_{\text{quasi-dev}} \Vert\,.
\end{equation} 
Here, $\Vert \; \Vert$ denotes an appropriate tensor norm.
From (\ref{eq: approximate structure permeability tensor quasi-developed flow}) and (\ref{eq: equivalence classical closure problem permeability and uniform macro-scale flow}) it follows that this section is given by $x_1 = x_{\text{approx}}$ with
\begin{equation}
\label{eq: point of approximate classical closure problem}
x_{\text{approx}} \simeq \frac{1}{\lambda} \ln 
\left( \frac{
\displaystyle \max_{x_2 \in I_2} \Vert
\boldsymbol{K}_{\text{approx}}
\boldsymbol{\cdot}  \boldsymbol{\mathsf{K}}^{-1} \boldsymbol{\cdot} \boldsymbol{\zeta}_{\text{ref}} \Vert C_{\velamp}^{+}
}{\varepsilon_{\boldsymbol{K}}
} \right) \,,
\end{equation}
and $I_2 \triangleq (\ell_{\text{sides}}, W-\ell_{\text{sides}})$.
To obtain the last result, it was assumed again that the dependence of the permeability tensor $\boldsymbol{K}_{\text{approx}}$ on the macro-scale velocity is sufficiently weak, so $\boldsymbol{K}_{\text{approx}} \left(\langle \boldsymbol{u} \rangle_m \right) \simeq \boldsymbol{K}_{\text{uniform}} \left( \boldsymbol{U} \right)$.
Expression (\ref{eq: point of approximate classical closure problem})
reveals that the point where the classical closure problem becomes accurate to within $\varepsilon_{\boldsymbol{K}}$ for some relative perturbation size $C_{\velamp}^{+}$, satisfies the scaling law 
\begin{equation}
\label{eq: scaling law point of approximate classical closure problem}
x_{\text{approx}} \simeq \frac{1}{\lambda} \left( \ln \frac{C_{\velamp}^{+}}{\varepsilon_{\boldsymbol{K}}} + c_3 \right)\,,
\end{equation}
with $c_3 \triangleq \ln \displaystyle  \max_{x_2 \in I_2} \Vert
\boldsymbol{K}_{\text{approx}}
\boldsymbol{\cdot}  \boldsymbol{\mathsf{K}}^{-1} \boldsymbol{\cdot} \boldsymbol{\zeta}_{\text{ref}} \Vert $,
provided that $x_{\text{approx}} \in ( x_{\text{quasi-dev}}, x_{\text{end}}-n_1\ell_1)$.
Of course, this scaling law is only of use once the eigenvalue $\lambda$ and the term $c_3$ have been determined by solving the conservation equations for quasi-periodically flow on one or two rows of the array. 

We remark that the term $c_3$, which is a measure for the difference between $\boldsymbol{\mathsf{K}}$ and $\boldsymbol{K}_{\text{approx}}$, 
is virtually independent of the macro-scale velocity $\boldsymbol{U}$ and hence the Reynolds number $\Rey$, if also the dependence of $\boldsymbol{\mathsf{K}}$ on the macro-scale velocity is weak, next to that of $\boldsymbol{K}_{\text{approx}}$.
Therefore, when $c_3$ is interpreted as a geometrical property of the channel and its array, the scaling law (\ref{eq: scaling law point of approximate classical closure problem}) yields the correct correlation between $x_{\text{approx}}/\ell_1$ and $\Rey$, as long as the inertial effects on the permeability tensors $\boldsymbol{\mathsf{K}}$ and $\boldsymbol{K}_{\text{approx}}$ (or $\boldsymbol{K}_{\text{uniform}}$) are not too strong. 
This correlation between $x_{\text{approx}}/\ell_1$ and $\Rey$ tends to be linear, due to the fact that the eigenvalue $\lambda$ scales inversely linear with the Reynolds number in the region of quasi-periodically developed flow (\cite{Buckinx2022}): $x_{\text{approx}}/\ell_1 \sim 1/(\lambda \ell_1) \sim  c_1 \Rey + 1 $.
So, at lower Reynolds numbers $\Rey$, the apparent permeability tensor  according to the classical closure problem, $\boldsymbol{K}_{\text{approx}}$, tends to match $\boldsymbol{K}_{\text{quasi-dev}}$ more upstream towards the channel inlet.
Nevertheless, for a given velocity profile at the inlet of the channel, also the relative perturbation size $C_{\velamp}^{+}$ will change when the Reynolds number $\Rey$ changes, as the flow will develop differently.
In particular, for a parabolic velocity profile at the channel inlet, the relative perturbation size $C_{\velamp}^{+}$ decreases at higher Reynolds number $\Rey$.
Yet, the influence of the relative perturbation size $C_{\velamp}^{+}$ on the point where the classical problem becomes valid, is less pronounced than that on the modelling error (\ref{eq: modelling error closure force quasi-developed macro-scale flow}) itself.
The reason is that $x_{\text{approx}}$ does not scale linearly with $C_{\velamp}^{+}$, but instead scales with its logarithm.

\subsection{Validity of the Classical Closure Problem for Quasi-Developed Flow \\Outside the Side-Wall Region -- Computational Study}
\label{subsec: Validity of the Classical Closure Problem for Quasi-Developed Flow Outside the Side-Wall Region -- Computational Study}

Thus far, we have shown that for quasi-developed macro-scale flow, the approximation errors in the classical closure problem, as well as the point where the classical closure problem becomes valid, depend on three factors: the relative perturbation size $C_{\velamp}^{+}$, which controls the mode amplitudes $ \velamp$ and $ \langle \velamp \rangle_m$, the eigenvalue $\lambda$, and lastly the difference between $\boldsymbol{\mathsf{K}}$ and $\boldsymbol{K}_{\text{uniform}}$.
So, a complete treatise on the validity of (\ref{eq: definition approx permeability tensor}) would require first an assessment of the relative perturbation size or magnitude of the mode amplitudes for a large set of relevant inlet conditions and channel geometries.
However, the formulation and characterization of physically realistic inlet conditions falls beyond the scope of the present work.
To get some idea of how large the mode amplitudes can be for the class of channel flows discussed in section \ref{sec: Channel Domain and Array Geometry}, albeit under the idealized case of a parabolic velocity profile at the channel inlet, we refer the reader to our preceding work (\cite{Buckinx2022}).
Here, we limit us to a discussion of the computational results for the macro-scale flow fields from figure \ref{fig: Quasi-developed macro-scale flow}, to support our main theoretical findings.
These computational results, which illustrate the accuracy of the classical closure problem as a model for the closure force in the developing flow region, are displayed in the next figures.

\begin{figure}
\begin{center}
\includegraphics[scale=0.55, trim=0.5cm 0cm 0.5cm 0cm, clip=true]{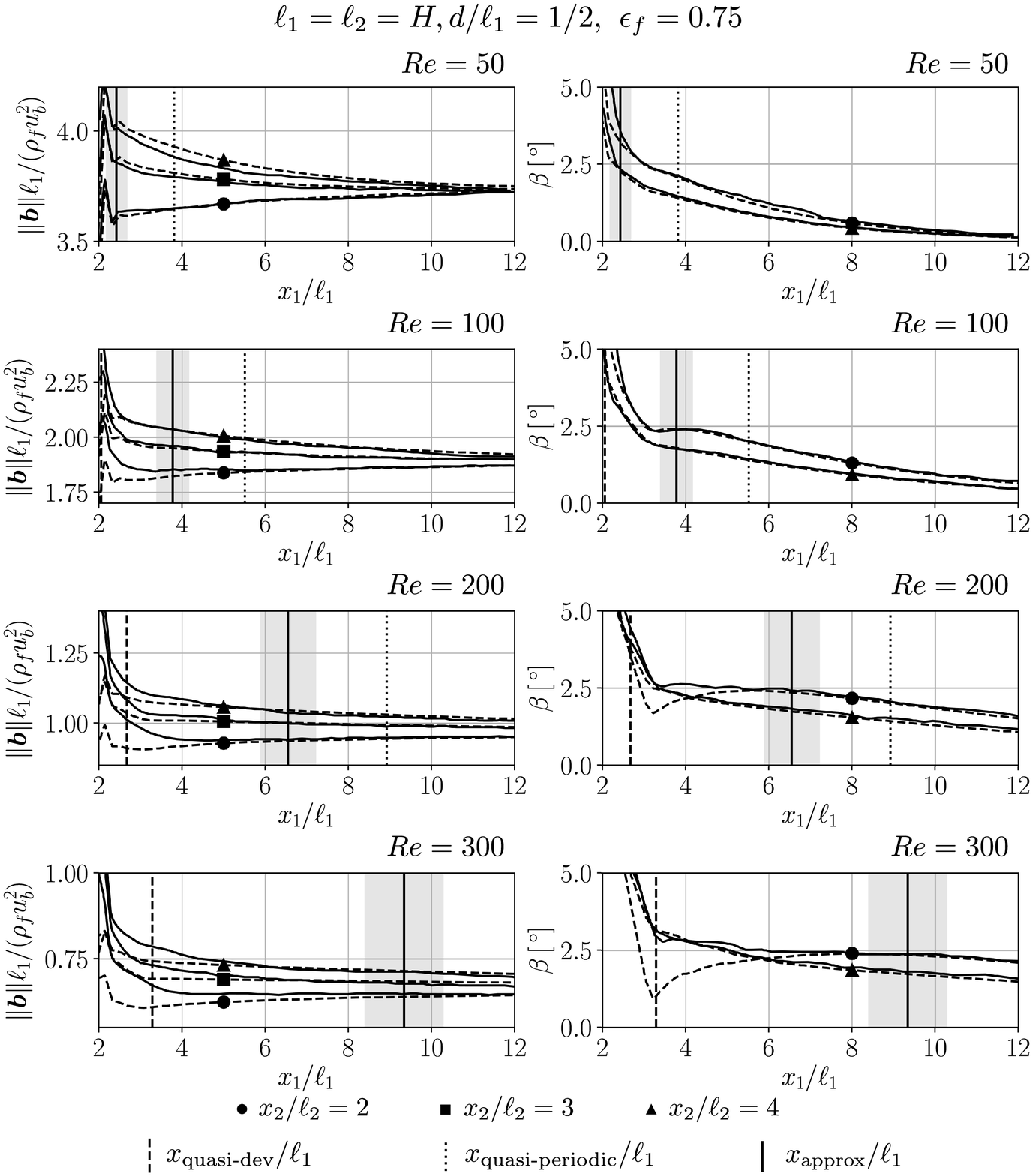}
\caption{Closure force in the developing flow region of a channel array with a porosity $\epsilon_f = 0.75$ ($N_1=60, N_2=10$, $s_0/\ell_1 =1, s_N/\ell_1 = 1$, $H/\ell_1 =1$, $\ell_1/\ell_2 =1$) for different Reynolds numbers $\Rey$ and different $x_2$-sections.
The solid lines ($-$) represent the magnitude and angle of the actual  closure force $\boldsymbol{b}$.
The dashed lines (-$\,$-) represent the solution of the classical closure problem, $\boldsymbol{b}_{\text{approx}}$. 
The onset points of quasi-periodic and quasi-developed flow, $x_{\text{quasi-periodic}}$ and  $x_{\text{quasi-dev}}$ have been indicated by a dotted and a dashed vertical line respectively.
The solid vertical line marks the point $x_{\text{approx}}$ from where on the classical closure problem is theoretically accurate to within $1\%$.}
\label{fig:Developing closure force comparison correlation porosity 0.75}
\end{center}
\end{figure}

In figure \ref{fig:Developing closure force comparison correlation porosity 0.75}, the solution  of the classical closure problem, $\boldsymbol{b}_{\text{approx}}$ from (\ref{eq: definition approx permeability tensor}), is compared with the actual closure force $\boldsymbol{b}$ in a channel, which consists of an array of $60 \times 10$ in-line equidistant square cylinders with a height $H=\ell_1=\ell_2$ and a porosity $\epsilon_f = 0.75$. 
The position of the first and last cylinder row have been chosen as  $s_0=\ell_1$ and  $s_N=\ell_1$.

The actual closure force, whose magnitude $\Vert \boldsymbol{b} \Vert$ and direction $\beta$ are given by the solid lines ($-$) in figure \ref{fig:Developing closure force comparison correlation porosity 0.75}, has been obtained from a direct numerical simulation of the flow in the channel, for the boundary conditions discussed in section \ref{sec: Channel Domain and Array Geometry}.
It has been calculated from the closure terms $\boldsymbol{b}_p$ and $\boldsymbol{b}_{\tau}$, as defined in section \ref{sec: Macro-Scale Flow Equations for Steady Channel Flow}, by explicitly filtering the pressure field, the pressure gradient, as well as the viscous stress tensor and its divergence.
The latter explicit filtering operation proved to be computationally very demanding, as it required an interpolation of the flow field onto a mesh twice as fine as the one used for the direct numerical simulation of the flow (thus containing up to 250 million mesh cells), in order to keep the average relative discretization error on $\Vert \boldsymbol{b}\Vert$ below $5\%$.

The approximation for the closure force according to the classical closure problem, $\boldsymbol{b}_{\text{approx}}$, which is indicated by the dashed lines (-$\,$-) in figure \ref{fig:Developing closure force comparison correlation porosity 0.75}, has been obtained in three steps.
First, the macro-scale velocity in the developing flow region was acquired by explicit filtering of the velocity field, to get the local angle of attack $\alpha$ and local Reynolds number $\Rey_{m}$ at each point of $\Omega_{\text{predev}}$. 
Then, the closure equations (\ref{eq: deviation momentum equation}) - (\ref{eq: zero averaged deviation velocity}) were solved to construct a data table for $F_{\boldsymbol{b}}$ and $\beta$ for an extensive set of angles of attack and local Reynolds numbers, covering the actual range of $\alpha$ and $\Rey_{m}$ in the developing flow region.
That way, the value of $\boldsymbol{b}_{\text{approx}}$ was already obtained for certain points in $\Omega_{\text{predev}}$.
Finally, two-dimensional interpolation based on univariate cubic splines and linear radial basis functions was used to evaluate $F_{\boldsymbol{b}}$ and $\beta$ for the intermediate values of $\alpha$ and $\Rey_{m}$ in $\Omega_{\text{predev}}$ that were not included in the data table.
The part of the data table for $F_{\boldsymbol{b}}$ and $\beta$ which is most relevant for reproducing the approximation $\boldsymbol{b}_{\text{approx}}$ in figure \ref{fig:Developing closure force comparison correlation porosity 0.75}, has been presented earlier in figure \ref{fig: Closure problem for developing macro-scale flow}.
Indeed, for all the flow conditions depicted in figure \ref{fig:Developing closure force comparison correlation porosity 0.75}, it holds that $\alpha(\boldsymbol{x}) \in (0, 5^{\circ})$ and $\Rey_m(\boldsymbol{x})  \in \left(0.4 Re, 1.3Re \right)$ if $\boldsymbol{x} \in\Omega_{\text{predev}}$, and $\Rey_{m} \simeq 1.07 \Rey$ in $\Omega_{\text{uniform}}$. 
Therefore, a complete overview of the data table used for the interpolation has been omitted here.
Besides, the data table is quite extensive, since the classical closure problem was solved numerically for more than 850 different combinations of $\alpha$, $\Rey_m$ and $\epsilon_f$, to keep the estimated interpolation error below $5\%$ (including the estimated maximum discretization error of $1.5\%$ on the values in the data table itself).
Almost 300 different simulations of the classical closure problem were carried out for just the geometry selected in figure \ref{fig:Developing closure force comparison correlation porosity 0.75}.

By comparing the actual closure force with its approximation according to the classical closure problem in figure \ref{fig:Developing closure force comparison correlation porosity 0.75}, we see that both converge downstream along the main flow direction.
As explained before, once the macro-scale velocity and the closure force have become uniform due to the onset of periodically developed flow, both are in exact agreement, apart from a small discretization error, which is in this case around $1\%$.
For the channel in figure \ref{fig:Developing closure force comparison correlation porosity 0.75}, this exact agreement between $\boldsymbol{b}$ and $\boldsymbol{b}_{\text{approx}}$ occurs around $x_1= x_{\text{dev}}$, with $  x_{\text{dev}} \simeq 12 \ell_1$ when $\Rey=50$, and  $x_{\text{dev}} \simeq 52 \ell_1$ when $\Rey=300$.

We also see that over the largest part of the developing flow region, the approximation based on the classical closure problem is already quite accurate.
For instance, when $\Rey \leq 100$, the solution of the classical closure problem deviates no more than $5\%$ in magnitude and $10\%$ in angle from the actual closure force, for $x_1/\ell_1 \geq 2.5$.
At higher Reynolds numbers, i.e. $200 \leq \Rey \leq 300 $, the same quantitative agreement is reached more downstream, for $x_1/\ell_1 \geq 6$.
If we take into account that the estimated discretization errors for $\Vert \boldsymbol{b} \Vert$ and $\beta$ are around $2.5\%$ and $10\%$ respectively (and certainly below $5\%$ and $15\%$), while the numerical solution of the classical closure problem has an estimated error of $1\%$ to $3\%$ (and certainly less than $5\%$) due to the interpolation, we can conclude that the classical closure problem yields an approximation which is accurate to within the margin of numerical uncertainty almost everywhere, except near the inlet region.
Close to the inlet region, around $x_1/\ell_1 = 2$, the relative difference between the solution of the classical closure problem and the actual closure force, is more than $25\%$ in terms of magnitude and angle, for all Reynolds numbers illustrated.

Despite the quantitatively good agreement between $\boldsymbol{b}_{\text{approx}}$ and $\boldsymbol{b}$ over most of the developing flow region, the difference $\boldsymbol{b}- \boldsymbol{b}_{\text{approx}}$  nowhere becomes zero in $\Omega_{\text{predev}}$: 
apart from locations with small discretization errors, there is everywhere some distance between the solid and dotted lines in figure \ref{fig:Developing closure force comparison correlation porosity 0.75}.
We do observe that the difference $\boldsymbol{b}- \boldsymbol{b}_{\text{approx}}$ decreases exponentially in the main flow direction after the section $x_1 = x_{\text{quasi-dev}}$, as predicted by (\ref{eq: modelling error closure force quasi-developed macro-scale flow}). 
Inevitably, this exponentially decreasing modelling error arises due to discrepancies between $\boldsymbol{K}_{\text{approx}}$ and $\boldsymbol{\mathsf{K}}$, whose largest components differ by as much as $20\%$, or even $40\%$, depending on the transversal position $x_2/\ell_2$.
These discrepancies between $\boldsymbol{K}_{\text{approx}}$ and $\boldsymbol{\mathsf{K}}$ result in just a minor modelling error $\Vert\boldsymbol{b}- \boldsymbol{b}_{\text{approx}} \Vert / \Vert \boldsymbol{b}\Vert$ of less than $1\%$ in $\Omega_{\text{quasi-periodic}}$, because the amplitude of the macro-scale velocity mode is rather small: $\Vert \langle \velamp \rangle_m\Vert < 0.25 U $ (see figure \ref{fig: Quasi-developed macro-scale flow modes}).
Therefore, the local solution of the classical closure problem can barely be distinguished from the actual closure force in $\Omega_{\text{quasi-periodic}}$. 

The point $\hat{x}_{\text{approx}}$ from where on the approximation $\boldsymbol{b}_{\text{approx}}$ deviates no more than $1\%$ in magnitude and $10\%$ in angle from the actual closure force $\boldsymbol{b}$ over the entire core of the channel, lies within the grey-coloured areas in figure \ref{fig:Developing closure force comparison correlation porosity 0.75}.
The width of these grey-coloured areas indicates the numerical uncertainty on the latter point, stemming from the fact that the gradients of $\boldsymbol{b}$ in the direction of the $x_1$-axis are so small.
Even though the point of agreement $\hat{x}_{\text{approx}}$ is located upstream of the region of quasi-periodically flow, hence to the left of the point $x_{\text{quasi-periodic}}$, it still lies in the region where the macro-scale velocity field can be considered quasi-developed, thus to the right of the point $x_{\text{quasi-dev}}$.
Therefore, the point $\hat{x}_{\text{approx}}$ still obeys the theoretical scaling law (\ref{eq: scaling law point of approximate classical closure problem}).
This scaling law corresponds to the vertical solid line in figure \ref{fig:Developing closure force comparison correlation porosity 0.75}, and is given by $x_{\text{approx}}/\ell_1 \simeq 1/(\lambda \ell_1) \left( \ln (1+c_2/Re) -  \ln(\varepsilon_{\boldsymbol{K}}) - 4 \right)$ with $\varepsilon_{\boldsymbol{K}} = 0.01$ and $1/(\lambda \ell_1) \simeq 0.05\Rey + 0.8$ (\cite{Buckinx2022}).
This follows from the fact that $C_{\velamp}^{+} \sim (1+ c_2/\Rey)$ with  $c_2=9.6$, as already appeared from (\ref{eq: scaling law amplitude macro-scale velocity mode}).

\begin{figure}
\begin{center}
\includegraphics[scale=0.5]{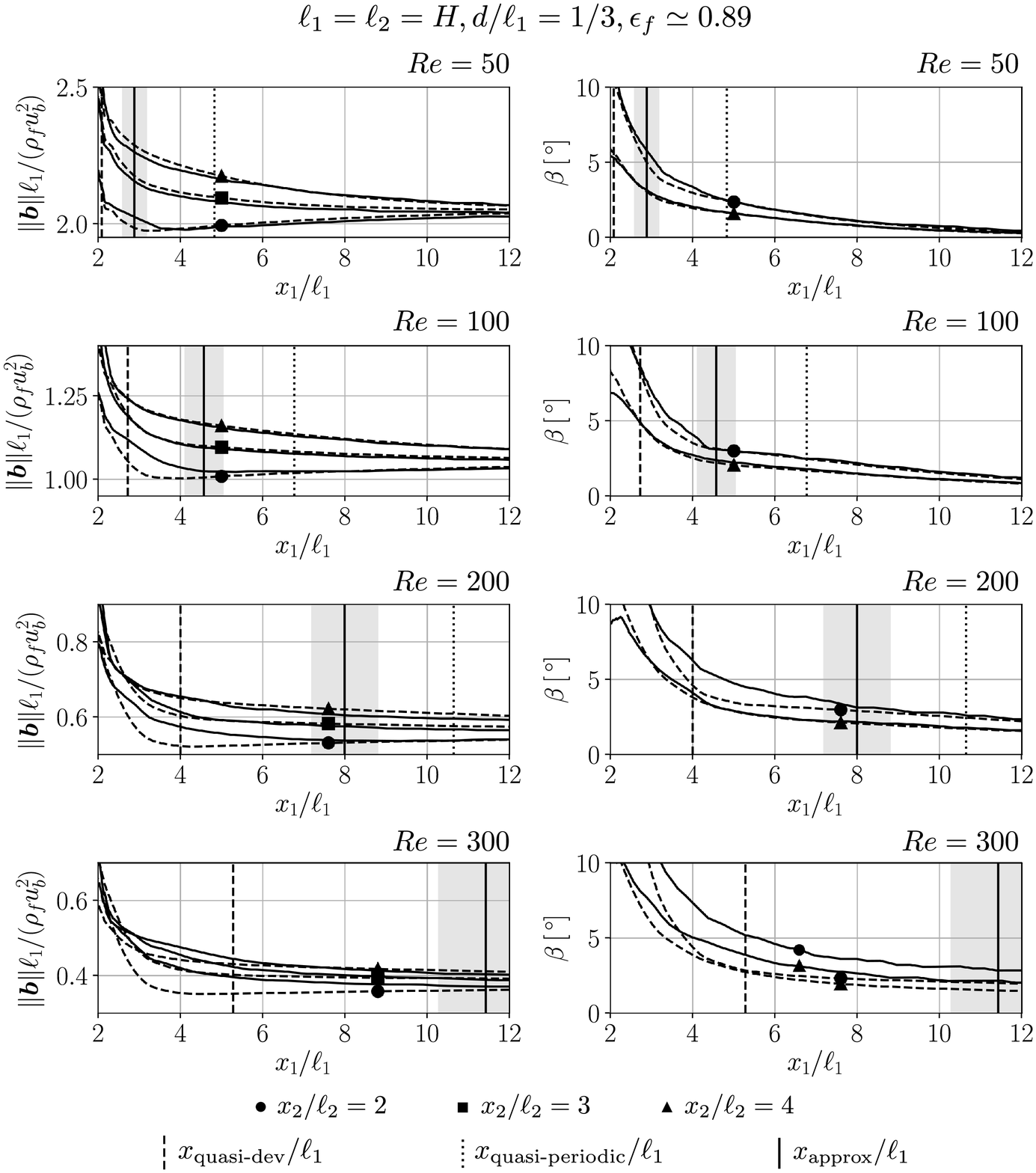}
\caption{
Closure force in the developing flow region of a channel array with a porosity $\epsilon_f \simeq 0.89$ ($N_1=60, N_2=10$, $s_0/\ell_1 =1, s_N/\ell_1 = 1$, $H/\ell_1 =1$, $\ell_1/\ell_2 =1$) for different Reynolds numbers $\Rey$ and different $x_2$-sections.
The solid lines ($-$) represent the magnitude and angle of the actual  closure force $\boldsymbol{b}$.
The dashed lines (-$\,$-) represent the solution of the classical closure problem, $\boldsymbol{b}_{\text{approx}}$. 
The onset points of quasi-periodic and quasi-developed flow, $x_{\text{quasi-periodic}}$ and  $x_{\text{quasi-dev}}$ have been indicated by a dotted and a dashed vertical line respectively.
The solid vertical line marks the point $x_{\text{approx}}$ from where on the classical closure problem is theoretically accurate to within $1\%$.
}
\label{fig:Developing closure force comparison correlation porosity 0.88}
\end{center}
\end{figure}

In figure \ref{fig:Developing closure force comparison correlation porosity 0.88}, another comparison between the actual closure force  and its approximation according to the classical closure problem is presented, yet for a  channel with a higher porosity $\epsilon_f \simeq 0.89$.
The actual closure force $\boldsymbol{b}$, indicated by the solid lines ($-$), has again been obtained through direct numerical simulation and explicit filtering of the flow field in the channel. 
Also the approximation according to the classical closure problem $\boldsymbol{b}_{\text{approx}}$, indicated by the dashed lines (-$\,$-), has been determined in a similar fashion as discussed before.
In this case, the classical closure problem was solved numerically for more than $250$ combinations of the angle of attack $\alpha$ and local Reynolds number $\Rey_m$, with respect to the selected geometry, i.e. $\ell_1=\ell_2=H$ and $d/\ell_1=1/3$. 
The solutions of the closure problem which are most relevant for reproducing figure \ref{fig:Developing closure force comparison correlation porosity 0.88} have been shown in figure \ref{fig: Closure problem for developing macro-scale flow}, as  $\alpha(\boldsymbol{x}) \in (0, 10^{\circ})$ and $\Rey_m(\boldsymbol{x})  \in \left(0.5 Re, 1.3Re \right)$ if $\boldsymbol{x} \in\Omega_{\text{predev}}$, and $\Rey_m \simeq 1.07 \Rey$ in $\Omega_{\text{uniform}}$.

From the profiles of the actual closure force in figure \ref{fig:Developing closure force comparison correlation porosity 0.88} it can again be seen that in a geometrically similar array with a higher porosity, the onset of the periodically developed flow region is delayed more downstream.
The point where the macro-scale flow becomes developed, is now given by 
$x_{\text{dev}} \simeq 14 \ell_1$  for $\Rey=50$, while $x_{\text{dev}} \simeq 62 \ell_1$ for $\Rey=300$, in agreement with the correlation presented in subsection \ref{subsec: Validity of the Local Closure Problem for Developed Macro-Scale Flow}.

Also in this case, we observe a quantitatively good agreement between the actual closure force $\boldsymbol{b}$ and its approximation $\boldsymbol{b}_{\text{approx}}$ once the macro-scale flow is quasi-developed, despite the fact that the largest components of $\boldsymbol{K}_{\text{approx}}$ and $\boldsymbol{\mathsf{K}}$ differ by as much as $20\%$ or $40\%$ at certain positions.
Nevertheless, the difference between $\boldsymbol{b}$ and $\boldsymbol{b}_{\text{approx}}$ in the region of quasi-developed flow is better visible than it was for the porosity $\epsilon_f =0.75$.
The first reason is that the amplitude of the macro-scale velocity mode is significantly larger, when the flows develops in an array with a higher porosity: $\Vert \langle \velamp \rangle_m\Vert < 0.35 U $ (see figure \ref{fig: Quasi-developed macro-scale flow modes}).
Secondly, the rate $\lambda$ at which the exponential velocity mode and thus the modelling error (\ref{eq: modelling error closure force quasi-developed macro-scale flow}) vanish after the point $x_{\text{quasi-dev}}$, is smaller: $1/(\lambda \ell_1) \simeq 0.06\Rey + 0.7$ (\cite{Buckinx2022}).

For the same two reasons, the section after which the classical closure problem is theoretically accurate within a relative error $\varepsilon_{\boldsymbol{K}} = 0.01$, is located more downstream:
$x_{\text{approx}}/\ell_1 \simeq 1/(\lambda \ell_1) \left( \ln (1+c_2/Re) -  \ln(\varepsilon_{\boldsymbol{K}}) - 4 \right)$, where $c_2=13$.
The position of the latter section, $x_{\text{approx}}$, could not be calculated without numerical uncertainty from the data itself in figure \ref{fig:Developing closure force comparison correlation porosity 0.88}, because the relative discretization error for the angle $\beta$ is around $10\%$ when $\Rey \leq 100$, and may become as large as $50\%$ when $\Rey\geq  200$.
Such a high upper bound for the discretization error on $\beta$ is attributed to the fact that in a high-porosity array, the macro-scale flow field will have almost no transversal component at higher Reynolds numbers: $\beta \simeq \alpha \simeq \langle \velampy \rangle_m/U_{\text{dev}} \exp(\lambda x_1) \simeq 0$, as $\langle \velampy \rangle_m/U_{\text{dev}} \sim \lambda (1+c_2/\Rey) \sim 1/\Rey$ for sufficiently high $\Rey$.
On the other hand, the discretization error for the magnitude $\Vert\boldsymbol{b}\Vert$ is comparable to that in figure \ref{fig:Developing closure force comparison correlation porosity 0.75}. 
Therefore, the closest numerical approximation for $x_{\text{approx}}$, denoted by $\hat{x}_{\text{approx}}$, has been indicated instead in figure \ref{fig:Developing closure force comparison correlation porosity 0.88}, by means of the grey areas, just like in the previous figure.
This position $\hat{x}_{\text{approx}}$ from where on $\boldsymbol{b}_{\text{approx}}$ deviates no more than $1\%$ in magnitude and $10\%$ in angle from $\boldsymbol{b}$ can be seen to obey the theoretical scaling law derived for $x_{\text{approx}}$.

\begin{figure}
\begin{center}
\includegraphics[scale=0.5]{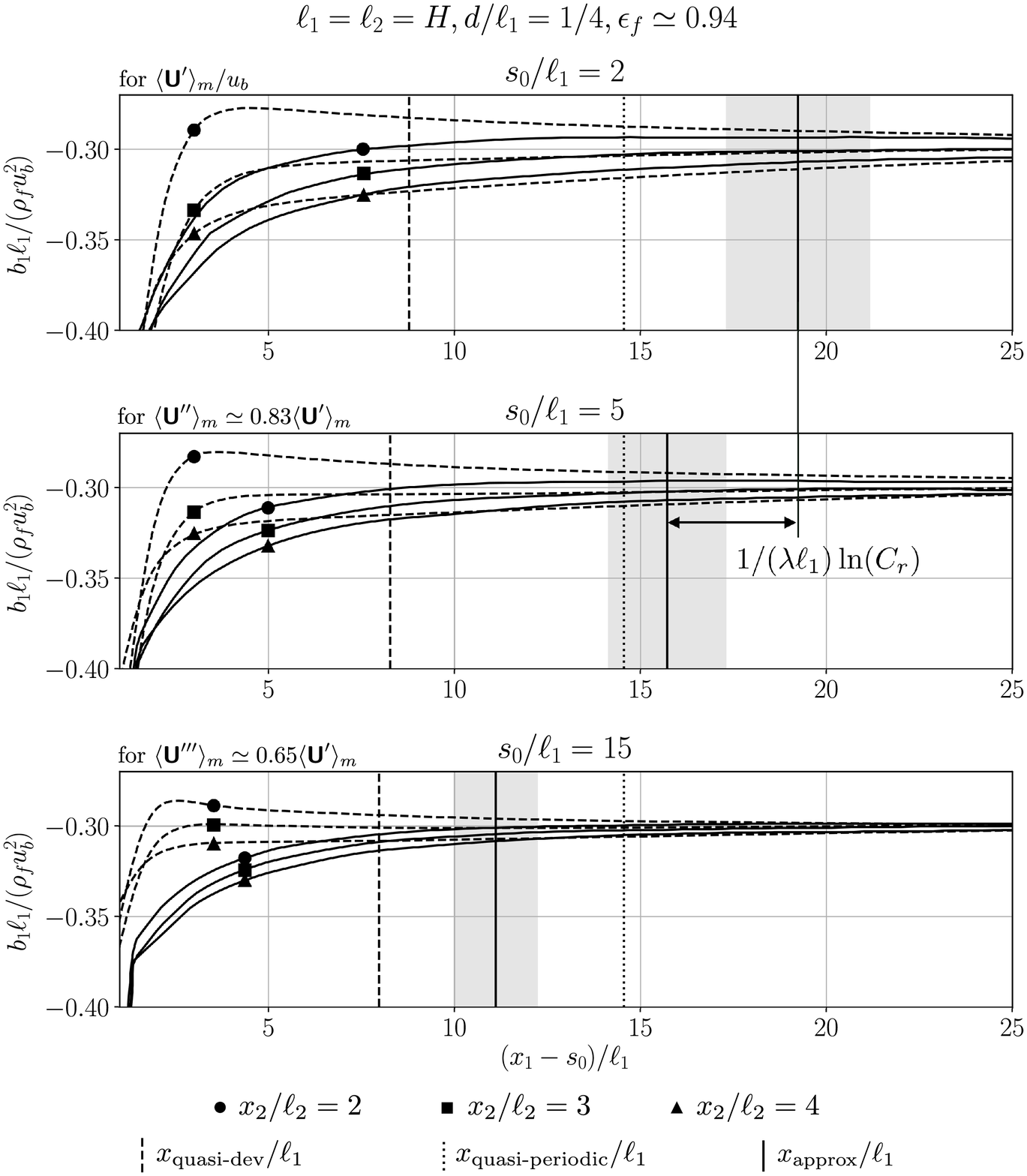}
\caption{
Closure force in the developing flow region of a channel array with a porosity $\epsilon_f \simeq 0.94$ ($N_1=90, N_2=10$, $s_N/\ell_1 = 1$, $H/\ell_1 =1$, $\ell_1/\ell_2 =1$) for different Reynolds numbers $\Rey$, different positions of the first cylinder row $s_0$, and different $x_2$-sections.
The solid lines ($-$) represent the component $b_1$ of the actual closure force along the main flow direction.
The dashed lines (-$\,$-) represent the solution of the classical closure problem, $b_{\text{approx},1}$. 
The onset points of quasi-periodic and quasi-developed flow, $x_{\text{quasi-periodic}}$ and  $x_{\text{quasi-dev}}$ have been indicated by a dotted and a dashed vertical line respectively.
The solid vertical line marks the point $x_{\text{approx}}$ from where on the classical closure problem is theoretically accurate to within $1\%$.}
\label{fig:Developing closure force comparison correlation porosity 0.93}
\end{center}
\end{figure}

The previous figures \ref{fig:Developing closure force comparison correlation porosity 0.75} and \ref{fig:Developing closure force comparison correlation porosity 0.88} confirm our initial expectation that especially in high-porosity arrays at higher Reynolds numbers ($\Rey > 100$), the classical closure problem fails to capture the macro-scale features of the developing flow.
However, in these figures, the relative perturbation size and thus the mode amplitudes were altered by each change in Reynolds number, as the dimensionless velocity profile $\boldsymbol{u}/u_b$ at the channel inlet was kept fixed. 
As a consequence, we observed a smaller mode amplitude at higher Reynolds numbers, which is a favourable condition with regard to the accuracy of the classical closure problem, even though it is outdone by the accompanying decrease of the eigenvalue $\lambda$.
It is therefore instructive to inspect the isolated effect of different mode amplitudes at a fixed (high) Reynolds number, as illustrated in figure \ref{fig:Developing closure force comparison correlation porosity 0.93} for an array with an even higher porosity $\epsilon_f \simeq 0.94$.

In figure \ref{fig:Developing closure force comparison correlation porosity 0.93}, the closure force $\boldsymbol{b}$ and its approximation $\boldsymbol{b}_{\text{approx}}$ are shown for each of the three mode amplitudes $\langle \velamp' \rangle_m$,  $\langle \velamp'' \rangle_m$ and  $\langle \velamp''' \rangle_m$ depicted in figure \ref{fig: Quasi-developed macro-scale flow} (c).
These three mode amplitudes, which all have the same shape but a different magnitude, have thus been obtained by varying the distance $s_0$ between the inlet and the first cylinder row, so that at the beginning of the array, at $x_1=s_0$, a different velocity profile was achieved.
Only the component $b_1$ along the main flow direction $x_1$ is shown in figure \ref{fig:Developing closure force comparison correlation porosity 0.93}, because the transversal component $b_2$ is so small that its numerical values suffer from significant discretization errors. 
We remark that the Reynolds number $\Rey$, which equals $300$ in this case, should be considered as high for a steady laminar flow, because it is quite close to the critical Reynolds number at which chaotic vortex shedding starts to occur in the selected channel and array geometry.
In fact, the critical Reynolds number lies somewhere between $320$ and $400$ according to our direct numerical simulations.

The grey-coloured areas in figure \ref{fig:Developing closure force comparison correlation porosity 0.93} indicate this time the location of the point $\hat{x}_{\text{approx}}$ from where on the approximation $b_{\text{approx,1}}$ deviates no more than $2\%$ in magnitude from the actual closure force $b_1$ over the entire cross section of the channel.
As before, their width reflects the numerical uncertainty on  $\hat{x}_{\text{approx}}$ due to the flatness of the $b_1$-curves.
The location of the grey-coloured areas  is well predicted by the theoretical scaling law for $x_{\text{approx}}$ (\ref{eq: scaling law point of approximate classical closure problem}), provided that we take $\varepsilon_{\boldsymbol{K}} = 0.01$ instead of $\varepsilon_{\boldsymbol{K}} = 0.02$.
This numerical inconsistency for $\varepsilon_{\boldsymbol{K}}$ is caused by two factors.
On the one hand, the estimated discretization error for $b_1$ lies between $1\%$ and $4\%$.
On the other hand, also the underlying assumption that $\boldsymbol{K}_{\text{approx}} \left(\langle \boldsymbol{u} \rangle_m \right) \simeq \boldsymbol{K}_{\text{uniform}} \left( \boldsymbol{U} \right)$ leads to an error on $b_1$ of at least $2\%$, as inertia effects are more important at  at the current Reynolds number and porosity.

Notwithstanding this small inconsistency, we clearly notice in  figure \ref{fig:Developing closure force comparison correlation porosity 0.93} that both points $\hat{x}_{\text{approx}}$ and $x_{\text{approx}}$ shift upstream over a distance $1/\lambda \ln C_r$, upon a reduction of the relative perturbation size by a factor $C_r \triangleq \langle \velampx' \rangle_m /\langle \velampx'' \rangle_m$, from the larger mode $\langle \velamp' \rangle_m $ to the smaller one, $\langle \velamp'' \rangle_m$.
Due to this shift, the points $\hat{x}_{\text{approx}}$ and $x_{\text{approx}}$ for the mode $\langle \velamp'' \rangle_m$ end up much closer to the onset point of quasi-periodically developed flow, $x_{\text{quasi-periodic}}$.
We also notice that a further decrease of the relative perturbation size, causing the mode amplitude to change from $\langle \velamp \rangle_m $ to $\langle \velamp''' \rangle_m $, eventually moves the points $\hat{x}_{\text{approx}}$ and $x_{\text{approx}}$ further downstream to the point $x_{\text{quasi-dev}}$, from where on the macro-scale flow can be treated as quasi-developed.
These findings illustrate that Whitaker's permeability tensor $\boldsymbol{K}_{\text{uniform}}$ suffices to accurately  describe quasi-developed flow, even at at higher Reynolds numbers and higher porosities, as long as the mode amplitude is small enough.
However, for larger mode amplitudes like $\langle \velamp' \rangle_m$, only the exact permeability tensor $\boldsymbol{K}_{\text{quasi-dev}}$ will accurately describe the macro-scale flow from the point $x_{\text{quasi-dev}}$ onwards.
How large the mode amplitude $\langle \velamp \rangle_m$ needs to be before the exact permeability tensor $\boldsymbol{K}_{\text{quasi-dev}}$ becomes of practical interest, depends on the tensor $\boldsymbol{\mathsf{K}}$, as it is the mode amplitude of the closure force, $\mu_f  \boldsymbol{\mathsf{K}}^{-1} \boldsymbol{\cdot} \langle \velamp \rangle_m$, which dictates the error of using $\boldsymbol{K}_{\text{uniform}}$ instead of $\boldsymbol{K}_{\text{quasi-dev}}$.
So, a relatively large mode amplitude $\langle \velamp \rangle_m$ does not necessarily imply a large error when replacing $\boldsymbol{K}_{\text{quasi-dev}}$  by $\boldsymbol{K}_{\text{uniform}}$.
For instance, in figure \ref{fig:Developing closure force comparison correlation porosity 0.93}, the relative size of the mode amplitude of the closure force, $\mu_f \Vert \boldsymbol{\mathsf{K}}^{-1} \boldsymbol{\cdot} \langle \velamp \rangle_m\Vert/\Vert \boldsymbol{b}^{\star} \Vert  $, is about a factor four smaller than $\langle \velamp \rangle_m/U$.
Vice versa, a small change in mode amplitude of the closure force can have a notable impact on the resulting mode amplitude of the macro-scale velocity field.

\subsection{Validity of the Classical Closure Problem for Quasi-Developed Flow \\ Inside the Side-Wall Region}
\label{subsec: Validity of the Classical Closure Problem for Quasi-Developed Flow Inside the Side-Wall Region}
In the side-wall region $\Omega_{\text{sides}}$, the classical closure problem tends to yield a very poor approximation of the actual closure force, due to the strong gradients of the (macro-scale) velocity field that form perpendicular to the side walls, as the flow is slowed down by viscous stresses near the solid boundaries. 
However, the closure problem for developed macro-scale flow presented in section \ref{subsec: Exact Local Closure in the Region of Developed Macro-Scale Flow}, which is an extension of the classical closure problem over $\Omega_{\text{sides}}$, can hold well enough to use the approximation 
$\boldsymbol{b}_{\text{approx}}= \langle p \rangle^f_m \boldsymbol{\nabla} \epsilon_{fm} - \mu_f \boldsymbol{K}_{\text{dev}}^{-1}(\langle \boldsymbol{u} \rangle_m) \boldsymbol{\cdot} \langle \boldsymbol{u} \rangle_m$ for quasi-developed flow in and outside the side-wall region.
The precondition is again that the relative perturbation size $C_{\velamp}^{+}$ and thus the amplitude of the macro-scale velocity mode are sufficiently small, and that $ \boldsymbol{K}_{\text{dev}}(\langle \boldsymbol{u} \rangle_m) \simeq  \boldsymbol{K}_{\text{quasi-dev}}(\boldsymbol{U})$, as it follows from (\ref{eq: approximate structure permeability tensor quasi-developed flow}).

\begin{figure}
\begin{center}
\includegraphics[scale=0.5]{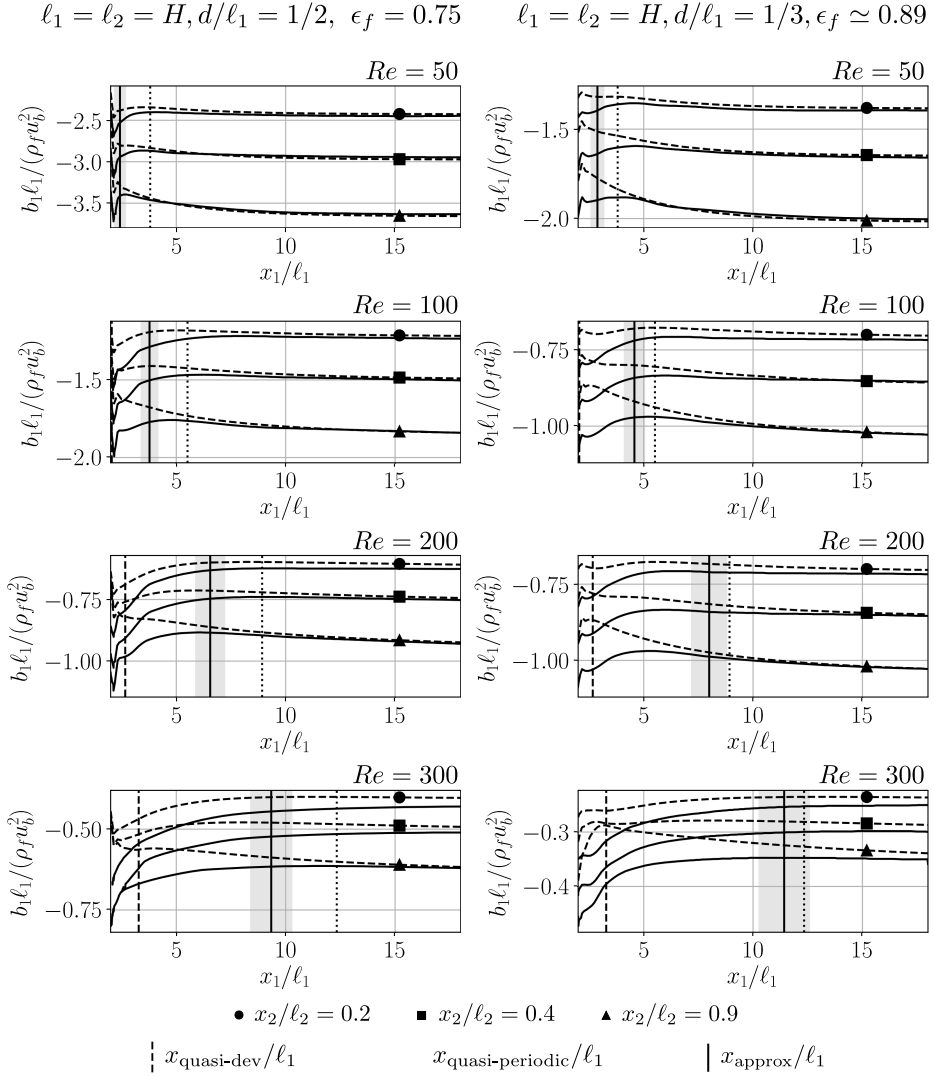}
\caption{Closure force in the side-wall region of a channel array with a porosity  $\epsilon_f =0.75$ (left) and $\epsilon_f \simeq 0.89$ (right) ($N_1=60, N_2=10$, $s_0/\ell_1 =1, s_N/\ell_1 = 1$, $H/\ell_1 =1$, $\ell_1/\ell_2 =1$) for different Reynolds numbers $\Rey$ and different sections $x_2/\ell_2$.
The solid lines ($-$) represent the actual closure force $b_1$.
The dashed lines (-$\,$-) represent the approximate solution of the closure problem for developed macro-scale flow, $b_{\text{approx},1}$, which is based on the permeability tensor $\boldsymbol{K}_{\text{uniform}}$ from the classical closure problem and a linear correlation for $\xi$. 
The onset points of quasi-periodic and quasi-developed flow, $x_{\text{quasi-periodic}}/\ell_1$ and  $x_{\text{quasi-dev}}/\ell_1$ have been indicated by a dotted and a dashed vertical line respectively.
The solid vertical line marks the point $x_{\text{approx}}/\ell_1$ from where on the classical closure problem is theoretically accurate to within $1\%$.}
\label{fig:Developing closure force comparison correlation side-wall region}
\end{center}
\end{figure}

The point $x'_{\text{approx}}$ from where on the latter approximation holds  with a relative accuracy $\varepsilon'_{\boldsymbol{K}} \triangleq \Vert\boldsymbol{I} - \boldsymbol{K}_{\text{dev}} \boldsymbol{\cdot} \boldsymbol{K}_{\text{quasi-dev}}^{-1} \Vert$ also obeys the scaling law (\ref{eq: scaling law point of approximate classical closure problem}):
\begin{equation}
\label{eq: scaling law point of approximate classical closure problem side-wall region}
x'_{\text{approx}}\simeq \frac{1}{\lambda} \left( \ln \frac{C_{\velamp}^{+}}{\varepsilon'_{\boldsymbol{K}}} + c'_3 \right)\,,
\end{equation}
if we define $c'_3 \triangleq  \ln \displaystyle \max_{x_2} \Vert
\boldsymbol{K}_{\text{dev}}
\boldsymbol{\cdot}  \boldsymbol{\mathsf{K}}^{-1} \boldsymbol{\cdot} \boldsymbol{\zeta}_{\text{ref}} \boldsymbol{\cdot} \boldsymbol{\xi}^{-1} \Vert $.
Often, the point $x'_{\text{approx}}$ is little affected by the shape of the developed macro-scale velocity profile $\xi$ itself, because $\boldsymbol{K}_{\text{dev}} \simeq \boldsymbol{\xi} \boldsymbol{\cdot} \boldsymbol{K}_{\text{dev, main}}$ (cf. (\ref{eq: main contribution for developed apparent permeability tensor})), so that the tensor $\boldsymbol{\xi}$ contained in $\boldsymbol{K}_{\text{dev}}$ almost cancels its inverse in $c_3$.
Conversely, it is strongly affected by the difference between $\boldsymbol{K}_{\text{dev, main}}$ and $\boldsymbol{\mathsf{K}}$, which can be significantly large in $\Omega_{\text{sides}}$.
Essentially, the larger this difference, the more downstream $x'_{\text{approx}}$ will lie, closer to the developed flow region.
Therefore, the point $x'_{\text{approx}}$ will lie downstream of $x_{\text{approx}}$, when the discrepancy between $\boldsymbol{K}_{\text{dev, main}}$ and $\boldsymbol{\mathsf{K}}$ in $\Omega_{\text{sides}}$ exceeds the discrepancy between $\boldsymbol{K}_{\text{approx}}$ and $\boldsymbol{\mathsf{K}}$ in $\Omega_{\text{core}}$.
This means that the closure problem for developed macro-scale flow then will become valid in the side-wall region for some criterion $\varepsilon'_{\boldsymbol{K}} = \varepsilon_{\boldsymbol{K}}$, after it has become valid in the core region.
On the other hand, when the variation of the tensor $\boldsymbol{K}_{\text{dev, main}}$ over the side-wall region is rather small, such that  (\ref{eq: side-wall region permeability approximately uniform}) applies, we find that $x'_{\text{approx}}\simeq x_{\text{approx}}$, as $c'_3 \simeq c_3$.
This situation, where the approximation $ \boldsymbol{b} \simeq \boldsymbol{b}_{\text{approx}}= \langle p \rangle^f_m \boldsymbol{\nabla} \epsilon_{fm} - \mu_f \boldsymbol{K}_{\text{dev}}^{-1}(\langle \boldsymbol{u} \rangle_m) \boldsymbol{\cdot} \langle \boldsymbol{u} \rangle_m$ with $\boldsymbol{K}_{\text{dev}} \simeq \boldsymbol{\xi} \boldsymbol{\cdot} \boldsymbol{K}_{\text{uniform}}$, has nearly the same  relative accuracy in $\Omega_{\text{sides}}$ as in $\Omega_{\text{core}}$ once the macro-scale flow is quasi-developed, occurs in figure \ref{fig:Developing closure force comparison correlation side-wall region}. 

In figure \ref{fig:Developing closure force comparison correlation side-wall region}, the approximation $\boldsymbol{b}_{\text{approx}}$
based on the permeability tensor from the classical closure problem (i.e.  based on $\boldsymbol{K}_{\text{dev}} \simeq \boldsymbol{\xi} \boldsymbol{\cdot} \boldsymbol{K}_{\text{uniform}}$), is compared with the actual closure force $\boldsymbol{b}$ in the side-wall region.
Only the components $b_1$ and $b_{\text{approx},1}$ along the main flow direction are set out, because the transversal component is known a priori for quasi-developed flow: $b_2 \simeq \langle p \rangle^f_m d\epsilon_{fm}/dx_2$.
The different channel flows considered in this figure correspond to two different porosities, $\epsilon_f=0.75$ and $\epsilon_f\simeq 0.89$.
The core regions of these flows have been shown before in figures \ref{fig:Developing closure force comparison correlation porosity 0.75} and \ref{fig:Developing closure force comparison correlation porosity 0.88}.
It must be remarked that $b_{\text{approx},1}$ is based on the linear approximation for $\xi$ given in (\ref{eq: linear macro-scale velocity shape profile}), instead the exact $\xi$-profile.
Therefore, we still observe deviations between $b_{\text{approx},1}$ and $b_1$ after the section $x_1=x_{\text{approx}}$, even though  $x'_{\text{approx}}$ is close to $x_{\text{approx}}$ within one to three unit cell lengths $\ell_1$ for the exact $\xi$-profile, over the displayed  Reynolds number range.

We remark that as the Reynolds number $\Rey$ increases in figure \ref{fig:Developing closure force comparison correlation side-wall region},  the local angle of attack $\alpha(\boldsymbol{x})$ in $\Omega_{\text{sides}}$ increases, while it was seen to decrease in $\Omega_{\text{core}}$. 
This is a consequence of the shape of the mode amplitudes illustrated in figure \ref{fig: Quasi-developed macro-scale flow modes}.

\section{Reconstruction of Quasi-Developed Macro-Scale Flow}
\label{sec: Reconstruction of Quasi-Developed Macro-Scale Flow}

The preceding theoretical considerations and empirical evidence suggest that the closure models for developed flow (\ref{eq: no-slip force permeability tensor developed flow region}) and quasi-developed flow (\ref{eq: no-slip force permeability tensor quasi-developed flow region}) are not able to capture the flow development before the point $x_{\text{quasi-dev}}$,
whether the mode amplitude  $\langle \velamp \rangle_m$ is small or large.
So, it seems that with the present closure models, accurate closure for the macro-scale flow equations can only be achieved when the point $x_{\text{quasi-dev}}$ is located close to the channel inlet -- that is, when the developing flow almost entirely can be treated as quasi-developed. 
Nevertheless, the validity of these closure models has been discussed so far only from an a-priori analysis, in which the exact macro-scale velocity field is known in advance.
Therefore, we will now discuss the validity of these closure models from an a-posteriori analysis.
By this we mean an analysis after the closure model has been employed to solve the macro-scale flow equations, and to reconstruct the quasi-developed macro-scale flow.

\begin{figure}
\begin{center}
\includegraphics[width=0.75\columnwidth]{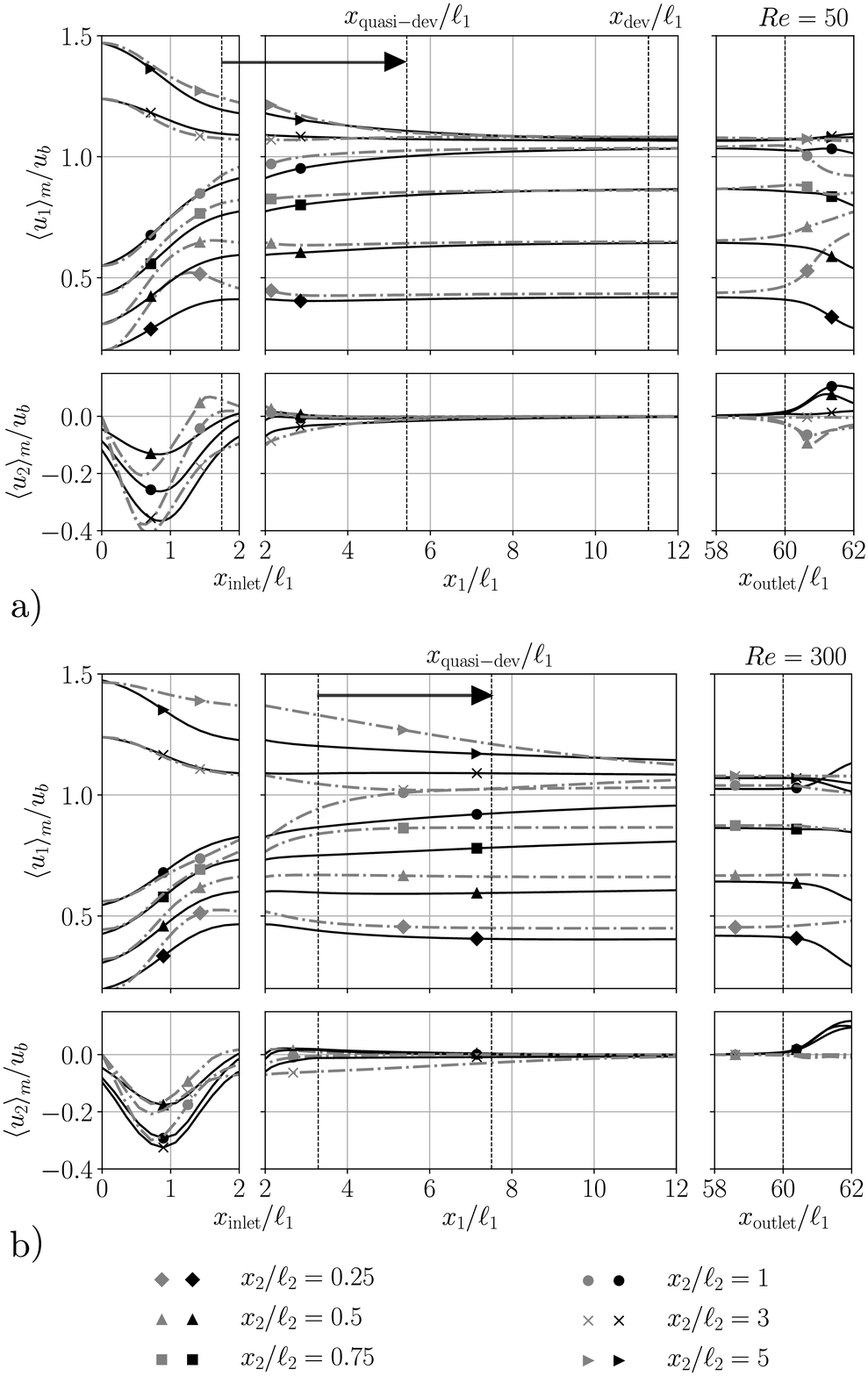}
\caption{Reconstructed macro-scale velocity field and flow regions in a channel array ($N_1 =60, N_2=10$, $s_0/\ell_1 =1, s_N/\ell_1 = 10$, $H/\ell_1 =1$, $\ell_1/\ell_2 =1$) with $\epsilon_f = 0.75$ at $\Rey =50$ (a) and $\Rey =300$ (b).
Black and grey lines correspond to the reconstructed and actual macro-scale velocity field respectively.
Note that the geometry is the same as in figure \ref{fig:Macro-Scale Flow Field and Regions Re 50 porosity 0.75}, although the number of cylinders is higher: $N_1 =60$ instead of $N_1 =20$.}
\label{fig: Reconstructed Macro-Scale Flow Field Re 50 porosity 0.75}
\end{center}
\end{figure}

In figure \ref{fig: Reconstructed Macro-Scale Flow Field Re 50 porosity 0.75}, the actual macro-scale velocity field in a channel, as obtained via direct numerical simulation and explicit filtering, has been reconstructed for $\Rey =50$ and $\Rey =300$, by solving the macro-scale flow equations (\ref{eq: macro-scale momentum equation}) and (\ref{eq: macro-scale continuity equation}) 
with the approximate closure model
\begin{equation}
\label{eq: reconstruction closure model}
\boldsymbol{b} =
\left\{
\begin{aligned}
&\boldsymbol{b}_{\vert \vert} 
&&\qquad \mbox{for} ~~ x_1 \in (0,s_0 - \ell_1) \cup (x_{\text{outlet}} +2\ell_1,L) \,, \\ 
& \boldsymbol{b}_{\text{dev}} 
&&\qquad \mbox{for} ~~ x_1 \in (x_{\text{inlet}},x_{\text{outlet}}) \,, \\ 
& \boldsymbol{b}_{\vert \vert} w_0(x_1) + \boldsymbol{b}_{\text{dev}} (1-w_0(x_1))
&&\qquad \mbox{for} ~~ x_1 \in (s_0 - \ell_1,x_{\text{inlet}})  \,, \\
& \boldsymbol{b}_{\vert \vert} w_N(x_1) + \boldsymbol{b}_{\text{dev}} (1-w_N(x_1))
&&\qquad \mbox{for} ~~ x_1 \in (x_{\text{outlet}},x_{\text{outlet}} +2\ell_1)  \,.
\end{aligned}
\right.
\end{equation}
In this approximate closure model, the closure force outside the array is given by $\boldsymbol{b}_{\vert \vert} \triangleq  \langle p \rangle^f_m \boldsymbol{\nabla} \epsilon_{fm} -\mu_f (12/H^2) \epsilon_{fm} \langle \boldsymbol{u}\rangle_m$, which is the value that would be attained for fully developed flow between the top and bottom wall, in the absence of side walls.
Inside the array, the closure force is given by the closure model for developed macro-scale flow, $\boldsymbol{b}_{\text{dev}} \triangleq \langle p \rangle^f_m \boldsymbol{\nabla} \epsilon_{fm} - \mu_f \boldsymbol{K}_{\text{dev}}^{-1}(\langle \boldsymbol{u} \rangle_m) \boldsymbol{\cdot} \langle \boldsymbol{u} \rangle_m$, with $\boldsymbol{K}_{\text{dev}} = \xi \boldsymbol{K}_{\text{uniform}}$.
Here, $\boldsymbol{K}_{\text{uniform}}$ is obtained from the solution of the classical closure problem (\cite{Buckinx2022}), while $\xi$ is given by (\ref{eq: linear macro-scale velocity shape profile}).
In the regions near the inlet and outlet where porosity gradients occur, the closure force is determined by an ad-hoc linear interpolation between $\boldsymbol{b}_{\vert \vert} $ and $\boldsymbol{b}_{\text{dev}}$, as suitable closure models for these regions are still lacking in the literature.
The linear interpolation functions in (\ref{eq: reconstruction closure model}) are defined as $w_0(x_1) \triangleq (s_0 + \ell_1/2-x_1)/\ell_1$ and $w_N(x_1) \triangleq (L-s_N + \ell_1/2-x_1)/\ell_1$.
The approximate closure model (\ref{eq: reconstruction closure model})
can thus be interpreted as a generalization of models which employ developed-friction-factor correlations to estimate the \textit{local}  macro-scale pressure drop (\cite{Buckinx2022}).

To obtain the reconstructed macro-scale velocity field in figure \ref{fig: Reconstructed Macro-Scale Flow Field Re 50 porosity 0.75} (a,b), the macro-scale flow equations were solved for the actual $\langle u_1 \rangle_m$-profile at the channel inlet.
Hereto, the actual profile of the macro-scale velocity at the channel inlet was described by the fitted function $\langle \boldsymbol{u} \rangle_m(x_1,x_2) = (0.57 x_2 - 0.057 x_2^2 + 0.05)u_b \boldsymbol{e}_1$ for $x_2 \in (0,W)$ and $x_1=0$.
Along the side walls of the channel, at $x_2= \pm W/2$, the slip condition (\ref{eq: slip length and slip condition for developed macro-scale flow}) for the velocity component $\langle u_1 \rangle_m$ was imposed.
The other component $ \langle u_2 \rangle_m $ was set to zero at both the channel inlet and side walls.
At the outlet of the channel, the macro-scale pressure was prescribed: $\langle p \rangle_m(x_1,x_2) = 0$ at $x_1=L$.
For the discretization of the macro-scale flow equations, a uniform triangular mesh of $500 000$ cells was chosen.
Further, a mixed-element variational formulation was used, in which the Taylor-Hood finite-element space was chosen for $\langle \boldsymbol{u} \rangle_m$ and $\langle p \rangle_m$.
The discretized macro-scale flow equations were solved as a coupled system using a Newton method with an exact linearization.

The slip condition (\ref{eq: slip length and slip condition for developed macro-scale flow}) at the side-walls is the most severe simplification of the boundary conditions satisfied by the actual macro-scale velocity field.
In fact, it may be replaced by the no-slip condition $\langle u_1 \rangle_m=0$ at $x_2= \pm W/2$, without significantly changing the reconstructed macro-scale velocity field.
Yet, the approximate closure model (\ref{eq: reconstruction closure model}) is the cause of the largest discrepancies between the actual macro-scale velocity field and the reconstructed one.
The poor approximation of the closure force outside the array, $\boldsymbol{b}_{\vert \vert}$, does not allow to accurately trace the magnitude of the actual closure force over the inlet and outlet region.
In addition, it fails to reproduce the strong misalignment between the directions of $\boldsymbol{b}$ and $\langle \boldsymbol{u} \rangle_m$ in these regions. 
The actual angle $\beta$ can be for instance more than five degrees larger than $\alpha$ in $\Omega_{\text{inlet}} \cap \Omega_{\text{core}}$, while the assumption $\boldsymbol{b} =\boldsymbol{b}_{\vert \vert}$ implies $\beta = \alpha$.
Therefore, we see in figure \ref{fig: Reconstructed Macro-Scale Flow Field Re 50 porosity 0.75} that the strongest deviations between the actual macro-scale velocity field and the reconstructed macro-scale velocity field occur in the inlet and outlet region.
As a result, the reconstructed macro-scale pressure gradient is up to four times smaller than the actual macro-scale pressure gradient $\partial \langle p \rangle^f_m / \partial x_1$ near $x_1=x_{\text{inlet}}$.

The most important message to take away from figure \ref{fig: Reconstructed Macro-Scale Flow Field Re 50 porosity 0.75} is that also the reconstructed macro-scale flow field exhibits a quasi-developed flow region.
The occurrence of quasi-developed solutions for channel flows is ultimately a mathematical property of the Navier-Stokes equations, as well as the quite similar macro-scale flow equations.
However, the onset point, eigenvalue and mode shape of the quasi-developed  flow after the reconstruction do not match that of the actual flow.
Hence, the classical closure problem leads to three types of reconstruction errors.

In figure \ref{fig: Reconstructed Macro-Scale Flow Field Re 50 porosity 0.75} (a), we see that the onset point of quasi-developed flow, $x_{\text{quasi-dev}}$, has shifted downstream after the reconstruction, due to the approximate closure model in $\Omega_{\text{predev}}$.
Also the point after which the reconstructed macro-scale flow can be considered developed, lies more downstream.
However, it is not visible in the figure, as the reconstructed macro-scale flow is just shown up to the point $x_{\text{dev}}$, where the actual macro-scale flow becomes developed.
Still, the distance over which the region of quasi-developed flow extends,  is nearly the same after the reconstruction. 
The reason is that both the perturbation size and eigenvalue of the reconstructed flow are a factor two larger than that of the original flow.
As such, the reconstruction errors in $C_{\velamp}^{+}$ and $\lambda$ cancel each other more or less, and the distance $x_{\text{quasi-dev}} - x_{\text{dev}}$ is barely affected.
For the same reason, the macro-scale velocity component $\langle u_1 \rangle_m$ in the quasi-developed flow region deviates less than $5\%$ from its actual value in the core of the channel $\Omega_{\text{core}}$, and less than $15\%$ in the side-wall region $\Omega_{\text{sides}}$.

In figure \ref{fig: Reconstructed Macro-Scale Flow Field Re 50 porosity 0.75} (b), the macro-scale flow field has been reconstructed for a higher Reynolds number $\Rey =300$.
It can be noticed that due to the higher Reynolds number, the onset point of quasi-developed flow has moved even further downstream after the reconstruction.
In addition, the differences between the reconstructed and actual macro-scale velocity field are significantly larger.
More specifically, the reconstructed macro-scale velocity field develops at a much higher rate than the actual macro-scale velocity field, since its eigenvalue is a factor four too large.

\begin{figure}
\begin{center}
\includegraphics[scale=0.5]{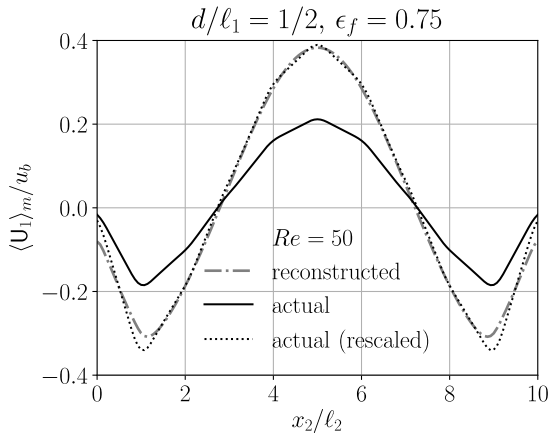}
\caption{
Reconstructed and actual macro-scale velocity mode for the channel flow of figure \ref{fig: Reconstructed Macro-Scale Flow Field Re 50 porosity 0.75} (a).
In order to compare the shape of the actual mode and the reconstructed mode, the former has been rescaled such that it has the same perturbation size as the latter.
}
\label{fig: Reconstructed Mode Macro-Scale Flow Field Re 50 porosity 0.75}
\end{center}
\end{figure}

If the reconstruction is repeated for a range of Reynolds numbers $\Rey \in \left\{50, 100, 200, 300 \right\}$, the reconstructed eigenvalue is found to obey the scaling law $1/(\lambda \ell_1) \simeq 0.009\Rey + 1.4$, whereas the actual eigenvalue scales as $1/(\lambda \ell_1) \simeq 0.05\Rey +0.8$.
This a-posteriori analysis shows that the classical closure problem is not able to capture the correct exponential evolution of quasi-developed flow, because it drastically changes the eigenvalues with respect the original flow.

In principle, the classical closure problem also alters the shape of the mode amplitude $\langle \velampx \rangle_m$ after the reconstruction.
Nonetheless, the reconstructed mode shape is almost identical to the actual mode shape for the flows from figure \ref{fig: Reconstructed Macro-Scale Flow Field Re 50 porosity 0.75}.
This is is shown in more detail in figure \ref{fig: Reconstructed Mode Macro-Scale Flow Field Re 50 porosity 0.75}, where the reconstructed mode shape $\langle \velampx \rangle_m$ for $\Rey=50$ is compared with the actual mode shape, before and after rescaling the latter to obtain the same perturbation size.
The reconstructed mode shape mainly differs in the side-wall region $\Omega_{\text{sides}}$, where the approximate permeability tensor $\boldsymbol{K}_{\text{dev}} = \xi \boldsymbol{K}_{\text{uniform}}$ was used and a constant slip length $\ell_{\text{slip}}$ was imposed.

Evidently, only the exact closure problem for the permeability tensor $\boldsymbol{K}_{\text{quasi-dev}}$, which contains the exact eigenvalue, allows for an exact reconstruction of quasi-developed flow.
Nevertheless, also with $\boldsymbol{K}_{\text{quasi-dev}}$, the onset point $x_{\text{quasi-dev}}$ and perturbation size  $C_{\velamp}^{+}$ may still be incorrect after the reconstruction, as long as an approximate closure model is used upstream of the actual region of quasi-developed flow.

\section{Conclusions}
\label{sec: Conclusions}
To obtain a macro-scale description of quasi-periodically developed flow in channels with arrays of periodic solid structures, the velocity field and pressure field can be filtered using a double volume-averaging operator.
That way, the macro-scale flow will display many of the features of quasi-developed flow in channels without solid structures, so we can speak of quasi-developed macro-scale flow.

A first feature is that the macro-scale velocity field evolves exponentially towards a developed velocity profile, whose shape remains the same at every cross section in the channel, until the end of the array.
A second feature is that the gradient of the macro-scale pressure evolves exponentially along the main flow direction, until it adopts a constant value in the core of the channel, once the macro-scale flow has become developed. 
The mode which determines this exponential evolution, inherits its eigenvalue (or decay rate) from the eigenvalue problem that defines quasi-periodically developed flow. 
The amplitude of the mode does not depend on the specific inlet conditions of the flow, apart from a single scaling factor called the perturbation size.
It is also the same over every cross section in the channel.

When the macro-scale flow is quasi-developed, the closure force has an exact representation in the form of an apparent permeability tensor, which consists of two contributions.
The first contribution is the apparent permeability tensor for developed macro-scale flow.
In the core of the channel, where also the macro-scale velocity is uniform, this tensor can be obtained from the periodic flow equations or the classical closure problem on a unit cell of the array.
Near the side walls of the channel, the developed permeability tensor becomes space-dependent, because it is affected by the local profile of the macro-scale velocity.
As such, it can be approximated by estimating the shape of the developed velocity profile near the side walls.
Theoretically, however, the exact developed permeability tensor is given by a specific closure problem on a unit cell in the side-wall region, when its uniform value in the core region is known.

The second contribution to the apparent permeability tensor for quasi-developed macro-scale flow expresses the resistance against the macro-scale velocity mode.
Therefore, it decays exponentially in the main flow direction at the same rate as the macro-scale velocity mode.
This additional permeability tensor can be obtained from a closure problem on a transversal row of the array.

The apparent permeability tensor for quasi-developed flow will approach asymptotically to the permeability tensor from the classical closure problem, if the channel is long enough.
The point where the classical closure problem becomes valid, scales inversely with the eigenvalue of the quasi-developed flow.
In addition, it scales with the logarithm of the perturbation size.
The reason is that all assumptions and approximations behind the classical closure problem are partly fulfilled, and partly violated by an exponentially vanishing error which is proportional to the perturbation size.

If the classical closure problem is used to model the closure force in the region of quasi-developed flow, the reconstructed macro-scale velocity field will have a mode shape and eigenvalue which differs from the actual macro-scale flow.
On the other hand, if the exact permeability tensor for quasi-developed macro-scale flow is used, only the onset point and perturbation size may no longer be exact after the reconstruction.

To support the former theoretical findings, we have studied the macro-scale flow in high-aspect ratio channels with high-porosity arrays of in-line equidistant square cylinders.
Hereto, we relied on direct numerical simulation and explicit filtering of the detailed flow in the channel.

We found that the shape of the macro-scale velocity profile in the developed region is nearly independent of the Reynolds number, as macro-scale inertia effects in the cylinder array are rather small.
The developed velocity profile is uniform everywhere, except over the cylinder rows closest to the side walls of the channel.
In the side-wall region, the profile is in good approximation linear, especially when the channel height is equal to or greater than the cylinder spacing.
The linear profile is affected by the slip length of the macro-scale velocity at the side walls, which correlates well with the displacement factor for the mass-flow rate in the side-wall region.
Both the slip length and the displacement factor become independent of the channel height, when the channel height is much larger than the cylinder spacing.
For smaller channel heights, they apparently obey a power-law scaling with the ratio of the channel height to cylinder spacing.
In addition, they are observed to increase when the porosity increases, because the friction at the cylinder surface becomes relatively smaller in that case.

We have illustrated the macro-scale velocity modes in the region of quasi-developed flow for different porosities of the cylinder array between 0.75 and 0.94, and Reynolds numbers up to 300.
Also the influence of the inlet conditions, the aspect ratio and the channel height on these modes has been discussed.
In particular, we have shown that the shapes of the macro-scale velocity modes resemble those of the velocity modes in quasi-developed Poiseuille flow.
In analogy, also their amplitudes scale inversely linear with the Reynolds number, as does the perturbation size.

To assess the accuracy of the classical closure problem in the region of quasi-developed macro-scale flow, its solutions have been compared with the actual closure force in the cylinder array.
To this end, an extensive set of closure solutions has been presented, covering a wide range of Reynolds numbers up to 600, and a variety of porosities between 0.2 and 0.95.
These solutions are restricted to the case where the channel height equals the cylinder spacing.
From these closure solutions, it was established that the developed closure force in the cylinder array is in good agreement with the Darcy-Forchheimer equation.
The Forchheimer contribution accounts for less than $10\%$ of the total closure force over the investigated parameter range.
The closure solutions predict a complex dependence between the direction of the closure force and the direction of the macro-scale velocity.
Furthermore, for larger angles of attack of the macro-scale velocity, stable steady solutions of the classical closure problem may not exist, especially at higher Reynolds numbers.

Our comparison of the former closure solutions with the actual closure force in different channel flows has revealed that the classical closure problem can be quite accurate over the entire region of quasi-developed macro-scale flow.
Our computational results for the cylinder array with a porosity of 0.75 and Reynolds numbers below 100, indicated that the classical closure problem is able to capture the actual closure force with relative errors of less than $5\%$ and $10\%$ in terms of magnitude and angle respectively, almost immediately after the inlet.
Nevertheless, for higher porosities or higher Reynolds numbers, the same quantitative agreement is reached more downstream, in accordance with the theoretically derived scaling laws.

Finally, our computational results suggest that on a macro-scale level, the flow in the cylinder array can be treated as (quasi-) developed, even before the detailed flow is (quasi-) periodically developed in the strict sense.
This observation suggests that the macro-scale description presented in this work applies to almost the entire channel flow at low to moderate Reynolds numbers, except the regions in near vicinity of the channel inlet and outlet.


\section{Acknowledgement}
The present work was supported by the Research Foundation — Flanders (FWO) through G. Buckinx's post-doctoral fellowship grant (12Y2919N).
The author would like to thank prof. dr. Michel Quintard and dr. Yohan Davit from IMFT, Toulouse, France for their preliminary remarks on parts of this draft article.

The resources and services used in this work were provided by the VSC (Flemish Supercomputer Center), funded by the Research Foundation - Flanders (FWO) and the Flemish Government.

\appendix 

\section{Form of the Macro-Scale Flow Equations}
\label{app: Form of the Macro-Scale Flow Equations}
In their most general form, the macro-scale flow equations may be written as
\begin{align}
\rho_f \boldsymbol{\nabla} \boldsymbol{\cdot} \left(
\epsilon_{fm}^{-1} \langle \boldsymbol{u}\rangle_m \langle \boldsymbol{u}\rangle_m \right)  &= 
- \boldsymbol{\nabla}\langle p \rangle_m + 
\mu_f \nabla^2  \langle \boldsymbol{u}\rangle_m
 + \boldsymbol{f}_{\text{closure}} \,, 
\label{eq: general macro-scale momentum equation}
\\
\boldsymbol{\nabla} \boldsymbol{\cdot} \langle
 \boldsymbol{u}\rangle_m  &= \varphi_{\text{closure}} \,,
\label{eq: general macro-scale continuity equation}
\end{align}
where $\boldsymbol{b}_{\text{closure}}$ is the total closure force, and $\varphi_{\text{closure}}$ the continuity closure source.
The total closure force in the macro-scale momentum equation (\ref{eq: general macro-scale momentum equation}) can be split into five closure terms: 
$\boldsymbol{f}_{\text{closure}} \triangleq  - \rho_f \boldsymbol{\nabla} \boldsymbol{\cdot} {\boldsymbol{M}} -\rho_f \boldsymbol{d}+ \boldsymbol{b}_p + \boldsymbol{b}_\tau + \boldsymbol{b}_\mu $. 
The first closure term stems from the macro-scale momentum dispersion tensor $\boldsymbol{M} \triangleq \langle \boldsymbol{u}\boldsymbol{u} \rangle_m - \epsilon_{fm}^{-1} \langle \boldsymbol{u}\rangle_m \langle \boldsymbol{u}\rangle_m $, while the second closure term is given by $\boldsymbol{d}\triangleq \langle  \boldsymbol{\nabla}^{\nu} \boldsymbol{\cdot}  \boldsymbol{u}\boldsymbol{u}\rangle_m - \boldsymbol{\nabla} \boldsymbol{\cdot} \langle \boldsymbol{u}\boldsymbol{u} \rangle_m$.
The other closure terms are defined as
$\boldsymbol{b}_p \triangleq -\langle \boldsymbol{\nabla}^{\nu} p \rangle_m +  \boldsymbol{\nabla} \langle p \rangle_m$,  $\boldsymbol{b}_{\tau} \triangleq \langle \boldsymbol{\nabla}^{\nu} \boldsymbol{\cdot} \boldsymbol{\tau} \rangle_m -  \boldsymbol{\nabla} \boldsymbol{\cdot} \langle \boldsymbol{\tau} \rangle_m$
and $\boldsymbol{b}_\mu \triangleq \boldsymbol{\nabla} \boldsymbol{\cdot} \langle \boldsymbol{\tau} \rangle_m $ -
$\mu_f \nabla^2  \langle \boldsymbol{u}\rangle_m$.
As a consequence of the spatial averaging gradient theorem (\cite{Slattery1999, Howes1985}), each of the previous four closure terms can also be expressed as an integral over all boundary parts in $\mathbb{R}^3$  where the distributions $\boldsymbol{u}$ and $p$ may exhibit a discontinuous jump:
$\boldsymbol{d} =  \langle \boldsymbol{n} \boldsymbol{\cdot}( \boldsymbol{u}_f \boldsymbol{u}_f - \boldsymbol{u}_e\boldsymbol{u}_e) \delta \rangle_m $, $\boldsymbol{b}_p = - \langle \boldsymbol{n}_{fs} p_f\delta_{fs} \rangle_m -  \langle \boldsymbol{n} (p_f-p_e) \delta \rangle_m$, $\boldsymbol{b}_{\tau} =  \langle \boldsymbol{n}_{fs} \boldsymbol{\cdot} \boldsymbol{\tau}_f \delta_{fs} \rangle_m +  \langle \boldsymbol{n} \boldsymbol{\cdot} (\boldsymbol{\tau}_f - \boldsymbol{\tau}_e) \delta \rangle_m$ and $\boldsymbol{b}_\mu = \mu_f \boldsymbol{\nabla} \boldsymbol{\cdot} \langle \boldsymbol{n} ( \boldsymbol{u}_f - \boldsymbol{u}_e) \delta \rangle_m +\mu_f  \boldsymbol{\nabla} \boldsymbol{\cdot} (\boldsymbol{\nabla} \langle \boldsymbol{u} \rangle_m^{\intercal} ) $.
Here, these boundary integrals have been represented by means of the Dirac surface indicators $\delta_{fs} $ and $\delta$ for the boundaries $\Gamma_{fs}$ and $\Gamma$ respectively, and their corresponding normals $\boldsymbol{n}_{fs}$ (pointing towards $\Omega_s$) and $\boldsymbol{n}$ (pointing outwards $\Omega$).
In a similar way, also the closure term in the macro-scale continuity equation 
can be expressed as $\varphi_{\text{closure}}\triangleq  \langle \boldsymbol{\nabla}^{\nu} \boldsymbol{\cdot} \boldsymbol{u}\rangle_m - \boldsymbol{\nabla} \boldsymbol{\cdot} \langle \boldsymbol{u} \rangle_m=  \langle \boldsymbol{n} \boldsymbol{\cdot}( \boldsymbol{u}_f - \boldsymbol{u}_e) \delta \rangle_m$.

The treatment of the boundary closure terms $\boldsymbol{d}$, $\boldsymbol{b}_\mu$ and $\varphi_{\text{closure}}$ is rarely discussed in the literature, although similar boundary closure terms or \textit{commutation errors} have been identified for LES filters, see for instance (\cite{Sagaut2001}).
Nevertheless, in order to consistently evaluate $\boldsymbol{b}_{\text{closure}}$ from the flow variables $\boldsymbol{u}_f$ and $p_f$, the latter closure terms must be rigorously defined through a careful choice of the velocity and pressure extensions $\boldsymbol{u}_e$ and $p_e$ outside of $\Omega$.
Therefore, in this work, $\boldsymbol{u}_e$, $p_e$, as well as $\boldsymbol{\tau}_e$, have been extrapolated from the inlet and outlet boundary conditions, in a direction normal to the boundary, such that for all $s>0$, we have $ \boldsymbol{u}_e(\boldsymbol{x}+s\boldsymbol{n}) = \boldsymbol{u}_f (\boldsymbol{x})$, $ p_e(\boldsymbol{x}+s\boldsymbol{n}) = p_f (\boldsymbol{x}) $ and $ \boldsymbol{\tau}_e(\boldsymbol{x}+s\boldsymbol{n}) = \boldsymbol{\tau}_f (\boldsymbol{x}) $ if $\boldsymbol{x}  \in (\Gamma_{\text{in}} \cup \Gamma_{\text{out}})$.
Furthermore, we define  $\boldsymbol{u}_e(\boldsymbol{r}) =0$, $\boldsymbol{\tau}_e(\boldsymbol{r}) =0$ and $p_e(\boldsymbol{r})=0$ if $\boldsymbol{r}  \notin \left\{ \boldsymbol{r}  \in \mathbb{R}^3 \vert \boldsymbol{r} =\boldsymbol{x} + s    , \boldsymbol{x}  \in (\Gamma_{\text{in}} \cup \Gamma_{\text{out}}), s>0 \right\}$.
That way, the macro-scale velocity becomes divergence-free as $\varphi_{\text{closure}}=0$, while the total closure force can be simplified into $ \boldsymbol{f}_{\text{closure}} = - \rho_f \boldsymbol{\nabla} \boldsymbol{\cdot} {\boldsymbol{M}} + \boldsymbol{b} $ with
$\boldsymbol{b} \triangleq \langle \boldsymbol{n}_{fs} \boldsymbol{\cdot}  (-p_f \boldsymbol{I} + \boldsymbol{\tau}_f) \delta_{fs} \rangle_m  + \langle \boldsymbol{n} \boldsymbol{\cdot}  (-p_f \boldsymbol{I} + \boldsymbol{\tau}_f) \delta_{\text{wall}} \rangle_m$.
This last expression is based on the Dirac surface indicator $\delta_{\text{wall}}$ of $\Gamma_{\text{wall}}$.
Eventually, we may write the closure force $\boldsymbol{b}$ in the more concise form (\ref{eq: final form macro-scale no-slip force}).

\section{Closure Problem for Developed Macro-Scale Flow}
\label{app: Closure Problem for Developed Macro-Scale Flow}
Substitution of the closure mapping $\boldsymbol{u}^{\star}(\boldsymbol{r}) = \closurevarvelperiodic(\boldsymbol{r}) \boldsymbol{\cdot} \boldsymbol{U}' $, $p^{\star}(\boldsymbol{r})= \mu_f \closurevarpressperiodic(\boldsymbol{r}) \boldsymbol{\cdot} \boldsymbol{U}'$ into the momentum equation and continuity equation for periodically developed flow (\cite{Buckinx2022}) gives rise to the following closure problem for developed macro-scale flow:
\begin{equation}
\label{eq: closure problem developed region}
\begin{aligned}
\rho_f \left(\closurevarvelperiodicf \boldsymbol{\cdot}  \boldsymbol{U}'\right) \boldsymbol{\cdot} \boldsymbol{\nabla} \closurevarvelperiodicf &= 
\mu_f \boldsymbol{K}_{\text{uniform}}^{-1}
- \mu_f\boldsymbol{\nabla} \closurevarpressperiodicf + 
\mu_f \nabla^2  \closurevarvelperiodicf \,, 
\\
\boldsymbol{\nabla}\boldsymbol{\cdot}  \closurevarvelperiodicf &= 0 \,,
\\
\closurevarvelperiodicf(\boldsymbol{x}) &= \closurevarvelperiodicf(\boldsymbol{x} + n_1 \boldsymbol{l}_1) \,, \qquad
\closurevarpressperiodicf (\boldsymbol{x}) = \closurevarpressperiodicf (\boldsymbol{x} + n_1 \boldsymbol{l}_1)\,, \\
\closurevarvelperiodicf(\boldsymbol{x}) & = 0 \qquad \mbox{for} ~~ \boldsymbol{x}  \in \Gamma_{0}  \,,
\\
\langle \closurevarvelperiodic \rangle^f(\boldsymbol{x}) &= \boldsymbol{I} \qquad \mbox{for} ~~ \boldsymbol{x} \in \Omega_{\text{uniform}}
\,.
\end{aligned}
\end{equation}
This closure problem relies on the equalities (\ref{eq: uniform interfacial force and permeability tensor}) and (\ref{eq: macro-scale momentum equation uniform}).

In the region of uniform macro-scale flow, the closure problem can be solved on a unit cell of the array, as $ \closurevarvelperiodicf(\boldsymbol{x}) =\closurevarvelperiodicf(\boldsymbol{x} + n_2 \boldsymbol{l}_2)$ for  $\boldsymbol{x} \in \Omega_{\text{uniform}}$, due to the periodicity of the velocity field (\ref{eq: flow periodicity transversal direction}).
This allows us to determine the relationship $\boldsymbol{K}_{\text{uniform}}\left(\boldsymbol{U} \right)$, as discussed in \S\ref{subsec: Local Closure for Developed Macro-Scale Flow}.

Once the relationship $\boldsymbol{K}_{\text{uniform}}\left(\boldsymbol{U} \right)$ is known, the former closure problem can be solved on one or two transversal rows in the region $\Omega_{\text{dev}} \setminus \Omega_{\text{uniform}}$, depending on whether $n_1=1$ or $n_1=2$.
It then yields the profile of the macro-scale velocity in the side-wall region, as its solution satisfies 
$\langle \closurevarvelperiodic \rangle^f_m = \boldsymbol{\xi}$ by definition (\ref{eq: shape developed macro-scale velocity profile}).
For channel flows which are symmetric with respect to the plane $x_2=W/2$, the closure problem would need to be solved only on a part of the side-wall region, $x_2 \in (0, \ell_{\text{sides}} + n_2 \ell_2)$, since $\closurevarvelperiodicf(\boldsymbol{x}) = \closurevarvelperiodicf(\boldsymbol{x} + n_2 \boldsymbol{l}_2)$ and $
\closurevarpressperiodicf(\boldsymbol{x}) = \closurevarpressperiodicf(\boldsymbol{x} + n_2 \boldsymbol{l}_2)$ for $x_2 > \ell_{\text{sides}}$.

\section{Closure Problem for Quasi-Developed Macro-Scale Flow}
\label{app: Closure Problem for Quasi-Developed Macro-Scale Flow}
When the closure mapping $\velamp(\boldsymbol{r}) = 
\boldsymbol{\Psi}(\boldsymbol{r}) \boldsymbol{\cdot} \boldsymbol{U}'$, $\pressamp(\boldsymbol{r}) = 
\mu_f \boldsymbol{\psi}(\boldsymbol{r}) \boldsymbol{\cdot} \boldsymbol{U}'$ (cf. (\ref{eq: closure mapping quasi-developed region})) is substituted into the momentum equation and continuity equation for quasi-periodically developed flow (\cite{Buckinx2022}), the following closure problem for quasi-developed macro-scale flow is obtained:
\begin{equation}
\label{eq: quasi-periodically developed flow equations - closure}
\begin{aligned}
\rho_f \boldsymbol{u}_f^{\star}  \boldsymbol{\cdot} \boldsymbol{\nabla} \boldsymbol{\Psi}_f +
\rho_f \boldsymbol{\Psi}_f \boldsymbol{\cdot} \boldsymbol{\nabla} \boldsymbol{u}_f^{\star} 
&= 
- \boldsymbol{\nabla} \boldsymbol{\psi}_f   + 
\mu_f \nabla^2  \boldsymbol{\Psi}_f +
\rho_f (\boldsymbol{u}_f^{\star}  \boldsymbol{\cdot} \boldsymbol{\lambda}) \boldsymbol{\Psi}_f\\
& \qquad - 
2\mu_f \boldsymbol{\lambda}  \boldsymbol{\cdot} \boldsymbol{\nabla}  \boldsymbol{\Psi}_f + 
\mu_f (\boldsymbol{\lambda}  \boldsymbol{\cdot} \boldsymbol{\lambda})  \boldsymbol{\Psi}_f +
\boldsymbol{\lambda} \boldsymbol{\psi}_f \,, \\
\boldsymbol{\nabla}\boldsymbol{\cdot}  \boldsymbol{\Psi}_f &= \boldsymbol{\lambda} \boldsymbol{\cdot} \boldsymbol{\Psi}_f \,, \\
\boldsymbol{\Psi}_f(\boldsymbol{x}) &= \boldsymbol{\Psi}_f(\boldsymbol{x} + n_1 \boldsymbol{l}_1) \,, \qquad
\boldsymbol{\psi}_f(\boldsymbol{x}) = \boldsymbol{\psi}_f(\boldsymbol{x} + n_1 \boldsymbol{l}_1)
\,, \\
\boldsymbol{\Psi}_f (\boldsymbol{x}) &= 0 \qquad \mbox{for} ~~ \boldsymbol{x}  \in \Gamma_{0}  \,. \\
\end{aligned}
\end{equation}
Due to the appearance of the periodically developed flow field $\boldsymbol{u}_f^{\star}  = \closurevarvelperiodic \boldsymbol{\cdot} \boldsymbol{U}' $, this closure problem can only be solved once the solution of the previous closure problem (\ref{eq: closure problem developed region}) is known.
In addition, the following constraint must be imposed to find a unique solution:
\begin{equation}
 \boldsymbol{e}_1    \boldsymbol{\cdot}  \langle \boldsymbol{\Psi} \rangle_{\text{row}} \boldsymbol{\cdot} \boldsymbol{U}'  = C_{\velamp} \,,
\end{equation}
in accordance with the definition of the perturbation size $C_{\velamp}$ given in (\cite{Buckinx2022}).

\section{Closure Mapping for Quasi-Developed Macro-Scale Flow}
\label{app: Closure Mapping for Quasi-Developed Macro-Scale Flow}
The closure problems (\ref{eq: closure problem developed region}) and (\ref{eq: quasi-periodically developed flow equations - closure}) yield an exact mapping for the deviation fields in the case of a matched filter:
\begin{align}
\devtilde{\boldsymbol{u}}_f &= \left( \devclosurevarvelperiodicf +  \devtilde{\boldsymbol{\Psi}}_f \exp(-\boldsymbol{\lambda} \boldsymbol{\cdot} \boldsymbol{x}) \right) \boldsymbol{\cdot} \boldsymbol{U}'\,,\\
\devtilde{p}_f &= \mu_f\left( \devclosurevarpressperiodicf +  \devtilde{\boldsymbol{\psi}}_f \exp(-\boldsymbol{\lambda} \boldsymbol{\cdot} \boldsymbol{x}) + \boldsymbol{m} \boldsymbol{\cdot}  \boldsymbol{K}_{\text{uniform}}^{-1} \right) \boldsymbol{\cdot}  \boldsymbol{U}' \,.
\end{align}
When we link this mapping to $\langle \boldsymbol{u} \rangle^f_m$ instead of $\boldsymbol{U}'$ via (\ref{eq: mapping macro-scale velocity mode to constant developed macro-scale velocity quasi-developed}), we find that the closure variables in the classical closure problem actually are given by 
\begin{align}
\devclosurevarvel_f &= \left( \devclosurevarvelperiodicf +  \devtilde{\boldsymbol{\Psi}}_f \exp(-\boldsymbol{\lambda} \boldsymbol{\cdot} \boldsymbol{x}) \right) \boldsymbol{\cdot} 
\left[
\boldsymbol{\xi} + \boldsymbol{\zeta} 
\exp \left(-\boldsymbol{\lambda} \boldsymbol{\cdot} \boldsymbol{x} \right) 
\right]^{-1} \,,
\\
\devclosurevarpress_f &= \left( \devclosurevarpressperiodicf +  \devtilde{\boldsymbol{\psi}}_f \exp(-\boldsymbol{\lambda} \boldsymbol{\cdot} \boldsymbol{x}) + \boldsymbol{m} \boldsymbol{\cdot}  \boldsymbol{K}_{\text{uniform}}^{-1} \right) \boldsymbol{\cdot}   \left[
\boldsymbol{\xi} + \boldsymbol{\zeta} 
\exp \left(-\boldsymbol{\lambda} \boldsymbol{\cdot} \boldsymbol{x} \right) 
\right]^{-1}\,.
\end{align}
So, we see that the classical closure problem is based on the assumption that the periodic closure variables are dominant:  $\devclosurevarvel_f \simeq \devclosurevarvelperiodicf$ and $\devclosurevarpress_f \simeq  \devclosurevarpressperiodicf$.

We remark that instead of the uniform macro-scale velocity $\boldsymbol{U}'$ also the constant pressure gradient $\boldsymbol{\nabla} P_{\text{dev}}$ could have been used to construct exact mappings for the deviation fields $\devtilde{\boldsymbol{u}}_f$ and $\devtilde{p}_f$ in quasi-developed macro-scale flow.
This equivalent approach, which would be in line with the closure problems from (\cite{Barrere1992, ValdesParada2021}), has been omitted in this work.

\section{Additional information}
\begin{figure}
\begin{center}
\includegraphics[scale=0.5]{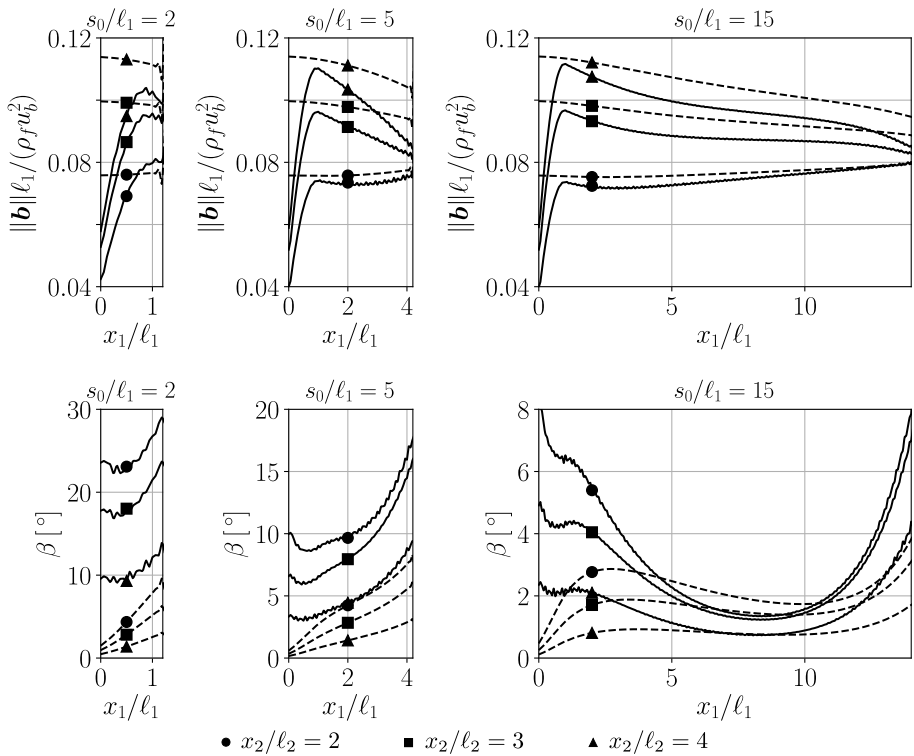}
\caption{closure force in the inlet region of a channel array with a porosity $\epsilon_f =0.9375$ ($N_2=10$, $H/\ell_1 =1$, $\ell_1/\ell_2 =1$) at a Reynolds number $\Rey=300$, for different positions of the first cylinder row $s_0$.
The solid lines ($-$) represent the actual no-slip force as obtained via direct numerical simulation and explicit filtering. 
The dashed lines (-$\,$-) represent the approximation $\boldsymbol{b}_{\Vert}$ for the closure force, in the case that the macro-scale flow would have been developed.}
\label{fig:Inlet closure force porosity 0.94}
\end{center}
\end{figure}

In figure \ref{fig:Inlet closure force porosity 0.94}, the closure force $\boldsymbol{b}$ is shown in the inlet region of the flows illustrated before in figure \ref{fig:Developing closure force comparison correlation porosity 0.93}.
The figure indicate that the macro-scale velocity field in the inlet region is determined by other modes than the one which dominates over the quasi-developed flow region.

\newpage
\section{Notes - To Do}
\subsection{Figures}
\begin{itemize}

\item Figure \ref{fig:Developed closure force uniform macro-scale flow}: Perhaps change $\Vert \boldsymbol{b} \Vert$ into $\Vert \boldsymbol{b}^{\star} \Vert$.

\item Figures \ref{fig: Quasi-developed macro-scale flow} (c) and figure \ref{fig:Developing closure force comparison correlation porosity 0.93}: Check whether $x_{\text{quasi-periodic}}$ is up to date with latest version of \citep{Buckinx2022}.

\end{itemize}

\subsection{Derivations and mathematics}
\begin{itemize}

\item Perhaps introduce new notation to distinguish the permeability tensor that depends on the local macro-scale velocity from the one that depends on the uniform macro-scale velocity: $\boldsymbol{K}_{\text{DEV}}(\langle\boldsymbol{u} \rangle_m) \triangleq \boldsymbol{K}_{\text{dev}}(\boldsymbol{\xi} \boldsymbol{\cdot} \boldsymbol{U}) $ and $\boldsymbol{K}_{\text{dev}} \triangleq \boldsymbol{K}_{\text{dev}}(\boldsymbol{U})$

\item Check whether the $ \langle\boldsymbol{\Lambda} \rangle^f_m = \boldsymbol{I} \xi^{-1}$, or whether only the contraction of $ \langle\boldsymbol{\Lambda} \rangle^f_m = \boldsymbol{I} \xi^{-1}$ on $\boldsymbol{U}_{\text{dev}}$ is a scalar.
Otherwise it might be better to introduce $\xi$ at a later stage, since it is not necessary to introduce it in the closure mapping.

\item Explain the link with Lasseux' closure problem for a boundary between a porous medium flow and a solid.

\item Perhaps we can also assume that $\boldsymbol{\nabla}\langle \boldsymbol{\Lambda}^{\star} \rangle^f_m (x_2) $ can be moved outside of the averaging operator in the definition of $\boldsymbol{K}_{\text{dev, main}}^{-1}$, so that $\boldsymbol{K}_{\text{dev, main}}^{-1} \simeq \boldsymbol{K}_{\text{uniform}}^{-1} + \epsilon_{fm}^{-1} \boldsymbol{\nabla} \epsilon_{fm} \boldsymbol{\cdot} \boldsymbol{\nabla} \boldsymbol{\xi}$, even though the last contribution is small anyway.

\item Verify that the mapping tensors $ \boldsymbol{\xi}$ and $ \boldsymbol{\zeta}$ are invertible (thus full rank, not projections).

\item Check current definition angle: Mention how $\beta$ is obtained for negative $\alpha$ 
Perhaps define angle of attack through $\alpha \triangleq \boldsymbol{e}_2 \boldsymbol{e}_s$, so that $\alpha <  0$ for $x_2 < N_2\ell_2/2$?
Also define $\beta$ to be negative for negative $\alpha$?

\end{itemize}

\subsection{Text}
\begin{itemize}

\item Check spelling: periodically-developed-flow region, periodically-developed-flow equations, midplane?

\item Mention discretisation and meshing aspects everywhere for all figures, also accuracy of discrete filter. Perhaps refer to DNS results from \citep{Buckinx2022}?

\item Perhaps mention possibility of row averaging to obtain 1D macro-scale flow and spatially independent permeability tensors?

\end{itemize}

\newpage

\bibliographystyle{jfm}
\bibliography{References}

\end{document}